\documentclass[12pt]{article}
\usepackage{amsmath,amsfonts,amssymb}

\usepackage{mathtools}
\usepackage{hyperref}
\usepackage{placeins}
\usepackage[bbgreekl]{mathbbol}
\usepackage{bbm,dsfont}
\usepackage{graphicx}
\usepackage{graphics}

\usepackage{color}
\usepackage{epsfig}
\usepackage{psfrag}	
\usepackage{tikz}
\usetikzlibrary{snakes}
\usepackage{slashed}
\usepackage{ulem}
\usepackage[font=small,labelfont=bf,width=1\textwidth]{caption}
\usepackage{comment}
\usepackage[all]{xy}
\usepackage{multirow}
\usepackage{float}
\usepackage{tensor}
\usepackage{multirow}
\usepackage{booktabs}
\usepackage{fancyhdr}
\usepackage{upgreek}
\usepackage[T1]{fontenc}
\usepackage[numbers,sort&compress]{natbib}

\usepackage{color}

\newcommand \be{\begin{eqnarray}}
\newcommand \ee{\end{eqnarray}}
\newcommand{\ft}[2]{{\textstyle\frac{#1}{#2}}}
\numberwithin{equation}{section}

\def\citare#1{{\color{blue} [citare: #1]}}

\def\ie{\textit{i.e.} }
\def\eg{\textit{e.g.} }

\DeclareMathOperator{\Tr}{Tr}

\DeclareMathOperator{\deltetha}{\partial_{\theta}}
\DeclareMathOperator{\deltethabar}{\Bar{\partial}_{\Bar{\theta}}}

\def\so{{\mathfrak{so}}}
\def\osp{{\mathfrak{osp}}}
\def\su{{\mathfrak{su}}}
\def\u{{\mathfrak{u}}}
\def\bF{{\mathbb{F}}}

\def\bG {\mathbb{G}}

\def\bD {\mathbb{D}}
\def\bP {\mathbb{P}}
\def\bO {\mathbb{O}}
\def\bR {\mathbb{R}}

\def\bT {\mathbb{T}}
\def\bZ {\mathbb{Z}}

\def\blambda{\mathbb{\Lambda}}

\def\bmF\mathbbm{F}
\def\bmD\mathbbm{D}

\newcommand{\gG}{{\boldsymbol G}}

\newcommand{\gP}{{\boldsymbol P}}
\newcommand{\gQ}{{\boldsymbol Q}}

\newcommand{\bea}{\begin{eqnarray}}
\newcommand{\eea}{\end{eqnarray}}
\newcommand{\beq}{\begin{equation}}
\newcommand{\eeq}{\end{equation}}
\newcommand{\bal}{\begin{equation}\begin{aligned}}
\newcommand{\eal}{\end{aligned} \end{equation}}

\newcommand{\vev}[1]{{\left< {#1} \right>}}

\newcommand{\address}[1]{\vbox{\center\em#1}}
\renewcommand{\title}[1]{\vbox{\center\huge{#1}}\vspace{5mm}}

\newcommand{\ii}{\mathrm{i}}
\newcommand{\ji}{\mathrm{j}}
\newcommand{\ki}{\mathrm{k}}
\newcommand{\li}{\mathrm{l}}

\newcommand{\cC}{{\mathcal C}}
\newcommand{\cD}{{\mathcal D}}

\newcommand{\cL}{{\mathcal L}}
\newcommand{\cM}{{\mathcal M}}
\newcommand{\cN}{{\mathcal N}}
\newcommand{\cP}{{\mathcal P}}

\newcommand{\cT}{{\mathcal T}}
\newcommand{\cR}{{\mathcal R}}
\newcommand{\cO}{{\mathcal O}}

\newcommand{\cW}{{\mathcal W}}
\newcommand{\cI}{{\mathcal I}}

\newcommand{\cZ}{{\mathcal Z}}

\renewcommand{\sl}{\mathfrak{s}\mathfrak{l}}

\newcommand{\ec}{\,,}

\renewcommand{\a}{\alpha}

\newif\ifshowexcursus

\begin{document}

\begin{titlepage}
\begin{center}

\vspace*{20mm}

\title{Universal sectors in superconformal defects}

\vspace{5mm}

\renewcommand{\thefootnote}{$\alph{footnote}$}

Riccardo Giordana Pozzi

\vskip 6mm

\address{
Departamento de Física Matemática, Instituto de Física, Universidade de São Paulo,\\
Rua do Matão 1371, São Paulo, SP 05508-090, Brazil\\ 
\&\\
Dipartimento di Scienze Fisiche, Informatiche e Matematiche, \\
Universit\`a di Modena e Reggio Emilia, via Campi 213/A, 41125 Modena, Italy\\
\&\\
INFN, Sezione di Bologna, via Irnerio 46, 40126 Bologna, Italy}

\vskip 5mm

\tt{riccardo.pozzi@unimore.it}

\renewcommand{\thefootnote}{\arabic{footnote}}
\setcounter{footnote}{0}

\end{center}

\vspace{8mm}
\abstract{
\normalsize{
\noindent
We study universal features of defect correlation functions in supersymmetric defect CFTs, focusing on four-point functions of the displacement supermultiplet. By perturbing the leading-order correlators at strong coupling, we derive constraints that identify the operators exchanged at next-to-leading order. From this, we determine the conditions under which four-point functions of defect insertions across different theories are equivalent, thereby establishing universality.
We confirm these features in examples including the $1/2$ BPS line in $\mathcal{N}=4$ SYM, $\mathcal{N}=2$ gauge theories and ABJM, up to the first subleading order in the strong-coupling expansion. We further analyze the $1/2$ BPS line in 3d $\mathcal{N}=2$ Chern-Simons-matter theories, identifying the preserved defect algebra and the superdisplacement multiplet. Exploiting universality, we find the superdisplacement four-point functions of defect operators and extract the conformal data at first subleading order. Finally, we comment on the $1/3$ BPS Wilson line in ABJM, presenting initial results from the strong-coupling analysis.

}}
\vfill

\end{titlepage}
\tableofcontents

\section{Introduction and summary}

The study of conformal field theories (CFTs) in the presence of extended defects has developed over the years into a thriving field, providing a powerful framework for modeling systems with applications across many areas of physics. Defects, which may appear as dynamical extended operators, generically break part of the symmetry of the bulk CFT while preserving a residual conformal subgroup. This structure defines a defect conformal field theory (dCFT) \cite{Billo:2016cpy, drukker}, describing the localized degrees of freedom on the defect and their interactions with the bulk CFT. The discussion is further enriched when supersymmetry is present. Superconformal defects, which preserve a portion of the bulk supercharges, exhibit enhanced protection of their data, such as operator dimensions and operator product expansion (OPE) coefficients, and impose stronger constraints on their correlators.

One of the most notable realizations of a line defect is the Wilson line in supersymmetric gauge theories \cite{Maldacena:1998im, Drukker:1999zq, Drukker:2006xg, Rey:1998ik,Drukker:2011za}, which provides a powerful laboratory for probing nonperturbative dynamics at strong coupling. In fact, when a holographic dual is available, correlation functions of local operator insertions along the line can be computed using Witten diagrams \cite{Giombi:2017cqn}. A well-known example is the $1/2$ BPS Wilson line in $\mathcal{N}=4$ super-Yang-Mills (SYM), where the bremsstrahlung function of the theory is determined by the coefficient of the two-point functions of protected operator insertions on the Wilson line \cite{Correa:2012at, Drukker:2011za}. Further insight into these operators has been gained through the computation of their four-point functions \cite{Giombi:2017cqn}. In the strong coupling regime, these correlation functions can be evaluated via the AdS/CFT correspondence \cite{Maldacena:1997re}. Here, the dual string model corresponds to an effective field theory on $\mathrm{AdS}_2$, enabling the calculation of correlation functions using Witten diagrams \cite{Giombi:2017cqn}.

However, a holographic realization of Wilson lines is not always available, and even in cases where it is known, computations via Witten diagrams quickly become demanding at higher orders. To address the limitations of standard perturbative holographic computations, which depend on detailed knowledge of the dual theory, the conformal bootstrap program \cite{Rattazzi:2008pe,El-Showk:2012cjh} has been extended to $1d$ dCFTs \cite{Ferrero:2019luz}. By constructing an ansatz and enforcing constraints such as crossing symmetry, internal consistency, and mild growth conditions on anomalous dimensions at large scaling dimensions \cite{Liendo:2018ukf, Heemskerk:2009pn, Fitzpatrick:2011dm}, one gains access to correlation functions at strong coupling. Moreover, this approach can be algorithmically applied to study higher-order subleading corrections.

The $1/2$ BPS Wilson line in $\cN=4$ SYM was the first case studied through an analytical bootstrap approach, carried out up to next-to-leading order (NLO) in the strong-coupling expansion \cite{Liendo:2018ukf}. This laid the groundwork for a more detailed understanding of Wilson lines in this theory \cite{Ferrero:2021bsb, Ferrero:2023znz, Ferrero:2023gnu} and paved the way for numerous applications to other defects, including the $1/2$ BPS Wilson line in $\cN=2$ gauge theories \cite{Gimenez-Grau:2019hez}. Then, the same framework was extended to lower-dimensional settings, such as the $1/2$ BPS Wilson line in ABJM theory \cite{Bianchi:2020hsz} and in $\cN=4$ Chern-Simons-matter (CSm) theories \cite{Pozzi:2024xnu}, which naturally serve as lower-dimensional analogs of the higher-dimensional examples.\newline

In this work, we investigate universal features in four-point functions of defect insertions and delineate the conditions under which such universality can be identified. By universal we mean that the expression for a four-point function, up to the tensorial structures given by the representations under the preserved symmetry, takes the same form regardless of the specific theory to which the defect operators belong.

The expectation of universal features comes from the fact that many dCFTs have a very similar structure when studied at strong coupling. A key role is played by correlation functions of the displacement supermultiplet, which contains the basic operators tied to the broken (super)currents. Because these operators are connected to the bulk currents through Ward identities, their scaling dimensions are protected. In the case of line defects, the displacement operator associated with broken transverse translations always has $\Delta = 2$. 

At weak coupling, these defect operators can be directly related to the fields in the weak-coupling Lagrangian. At strong coupling, however, this connection is no longer accessible, and the defect operators are characterized only by their quantum numbers under the preserved symmetry. This shows that, at strong coupling, the particular weak-coupling realization of the bulk theory does not play a central role. The dependence on the particular theory is captured entirely by certain constants that encode information about the bulk dynamics. One example is the overall normalization of the two-point function of the superdisplacement, which is related to the bremsstrahlung function \cite{Correa:2012at}.

A second observation suggesting universality is that, at leading order in strong coupling, all these defect Wilson lines are described by a generalized free field (GFF), as consequence of the large $N$ limit \cite{tHooft:1973alw}. It then follows that all four-point functions of operators with a given scaling dimension take the same functional form in a GFF, up to possible differences in tensor structures.

Thus, the interplay between protected conformal dimensions, the loss of weak coupling realizations at strong coupling, and the emergence of a GFF description naturally leads to the universality of defect correlators at leading order.

Evidence that this general structure persists at subleading orders comes from the study of the dual string theory realizations of these setups. In particular, when analyzing the bulk string description of Wilson lines via Witten diagrams up to NLO, one finds that the structures arising from the tree-level expansion of the Nambu-Goto action are equivalent. This is, for example, the case for the $1/2$ BPS Wilson line in SYM and ABJM, \cite{Giombi:2017cqn, Bianchi:2020hsz}. A concrete example is provided by the four-point function of the orthogonal fluctuations, which are dual to displacement insertions. In these cases, the tree-level interactions never mix with the elementary excitations associated with fluctuations along the compact manifold. This shows that different theories produce equivalent structures, \ie equivalent four-point functions, see Sec. \ref{sec:The holographic perspective}.

Motivated by this evidence, we investigate how far these observations can be extended to more general cases. Since holographic arguments are not always available, we focus on the study of universal sectors purely from the dCFT perspective, finding, interestingly, that many defect theories exhibit universality.

\subsection*{Results}

Most of our analysis relies on the observation that, when perturbing around a given leading-order theory, the spectrum of operators that can be exchanged at the first subleading order can be constrained by studying mixed correlation functions. We briefly introduce here the line of reasoning that is discussed completely in Sec. \ref{subsec: Bootstrapping the epsilon-order of OPE Coefficients}. In this regard, we consider a CFT$_1$ toy model with two protected operators, that we call $\bD(t)$ and $\bO(t)$, and towers of higher-particle operators built by these fundamental words, \eg $[\bD\bD]^{L=2}_n$. As we will see these will play the role of defect operators. 
We are interested in studying the four-point function of a given operator such as

\begin{equation*}
\langle\bD(t_1)\bD(t_2)\bD(t_3)\bD(t_4)\rangle = \frac{C^2}{t_{12}^{2\Delta_\bD}t_{34}^{2\Delta_\bD}}f_{\bD^4}(z)\,,
\end{equation*}
where $C$ is an overall normalization constant and $z=\frac{t_{12}t_{34}}{t_{14}t_{32}}$ is  the cross-ratio. We also assume, for later convenience, that the function $f(z)$ is expressed as an expansion about a given leading-order theory by
\begin{equation*}
    f(z)=f_{\bD^4}^{(0)}(z)+\epsilon\,f_{\bD^4}^{(1)}(z) + \cO(\epsilon^2)\,,
\end{equation*}
where $f^{(0)}(z)$ represent the leading-order theory and $f^{(1)}(z)$ represent a small deviation from that. In the following, we take $f^{(0)}(z)$ to be that of a GFF, which thus allows us to consider

\begin{equation*}
    f_{\bD^4}^{(0)}(z) = \sum_\Delta (\lambda^{(0)}_\Delta)^2 \,g_\Delta(z)\,,
\end{equation*}
where $g_\Delta(z)$ are the $1d$ conformal blocks and $\lambda^{(0)}_\Delta = \lambda^{(0)}_{\bD\bD\bT_\Delta}$ denote the OPE coefficients associated with the exchanged operators, here generically called $\bT_\Delta$. The spectrum of operators that can appear at this order can be easily identified by performing Wick contractions, which show that only $[\bD\bD]^{L=2}_n$ operators are exchanged at leading-order. One can then move to the more interesting case in which interactions are included. In this case, the OPE coefficients receive next-to-leading-order corrections parametrized by $\lambda^{(1)}_\Delta$ and the conformal dimensions acquire anomalous contributions $\gamma_\Delta$. The conformal block expansion of $f^{(1)}(z)$ can therefore be obtained by perturbing the GFF expansion, leading to
\begin{equation*}
    \epsilon\, f_{\bD^4}^{(1)}(z)=\sum_\Delta\Bigl(2\lambda_\Delta^{(0)}\lambda_\Delta^{(1)} + (\lambda_\Delta^{(1)})^2\Bigr)g_\Delta(z)+(\lambda_\Delta^{(0)})^2\,\gamma^{(1)}_\Delta(g_\Delta(z)\log(-z)+\mathrm{g}_\Delta(z))\,,
\end{equation*}
with $\mathrm{g}_\Delta(z)$ representing a particular derivative contribution of the conformal block $g_\Delta(z)$, \eqref{eq: derivative of the block}.

However, at this order the identification of the exchanged operator is not straightforward. In this respect, the issue can be addressed by analyzing mixed four-point functions and reinterpreting the resulting expressions as constraints on the order in $\epsilon$ of the OPE coefficients. We introduce this idea by studying the $1d$ mixed four-point function of $\bD$ and $\bO$. At NLO one finds
\begin{equation*}
    \epsilon\, f_{\bD^2\bO^2}^{(1)}(z)=\sum_\Delta(\lambda_\Delta^{(0)}\tilde\lambda_\Delta^{(1)}+\lambda_\Delta^{(1)}\tilde\lambda_\Delta^{(0)})g_\Delta(z) + \lambda_\Delta^{(1)}\tilde\lambda_\Delta^{(1)}g_\Delta(z)+\lambda_\Delta^{(0)}\tilde\lambda_\Delta^{(0)}\gamma^{(1)}_\Delta(g_\Delta(z)\log(-z)+\mathrm{g}_\Delta(z))\,,
\end{equation*}
arising from perturbing the leading order conformal data. In order to distinguish the OPE coefficients, we further introduced $\tilde\lambda^{(i)}_\Delta=\lambda_{\bO\bO\bT_{\Delta}}^{(i)}$.
As anticipated, this expression can be read as a constraint on the OPE data, allowing us to identify the minimum order in $\epsilon$ at which these contributions can appear consistently with the expansion.

For instance, consider a single contribution associated with a particular $\bT_{\Delta^\star}$ such that the two leading OPE data satisfy $\lambda_{\Delta^\star}^{(0)} = 0$ while $\tilde\lambda_{\Delta^\star}^{(0)} \neq 0$. This is the case, for example, for the multiparticle operators $[\bO\bO]^{L=2}_n$. In this situation, consistency requires $\lambda_{\Delta^\star}^{(1)} \sim \mathcal{O}(\epsilon)$, as follows from matching the orders in $\epsilon$ on both sides of the relation. We can thus, in a sense, bootstrap the order of the individual OPE coefficients.

Although this conclusion may appear innocuous, this information becomes particularly valuable when inserted into the expansion of $\epsilon\, f_{\bD^4}^{(1)}(z)$ discussed above.
This is because, given the hypothesis $\lambda_{\Delta^\star}^{(0)} = 0$, the only contribution that can appear associated to the exchange of $\bT_{\Delta^\star}$ is the quadratic term $(\lambda_{\Delta^\star}^{(1)})^2$. Since we have identified $\lambda_{\Delta^\star}^{(1)} \sim \mathcal{O}(\epsilon)$, such contributions are of order $\mathcal{O}(\epsilon^2)$ and therefore do not appear up to NLO.
See Sec. \ref{subsec: Bootstrapping the epsilon-order of OPE Coefficients} for a more detailed discussions.

This approach provides a powerful tool for identifying universal properties within a given system. At the same time, it is not restricted to the specific setup considered here and can be easily extended and applied to other contexts. We therefore expect it to remain applicable in more general settings and to provide useful insights elsewhere as well, as discussed later.
In the present case, we apply these methods to supersymmetric defect CFTs at strong coupling. Under additional assumptions, one can show that four-point functions built from repeated insertions of a given component of a supermultiplet do not exchange operators from different families up to NLO.

In this sense, the correlator factorizes: in a four-point function with identical external operators---such as four tilt or four displacement operators---the exchanged spectrum consists only of multiparticle operators constructed from the external one.

Consider, for example, two defect operators with the same protected scaling dimension in two different theories. If their four-point functions receive no contributions from multiparticle operators or from fundamental operators associated with other fields in the respective supermultiplet, the resulting four-point structures would be equivalent, provided these have similar three-point interactions. In this sense, the corresponding correlation functions are universal.
See Sec. \ref{subsec: Correlators of the displacement multiplet} for the full discussion.

Beyond providing the physical insight of uncovering such features, the main advantage of identifying universal sectors is that it allows for the determination of a four-point function up to NLO directly, bypassing the standard bootstrap strategy of postulating and constraining an ansatz. Once universal sectors have been identified, subsequent analyses can build on established results rather than rederiving them from first principles. This way, it is possible to obtain four-point functions even in systems that cannot be fully constrained, either due to limited supersymmetry or an insufficient set of constraints. For recent developments in this regards see \cite{Girault:2025kzt,Drukker:2025dfm}.

Moreover, identifying the parameter $\epsilon$ is in general a complicated task, as it cannot be determined by bootstrap arguments alone. Its identification can therefore be achieved only by explicitly matching to holographic theories, when available \cite{Liendo:2018ukf, Bianchi:2020hsz}, or alternatively through purely CFT-based arguments \cite{Drukker:2022pxk, Gabai:2025zcs}. In the present case, we exploit the identification of the universal sector to determine the form of the parameter $\epsilon$ by relying on previously studied examples in which it has already been fixed. Indeed, once a correlation function is identified in a given case, all correlators within the same universal class can be straightforwardly determined. In this way, the problem of explicitly computing these quantities is reduced to demonstrating that two different theories belong to the same universal class.

We verify these arguments by showing that universal features are indeed present in a variety of known defect theories. When the defects have the same codimension, the identification for the four-point function of the displacement is straightforward. All the others follows by requiring consistency with supersymmetry. For instance, the displacement four-point functions of the $1/2$ BPS line in ABJM \cite{Bianchi:2020hsz} and in $\mathcal{N}=4$ CSm theories \cite{Pozzi:2024xnu} fall within the same universality class, as shown in Sec. \ref{subsec: universality ABJM N4 SCSm Th}. Up to NLO one finds the  universal expression
\begin{align*}
\langle\mathbb{D}(t_1)\bar{\mathbb{D}}(t_2)\mathbb{D}(t_3)\bar{\mathbb{D}}(t_4)\rangle_{\cT}=&
\frac{(\mathbb{n}_{\cT} C_{\cT})^2}{t_{12}^4 t_{34}^4}\, 
\Bigl[1+z^4+\frac{1}{4\pi^2 C_{\cT}}\Bigl(-16 -2z -\frac{7z^2}{3} -2z^3-16z^4
\cr
+&\Bigl(6 - \frac{16}{z}+6 z^4 -16z^5\Bigr)\log(1-z) + 2z^4 (8z-3)\log(-z)\Bigr)\Bigr]\,,
\end{align*}
where now $\bD(t)$ is the actual displacement operator while $C_\cT$ and $\mathbb{n}_\cT$ are theory-specific constants, related to the particular bremsstrahlung function and to the preserved supersymmetry, respectively. By $\cT$ we mean the $1/2$ BPS line ether in ABJM or in $\cN=4$ CSm theories. 

When considering defects of different codimension, one must also account for the distinct representation of the displacement operator under the preserved rotational symmetry. Once these considerations are included, one can relate correlation functions of defect theories with different codimensionality, such as the $1/2$ BPS Wilson line in ABJM and in $\cN=4$ SYM. Explicitly, one can consider the four-point displacement insertions on the Wilson line in SYM
\begin{equation*}
    \langle\mathbb{D}^{i}(t_1){\mathbb{D}}^{j}(t_2)\mathbb{D}^{k}(t_3){\mathbb{D}}^{l}(t_4)\rangle_{\cW_\text{SYM}} = \frac{\cC_{\text{SYM}}^2}{t_{12}^4t_{34}^4}G^{ijkl}(z)\,,
\end{equation*}
where $i,j,k,l=1,2,3$ and all the overall constant and the superprimary normalization $C_{\text{SYM}}$ have been collected in $\cC_{\text{SYM}}$. By studying the function up to NLO \cite{Giombi:2017cqn, Liendo:2018ukf} one finds
\begin{align*}
    G^{ijkl}(z)=&\Bigl[\delta^{ij}\delta^{kl} \,G_S^{(0)}(z)+(\delta^{ij}\delta^{kl}-\delta^{il}\delta^{jk}) G_A^{(0)}(z) + (\delta^{ik}\delta^{jl}+\delta^{il}\delta^{jk}-2/3\,\delta^{ij}\delta^{kl})G_T^{(0)}(z)+\\
    +&\frac{1}{4\pi^2 C_{\text{SYM}}}\Bigl(\delta^{ij}\delta^{kl} \,G_S^{(1)}(z)+(\delta^{ij}\delta^{kl}-\delta^{il}\delta^{jk}) G_A^{(1)}(z) + (\delta^{ik}\delta^{jl}+\delta^{il}\delta^{jk}-2/3\,\delta^{ij}\delta^{kl})G_T^{(1)}(z)\Bigr)\Bigr]\,.
\end{align*}
By properly dealing with the tensorial structure, \ie taking the embedding, one obtains an expression for the displacement insertions in the ABJM Wilson line given by
\begin{align*}
    \langle\mathbb{D}(t_1)\bar{\mathbb{D}}(t_2)\mathbb{D}(t_3)\bar{\mathbb{D}}(t_4)\rangle_{\cW_\text{ABJM}} =& \frac{\cC_{\text{ABJM}}^2}{t_{12}^4t_{34}^4}\Bigl[G_S^{(0)}(z)-G_A^{(0)}(z)+\frac{G_T^{(0)}(z)}{3}+\nonumber\\
    +&\frac{1}{4\pi^2 C_{\text{ABJM}}}\Bigl(G_S^{(1)}(z)-G_A^{(1)}(z)+\frac{G_T^{(1)}(z)}{3}\Bigr)\Bigr]\,,
\end{align*}
which, when explicitly evaluated, agrees with the previous result as well as with the findings of \cite{Bianchi:2020hsz}, as discussed in more details in Sec. \ref{subsec: Universality in 1/2 BPS wilson line N=4 SYM}.

Although the above arguments were illustrated for the displacement operator, they extend to any four-point function of the supermultiplet components. An especially interesting case arises when considering the four-point function of the superprimary of the $1/2$ BPS Wilson line in $\cN=2$ gauge theories in $4d$, discussed in \cite{Gimenez-Grau:2019hez}. In this setup, universality does not extend to the entire displacement supermultiplet, but one can still identify subsectors in which it applies. In Sec. \ref{subsec: universal sector in N=2 4d}, we show that this is the case for the superprimary four-point function. By showing that this correlator does not receive contributions from other sectors up to NLO, we are able to identify it explicitly. Furthermore, we find consistency with the results of \cite{Gimenez-Grau:2019hez} by fixing the free parameters left undetermined in the bootstrap approach to 
\begin{equation*}
    c_1 = c_2 = -4\,,
\end{equation*}

thereby fully determining the superprimary four-point function to be
\begin{align*}
        \langle\mathbb{P}(t_1){\mathbb{P}}(t_2)\mathbb{P}(t_3)&{\mathbb{P}}(t_4)\rangle_{\cW_{\cN=2}} = \frac{C_{\Phi}^2}{t_{12}^2 t_{34}^2}\Bigl[1+z^2+\frac{z^2}{(z-1)^2}+\Bigl(\frac{1}{\pi^2 C_{\Phi}}\Bigr)\Bigl(-\frac{(1-z+z^2)^2}{(z-1)^2}+\nonumber\\
        &+(-4+2z+z^3-4z^4)\frac{\log(1-z)}{4z}+\frac{z^2(2-4z+9z^2-7z^3+2z^4)}{2(z-1)^3}\log(-z)\Bigr)\Bigr]\,.
\end{align*}
Through these explicit values, we finally fix the anomalous dimensions \cite{Gimenez-Grau:2019hez} to
\begin{align}
    \gamma^{[0,0]}&=4-\Delta-\Delta^2\nonumber\,,\\
    \gamma^{[1,0]} &= 2-\Delta-\Delta^2\nonumber\,.
\end{align}

In Sec. \ref{subsec: the 1/2 BPS line in AdS3 T3 S4}, we use a similar approach to identify a universal sector for the $1/2$ BPS line in $AdS_3 \times S^3 \times T^4$ \cite{Bliard:2024bcz, Correa:2021sky} with purely Ramond-Ramond flux, finding consistency with previous analysis and fixing the overall two-point normalization of the displacement supermultiplet in terms of the bulk coupling.

We then provide brief general arguments supporting universality from the holographic perspective, in Sec. \ref{sec:The holographic perspective}.

In line with previous analyses of $1/2$ BPS Wilson lines at strong coupling in $3d$ SCFTs, such as ABJM \cite{Bianchi:2020hsz} and $\cN=4$ CSm theories \cite{Pozzi:2024xnu}, we investigate the case of $\cN=2$ CSm theories, in Sec. \ref{sec: The 1/2 BPS line in N=2 Chern-Simons matter theories}. We examine the preserved defect algebra, identify the superdisplacement multiplet ($\cD$), and employ superspace methods to express the four-point functions in terms of cross-ratios. We study the OPE structure for the superdisplacement in the two different channels: the chiral-chiral $\cD\times\cD$ channel, which contains one short multiplet and infinitely many long operators with unprotected, higher dimensions, and the chiral-antichiral $\cD\times\bar\cD$ channel, where only long multiplets appear \cite{Bianchi:2020hsz, Bianchi:2018scb}. We thus derive the corresponding superblocks by diagonalizing the superconformal Casimir.

By identifying the universal sectors, we determine the four-point functions of the superdisplacement up to leading- and next-to-leading order at strong coupling. For the superprimary we find
\begin{align*}
\langle\mathbb{\Lambda}(t_1)\bar{\mathbb{\Lambda}}(t_2)\mathbb{\Lambda}(t_3)\bar{\mathbb{\Lambda}}(t_4)\rangle_{\cW_{\cN=2\,CSm}}=&
\frac{C_{\cD}^{2}}{t_{12}^3 t_{34}^3}\, 
\Bigl[1-z^3+\frac{1}{4\pi^2C_\cD}\Bigl(-9+\frac{z}{2}-\frac{z^2}{2}+9z^3 +\nonumber\\
+&\Bigl( 5-\frac{9}{z}-5z^3+9z^4\Bigr)\log(1-z) +(5-9z)z^3\log(-z)\Bigr)\Bigr]\,,
\end{align*}
with $C_\cD$ the normalization of the superdisplacement two-point function.
We then extract the CFT data of the exchanged contributions. For the long multiplets appearing in the chiral-antichiral channel, we find
\begin{equation*}
    \Delta_n=3+n-\frac{ 5+6n+n^2}{4\pi^2 C_\cD}\,,
\end{equation*}
where $n$ is associated with the derivative contributions of the GFF length-two multiparticle states $[\blambda\bar\blambda]_{n}^{L=2}$. For the chiral-chiral channel we find instead
\begin{equation*}
    \Delta_n=3+n-\frac{4+5n+n^2}{4\pi^2 C_\cD}\,,\qquad \qquad n\quad\text{odd}\,,
\end{equation*}
with $n$ associated with the multiparticle states $[\blambda\blambda]_{n}^{L=2}$.

Sec. \ref{Bootstrapping the 1/3 BPS Wilson line in ABJM}  concludes this work with preliminary results for the $1/3$ BPS Wilson line in ABJM at strong coupling, discussed in detail in \cite{Drukker:2022txy}. In particular, we argue that universal features can also be identified in this case, allowing for the determination of the four-point functions of the displacement and tilt (tlit) supermultiplets, respectively. The analysis of the displacement will be equivalent to that of the Wilson line in $\cN=4$ CSm theory. Therefore, we focus on the tilt multiplet which four-point function is specified by that of the superprimary
\begin{align*}
    \langle \mathbbm{F}(t_1)\bar{\mathbbm{F}}(t_2)\mathbbm{F}(t_3)\bar{\mathbbm{F}}(t_4)\rangle_{\cW_{1/3}}=\frac{C_{T}^2}{t_{12}t_{34}}f(z)\,.
\end{align*}
Exploiting universality we can identify $f(z)$ up to NLO to be
\begin{align*}
    f(z)=\Bigl[1-z+\frac{1}{4\pi^2 C_{T}}\Big(z-1+z (3-z) \log (-z)-\frac{(1-z)^3 }{z}\log (1-z)\Big)\Bigr]\,,
\end{align*}
where $C_T$ is the tilt two-point normalization, \cite{Drukker:2022txy}. While the expansion in the chiral-chiral channel can be related to the results of the ABJM superprimary \cite{Bianchi:2020hsz}, being in the same universal class, the chiral-antichiral lead to a new expression, essentially due to the different preserved algebra, and hence different superblocks. Extracting the conformal data, we find for the long multiplets
\begin{equation*}
    \Delta_n= 1+n-\frac{3-3n-n^2}{4\pi^2C_T}\,,
\end{equation*}
with $n$ the label of the GFF multiparticle states $[\mathbbm{F} \bar{\mathbbm{F}}]_n^{L=2}$.

\subsection*{Outlook}

A natural extension of this work would be to study higher-point correlation functions at strong coupling. By applying procedures similar to those discussed here, one can investigate the existence of universal sectors and relate the $n$-point functions of different theories in a similar ways as discussed for four-point functions. This approach has the advantages of being computationally convenient, avoiding some of the difficulties of direct bootstrap computations, while also providing access, in principle, to a variety of interesting configurations also for cases in which the bootstrap would not lead to a completely fixed answer.

While universality is not expected to hold beyond NLO, it is still plausible that higher-order corrections to four-point functions contain universal contributions mixed with theory-specific ones. In such cases, one can decompose a given correction into a universal part and a nonuniversal remainder. Identifying and isolating these universal sectors can provide valuable input for bootstrapping higher-order four-point functions.  

From an applications perspective, a natural next step is to apply these techniques and the identification of universal sectors to a broader range of line defects, such as lower-BPS examples, with the aim of analyzing their four-point functions up to NLO in the strong coupling expansion.

Finally, while the present discussion has focused primarily on Wilson lines and line defects, the underlying philosophy of constraining the order of possible exchanges is more general and can be generalized and applied to higher-dimensional defects, as, for example, surface defects \cite{Drukker:2020swu}, or perhaps even to strongly coupled CFTs beyond defect theories. 

We leave these questions for future investigations.\\

The paper is organized as follows. Sec. \ref{sec: The CFT perspective} introduces the general structures of supersymmetric defect CFTs and sets the stage for constraining the order in which OPE coefficients enter the block expansion.
In Sec. \ref{subsec: Correlators of the displacement multiplet} we discuss what are the assumptions under which a defect CFT may feature universal properties. We comment on the universal sectors when relaxing some of these assumptions and, finally, we provide the structure for the full NLO order contributions. 
We verify the universal property in a wide spectrum of supersymmetric defect CFTs realized by BPS Wilson lines in Sec. \ref{sec: Universal sectors in 1/2 BPS Wilson lines} and we give a brief argument of universality from an holographic  point of view in Sec. \ref{sec:The holographic perspective}.
We conclude the work with a full analysis regarding the $1/2$ BPS Wilson line in Chern-Simons theories in Sec. \ref{sec: The 1/2 BPS line in N=2 Chern-Simons matter theories} and in Sec. \ref{Bootstrapping the 1/3 BPS Wilson line in ABJM} we initiate the study of defect correlators of the $1/3$ BPS Wilson lines in ABJM at strong coupling. Appendix \ref{app:algebra} contains further details on the $\osp(2\vert4)$ superalgebra, the symmetry-breaking pattern associated with the $1/2$ BPS line, and the orthogonality conditions entering the (super)block expansions.

\section{Defect correlation functions}
\label{sec: The CFT perspective}
We begin by setting our conventions and outlining the general structure of defect supermultiplets that typically arise for $1/2$ BPS line defects. In particular, we focus on their realizations as Wilson lines and review the features common to this class of defects.

\subsection{Wilson lines as defect CFTs}
\label{subsec: Wilson lines as defect CFTs}
Wilson lines are a central example of line defects in supersymmetric CFTs, and are generically defined at weak coupling by
\begin{equation*}
    \cW = \Tr\Bigl[\cP\exp\Bigl(-i\int_{-\infty}^{\infty} \cL(t)dt\Bigr)\Bigr]\,,
\end{equation*}
where $\cL(t)$ denotes a theory-dependent (super)connection\footnote{In $3d$ theories, constructing $1/2$ BPS Wilson lines requires superconnections. For instance, in ABJM the $1/2$ BPS Wilson line is defined as the holonomy of a  $\u(N|N)$ superconnection \cite{Drukker:2009hy}. For other realizations, \eg in $\cN=4$, see \cite{Cooke:2015ila, Drukker:2020dvr, Drukker:2022bff} } and $t=x^0\in(-\infty,\infty)$ the Euclidean time direction parametrizing an infinite straight line in $\bR^d$.
The presence of such a line defect breaks the superconformal symmetry of the ambient theory, reducing it to a smaller subgroup. This residual symmetry generically includes, at least, the bosonic subalgebra $\so(1,2)_{\text{conf}}\oplus \so(d-1)_{\text{rot}}$, where $d$ is the spacetime dimension of the ambient Euclidean theory. In supersymmetric settings, only a part of the full supersymmetry algebra remains intact. Defects of this type are classified as $1/n$ BPS according to the fraction of supercharges they preserve.

Central features in the study of any CFT are the correlation functions. In the presence of a line defect, this naturally extends to the study of defect correlators, which involve the local operators $\mathcal{O}_i(t_i)$ inserted along the line, defined by
\begin{equation*}
\langle\mathcal{O}_1(t_1)\mathcal{O}_2(t_2)\ldots \mathcal{O}_n(t_n)\rangle_\mathcal{W}
\equiv  
\frac{\langle \Tr \mathcal{P}[ \mathcal{W}_{-\infty,t_1}\mathcal{O}_1(t_1) \mathcal{W}_{t_1,t_2}\mathcal{O}_2(t_2) \ldots \mathcal{O}_n(t_n)\mathcal{W}_{t_n,\infty}]\rangle}{\langle \cW \rangle},
\end{equation*}
where $\mathcal{W}_{t_n,t_{n+1}}$ is the untraced Wilson link  $\mathcal{W}_{t_n,t_{n+1}}=\exp\Bigl(-i\int_{t_n}^{t_{n+1}}\mathcal{L}(t)dt \Bigr)$. It is common to shorten the notation  by defining the insertions by 
\begin{equation*}
\langle\mathcal{O}_1(t_1)\mathcal{O}_2(t_2)\ldots \mathcal{O}_n(t_n)\rangle_\mathcal{W}
\equiv  
\frac{\langle \cW[\mathcal{O}_1(t_1) \mathcal{O}_2(t_2) \ldots \mathcal{O}_n(t_n)]\rangle}{\langle \cW \rangle}.
\end{equation*}

In defect CFTs, a particularly distinguished set of defect operators are those related to the breaking of the bulk symmetries. When considering supersymmetric defects CFT, the breaking pattern of supersymmetric currents leads to additional operators. Crucially, one can identify them by exploiting the Ward identity
\begin{equation}
\label{Ward-id}
    [\mathbf{G}, \cW ]={\delta_{\mathbf{G}}} \cW \equiv \int \mathcal{W}[\mathbb{G}(t)]dt ,
\end{equation}
that explicitly relates broken generators to local insertions on the line. We will refer to this set of defect operators obtained from Ward identities as fundamental (or elementary).
One can then further characterize the local operators by finding the representations under the preserved supergroup. By exploiting (super-)Jacobi identities like
\begin{equation*}
    [G,[\gG, \cW ]]+[\gG,[\cW,G]]+[\cW,[G,\gG]]=0\implies [G,[\gG, \cW ]]=[[G,\gG],\cW]\,,
\end{equation*}
one can explicitly read the charges of defect insertion $\bG(t)$ by the algebra involving preserved $G$ and broken $\gG$ generators.

A particular example that will be always present in the line defects is the displacement operator which is associated with the breaking of the translations orthogonal to the line. This is thus given by
\begin{equation}
    [P^i,\cW]=\int \cW[\bD^i(t)]dt\,,
\end{equation}
where the index runs up to the codimension of the defect $i=1,..., d-1$, as expected from the $\so(d-1)_{\text{rot}}$\footnote{ For related discussions and a manifestly local realization of the displacement operator, see \cite{Correa:2012at}.}.  

In general, defect operators arrange themselves into multiplets according to the preserved supersymmetry. Of particular importance is the displacement supermultiplet (or superdisplacement), which contains the displacement operator among its components. In what follows, we will denote this multiplet, as well as its superfield realization, by $\cD$.

To keep the discussion as general as possible, we denote by $\mathbb{D}(t)$ the displacement operator, and by $\mathbb{O}(t)$ a generic fundamental operator, distinct from the displacement, appearing in $\cD$. When $\mathbb{O}(t)$ is specifically the superprimary, we will instead write $\mathbb{P}(t)$. 

Crucially, in these preliminary remarks we refrain from making explicit reference to symmetries, as their precise realization is theory dependent. In specific models, symmetries naturally impose further constraints. 

Our purpose here is instead to present a general and schematic description of the correlation functions of the displacement supermultiplet, particularly the four-point function. This is done in order to highlight the universal features common to all such defect theories. The following discussion should therefore be understood as theory independent and schematic, to be refined when applied to concrete cases.

In a superfield representation we can schematically consider the two-point function in superspace as
\begin{equation*}
    \vev{\cD(y_1,\theta_1)\cD(y_2,\theta_2)}=\frac{C_\cT}{\vev{12}^{2\Delta_\cD}}\,,
\end{equation*}
Here, the $y_i$ denote the supersymmetric coordinates constructed from $t_i$, the coordinate along the line and $\theta_i$, the fermionic coordinates. The quantity $\langle ij \rangle$ is a supertranslation-invariant distance, whose precise form depends on the superspace, and thus on the specific theory, that the correlation function describes. All components except $t_i$ are theory dependent; for an explicit example see Sec. \ref{subsec: Superfield realization and correlation function in superspace}. The overall $C_\cT$ is a physical theory-dependent constant related to the bremsstrahlung function \cite{Correa:2012at}. 

For the four-point function one schematically has
\begin{equation}
\label{eq: generic four-point of the superdisplacement}
    \vev{\cD(y_1,\theta_1)\cD(y_2,\theta_2)\cD(y_3,\theta_3)\cD(y_4,\theta_4)}=\frac{C_\cT^2}{\vev{12}^{2\Delta_\cD}\vev{34}^{2\Delta_\cD}}f(\cZ)\,.
\end{equation}
Here, $\mathcal{Z}$ denotes a superconformal cross-ratio. For chiral correlators, this is, for example, \eqref{eq: super cross ratio}. More generally, the function can be expressed in terms of nilpotent invariants constructed from the Gra\ss mann coordinates. Expanding in this basis yields a larger set of independent functions contributing to the correlators; see for example \cite{Gimenez-Grau:2019hez}. By expanding the left-hand side in superfield components and the right-hand side as a Taylor series in the fermionic coordinates, one can relate, term by term, the correlation functions of the multiplet components to derivatives of the superconformal invariant function.

The first contribution comes from the four-point function of the superprimary, \ie $\bP(t)$, for which the identification is immediate being the top component of the displacement multiplet. Generically, this is given by
\begin{equation*}
    \vev{\bP(t_1)\bP(t_2)\bP(t_3)\bP(t_4)} = \frac{C^2_\cT}{t_{12}^{2\Delta_{\bP}}t_{34}^{2\Delta_{\bP}}}f_{\bP^4}(z)\,,
\end{equation*}
where $\Delta_\bP=\Delta_\cD$ and $z$ is the bosonic contribution of $\cZ$. In one dimension, the two invariant cross-ratios are not independent. Defining $t_{ij} = t_i - t_j$, they are given by
\begin{equation}
\label{eq: cross ratios}
z=\frac{t_{12}t_{34}}{t_{14}t_{32}}\in(-\infty,0)\,,\qquad \chi=\frac{t_{12}t_{34}}{t_{13}t_{24}}\in(0,1)
\end{equation}
and are related through
\begin{equation*}
     z=\frac{\chi}{\chi-1}\,.
\end{equation*}

In the following, we will be interested also in dealing with all the other four-point functions of the multiplet components. For example, we will deal with the four-point of the displacement operators
\begin{equation}
\label{eq:4D corr}
    \vev{\bD(t_1)\bD(t_2)\bD(t_3)\bD(t_4)}=\frac{(\mathbb{n}_{\bD}C_{\cT})^2}{t_{12}^4t_{34}^4}f_{\bD^4}(z)
\end{equation}
and with the mixed four-point with two displacement and two other generic operators in the multiplet
\begin{equation}
\label{eq:2D2O corr}
    \vev{\bD(t_1)\bD(t_2)\bO(t_3)\bO(t_4)}=\frac{(\mathbb{n}_{\bD\bO}C_{\cT})^2}{t_{12}^4t_{34}^{2\Delta_{\bO}}}f_{\bD^2\bO^2}(z)\,.
\end{equation}
Here $\mathbb{n}_{\bD}$ and $\mathbb{n}_{\bD\bO}$ are theory dependent constants that originate from the supersymmetric structure of the problem. While the two-point function of the superprimary is fixed solely in terms of $\cC_\cT$, with $\mathbb{n}_{\bP}=1$, the inclusion of $Q$-descendants generally introduces additional constants, which arise from the action of the supercharges.

As anticipated, the functions $f_{\bP^4}(z)$, $f_{\bD^4}(z)$, and $f_{\bD^2\bO^2}(z)$ are all related by supersymmetry. Particularity for chiral theories, both $f_{\bD^4}(z)$ and $f_{\bD^2\bO^2}(z)$ are specific combinations of polynomials and derivatives acting on $f_{\bP^4}(z)$. For example, one has generically
\begin{equation}
\label{eq: function of displacement from primary}
    f_{\bD^4}(z) = \sum_n p_{n}(z)\partial_z^{n}\,f_{\bP^4}(z)\,,
\end{equation}
for some theory-dependent set of polynomials $\{p_n(z)\}$; see for example \eqref{eq: four-point of the displacement in 1/2 BPS N=2 as derivatives}. When \eqref{eq: generic four-point of the superdisplacement} has nilpotent invariants, one has also additional contributions coming from the higher number of independent functions.
These types of relations are exact and therefore hold at any order in perturbation theory. In particular, they remain valid when expanding around the generalized free-field limit at strong coupling. In this regime, the leading behavior of the correlators is determined by the GFF, while higher-order corrections account for deviations.

Consequently, the functions appearing in the correlators can be systematically expanded as
\begin{equation}
\label{eq: 4point P}
    \vev{\bP(t_1)\bP(t_2)\bP(t_3)\bP(t_4)} = \frac{C^2_\cT}{t_{12}^{2\Delta_{\bP}}t_{34}^{2\Delta_{\bP}}}\Bigl(f_{\bP^4}^{(0)}(z)+\epsilon_\cT f_{\bP^4}^{(1)}(z)+\cO(\epsilon_\cT^2)\Bigr)\,,
\end{equation}
where, as we will see, $f_{\bP^4}^{(0)}(z)$ is the GFF contribution, $f_{\bP^4}^{(1)}(z)$ is the next-to-leading order one and $\epsilon_\cT$ is a theory-dependent perturbation parameter. Similarly, one can consider the expansions for \eqref{eq:4D corr} and \eqref{eq:2D2O corr} given by
\begin{equation}
\label{eq: 4point D}
    \vev{\bD(t_1)\bD(t_2)\bD(t_3)\bD(t_4)}=\frac{(\mathbb{n}_{\bD}C_{\cT})^2}{t_{12}^4t_{34}^4}\Bigl(f_{\bD^4}^{(0)}(z)+\epsilon_\cT f_{\bD^4}^{(1)}(z)+\cO(\epsilon_\cT^2)\Bigr)
\end{equation}
and 
\begin{equation}
\label{eq: 4point 2D2O}
    \vev{\bD(t_1)\bD(t_2)\bO(t_3)\bO(t_4)}=\frac{(\mathbb{n}_{\bD\bO}C_{\cT})^2}{t_{12}^4t_{34}^{2\Delta_{\bO}}}\Bigl(f_{\bD^2\bO^2}^{(0)}(z)+\epsilon_\cT f_{\bD^2\bO^2}^{(1)}(z)+\cO(\epsilon_\cT^2)\Bigr)\,.
\end{equation}
Crucially, supersymmetry ties these correlators together, so consistency requires that the constants $C_\mathcal{T}$ and $\epsilon_\mathcal{T}$ appearing in \eqref{eq: 4point P}, \eqref{eq: 4point D}, and \eqref{eq: 4point 2D2O} are identical.\\

So far, our discussion has focused on the expansion of the four-point function. One can also analyze the consequences of this expansion within the conformal partial wave decomposition in terms of conformal blocks \cite{Dolan:2003hv}. In particular, given the preserved subalgebra $\mathfrak{so}(1,2)_\text{conf}$, each contribution can be expanded in $1d$ conformal blocks 
\begin{equation*}
    g_{\Delta}(z)=(-z)^\Delta\,_2F_1(\Delta,\Delta,2\Delta;z)\,.
\end{equation*}
At leading order one would find for \eqref{eq: 4point D}
\begin{equation}
\label{eq: block expansion 4D at leading order}
    f^{(0)}_{\bD^4}(z)= \sum_\Delta \lambda_\Delta^{(0)}\lambda_\Delta^{(0)}\,g_\Delta(z)\,,
\end{equation} 
for some $\lambda_\Delta^{(0)}=\lambda_{\bD\bD\bT_\Delta}^{(0)}$ representing the leading order three-point coefficients\footnote{ In the CPW expansion, the coefficients appearing are related to the three-point one by $\lambda_\Delta^{(0)}=k_{\bD\bD\bT_\Delta}\lambda_{\bD\bD\bT_\Delta}^{(0)}/(c_{\bD\bD}c_{\bT\bT})$, where $k_{\bD\bD\bT_\Delta}$ denotes the normalization of the three-point function, and $c_{\bD\bD}$, $c_{\bT\bT}$ those of the two-point functions. However, for brevity, we neglect these normalization constants,  specifying them explicitly only when necessary. } of a generic conformal primaries $\bT_\Delta$ having dimensions $\Delta$. In order to be exchanged we assume that it lies in the displacement OPE, \ie $\bT_\Delta\in\bD\times\bD$. The sum runs over all the conformal dimensions allowed by the OPE.
Similarly one could do the same for \eqref{eq: 4point 2D2O}, thus obtaining
\begin{equation}
\label{eq: block expansion 2D2D at leading order}
    f^{(0)}_{\bD^2\bO^2}(z)= \sum_\Delta \lambda_\Delta^{(0)}\tilde\lambda_\Delta^{(0)}\,g_\Delta(z)\,,
\end{equation} 
with $\tilde\lambda_\Delta^{(0)}=\lambda_{\bO\bO\bT_\Delta}^{(0)}$. 
Exploiting the orthogonality for the conformal blocks, one can obtain explicitly the LO conformal data.

The analysis can then be extended to the study of the NLO, which is obtained by perturbing the leading order conformal data. In the case of the four-point function discussed above, this implies that
\begin{align}
\label{eq: NLO expansion of CFT data}
    \Delta&\rightarrow\Delta+\gamma^{(1)}_\Delta\nonumber\,,\\
    \lambda^{(0)}_\Delta&\rightarrow \lambda^{(0)}_\Delta+\lambda^{(1)}_\Delta\,,\\
    \tilde\lambda^{(0)}_\Delta&\rightarrow \tilde\lambda^{(0)}_\Delta+\tilde\lambda^{(1)}_\Delta\,,\nonumber
\end{align}
where $\gamma^{(1)}_\Delta$ are the anomalous dimensions corresponding to the exchanged operators of dimension $\Delta$. In this first instance, we keep the order in $\epsilon_\cT$ in the definitions of the NLO data.
This leads to the identifications
\begin{align*}
    f^{(0)}_{\bD^4}(z)+\epsilon_\cT f^{(1)}_{\bD^4}(z)=\sum_\Delta (\lambda_\Delta^{(0)}+\lambda^{(1)}_\Delta)(\lambda_\Delta^{(0)}+\lambda^{(1)}_\Delta)\,g_{\Delta+\gamma_\Delta^{(1)}}(z)\,,\\
    f^{(0)}_{\bD^2\bO^2}(z)+\epsilon_\cT f^{(1)}_{\bD^2\bO^2}(z)=\sum_\Delta (\lambda_\Delta^{(0)}+\lambda^{(1)}_\Delta)(\tilde\lambda_\Delta^{(0)}+\tilde\lambda^{(1)}_\Delta)\,g_{\Delta+\gamma_\Delta^{(1)}}(z)\,,
\end{align*}
By expanding the conformal blocks as
\begin{equation*}
    g_{\Delta+\gamma_\Delta^{(1)}}(z)= g_{\Delta}(z) + \gamma_\Delta^{(1)}\Bigl( g_\Delta(z)\log(-z)+\mathrm{g}_\Delta(z)\Bigr)+\cO((\gamma_\Delta^{(1)})^2)\,,
\end{equation*}
where we defined for brevity
\begin{equation}\label{eq: derivative of the block}
    \mathrm{g}_\Delta(z) = (-z)^\Delta\Bigl(\partial_{\tilde\Delta} g_{\tilde\Delta}(z)(-z)^{-\tilde\Delta}\Bigr)_{\tilde\Delta=\Delta}\,,
\end{equation}
we obtain the expressions
\begin{align}
    f^{(0)}_{\bD^4}(z)+\epsilon_\cT f^{(1)}_{\bD^4}(z)=\sum_\Delta \lambda_\Delta^{(0)}\lambda_\Delta^{(0)}\,g_{\Delta}(z)+\sum_\Delta &(\lambda_\Delta^{(0)}\lambda_\Delta^{(1)}+\lambda_\Delta^{(1)}\lambda_\Delta^{(0)})g_\Delta(z) + \lambda_\Delta^{(1)}\lambda_\Delta^{(1)}g_\Delta(z)+\nonumber\\
    &+\lambda_\Delta^{(0)}\lambda_\Delta^{(0)}\gamma^{(1)}_\Delta(g_\Delta(z)\log(-z)+\mathrm{g}_\Delta(z))\label{eq: four-point 4D at NLO in block expansion}\,,\\[0.5cm]
    f^{(0)}_{\bD^2\bO^2}(z)+\epsilon_\cT f^{(1)}_{\bD^2\bO^2}(z)=\sum_\Delta \lambda_\Delta^{(0)}\tilde\lambda_\Delta^{(0)}\,g_{\Delta}(z)+\sum_\Delta &(\lambda_\Delta^{(0)}\tilde\lambda_\Delta^{(1)}+\lambda_\Delta^{(1)}\tilde\lambda_\Delta^{(0)})g_\Delta(z) + \lambda_\Delta^{(1)}\tilde\lambda_\Delta^{(1)}g_\Delta(z)+\nonumber\\
    &+\lambda_\Delta^{(0)}\tilde\lambda_\Delta^{(0)}\gamma^{(1)}_\Delta(g_\Delta(z)\log(-z)+\mathrm{g}_\Delta(z))\label{eq: four-point 2D2O at NLO in block expansion}\,.
\end{align}
If the functions of the left-hand side are known explicitly, one can obtain the NLO conformal data just by projecting on the conformal blocks.

\subsection{Consistency constraints}
\label{subsec: Bootstrapping the epsilon-order of OPE Coefficients}
In this section, we highlight some general consequences that follow from the identifications in \eqref{eq: four-point 4D at NLO in block expansion} and \eqref{eq: four-point 2D2O at NLO in block expansion}. This line of reasoning will be systematically exploited in the next section to study the possible exchanged operators that can appear at NLO where, with additional assumptions, one can identify universal features. Before this, however, we first derive constraints that can be obtained from more general considerations.

We begin by noting that the expansion of correlation functions, together with their conformal block decompositions, imposes nontrivial constraints order by order. In particular, the generic form
\begin{equation}
\label{eq: constrain on the order}
    \epsilon_\cT f^{(1)}(z)=\sum_\Delta(\lambda_\Delta^{(0)}\tilde\lambda_\Delta^{(1)}+\lambda_\Delta^{(1)}\tilde\lambda_\Delta^{(0)})g_\Delta(z) + \lambda_\Delta^{(1)}\tilde\lambda_\Delta^{(1)}g_\Delta(z)+\lambda_\Delta^{(0)}\tilde\lambda_\Delta^{(0)}\gamma^{(1)}_\Delta(g_\Delta(z)\log(-z)+\mathrm{g}_\Delta(z))
\end{equation}
can be interpreted as a constraint on the $\epsilon_\cT$ order of $\lambda_\Delta^{(1)}$ and $\tilde\lambda_\Delta^{(1)}$. Crucially, as discussed above, the sum runs over all operator dimensions that may appear in the OPE. Operators that do not contribute at a given order have vanishing OPE coefficients, although they are still formally included in the sum.

Consider, for example, the case in  which we look only at a given primary $\bT_{\Delta^\star}$ for which the three-point coefficients at leading order are
\begin{equation*}
    \lambda^{(0)}_{\Delta^\star}=0\,,\qquad\ \tilde\lambda^{(0)}_{\Delta^\star}\neq0\,.
\end{equation*}
If we compare with the previous discussion by taking $\lambda_{\Delta^*}^{(0)}=\lambda_{\bD\bD\bT_{\Delta^\star}}^{(0)}$ and $\tilde\lambda_{\Delta^*}^{(0)}=\lambda_{\bO\bO\bT_{\Delta^\star}}^{(0)}$ this means that, in principle, $\bT_{\Delta^\star}$ appears in both the $\bD\times\bD$ and $\bO\times\bO$ OPEs, but it is actually exchanged at GFF only in the $\bO$ sector, and not in the $\bD$ one \eqref{eq: block expansion 4D at leading order}. 
For this particular choice of leading order coefficients, the expression \eqref{eq: constrain on the order} simply reads
\begin{equation*}
    \epsilon_\cT f^{(1)}(z) \ni\lambda_{\Delta^\star}^{(1)}\tilde\lambda_{\Delta^\star}^{(0)}g_{\Delta^\star}(z) + \lambda_{\Delta^\star}^{(1)}\tilde\lambda_{\Delta^\star}^{(1)}g_{\Delta^\star}(z)\,.
\end{equation*}
Excluding the trivial case in which the function vanishes, consistency requires that the $\epsilon_\cT$-order of the products must be
\begin{equation}
\label{eq: orders of the ope}
    \lambda_{\Delta^\star}^{(1)}\tilde\lambda_{\Delta^\star}^{(0)}\sim \cO(\epsilon_\cT)\,\qquad\text{and}\qquad \lambda_{\Delta^\star}^{(1)}\tilde\lambda_{\Delta^\star}^{(1)}\sim \cO(\epsilon_\cT)\,,
\end{equation}
where by $\cO(\epsilon_\cT)$ we mean that the order of the OPE can be either $\epsilon_\cT$ or higher.\footnote{For clarity, with a slight abuse of notation these expressions can be interpreted as $\lambda_{\Delta^\star}^{(1)}\tilde\lambda_{\Delta^\star}^{(0)}\gtrsim\cO(\epsilon)$ and $\lambda_{\Delta^\star}^{(1)}\tilde\lambda_{\Delta^\star}^{(1)}\gtrsim \cO(\epsilon)$ to stress that the order can be higher. This is equivalent to say that, for example, if a function is bounded by $\cO(\epsilon^2)$, then it is also bounded by $\cO(\epsilon)$ contributions}
Given that $\tilde\lambda_\Delta^{(0)}$ is constrained by the nonvanishing condition at GFF, $\tilde\lambda_\Delta^{(0)}\sim\cO(1)$, consistency with the order in $\epsilon_\cT$ of the left-hand-side leads to the requirement that either
\begin{equation*}
    \lambda_{\Delta^\star}^{(1)}\sim\cO(\epsilon)\qquad \text{or}\qquad \lambda_{\Delta^\star}^{(1)}=0\,.
\end{equation*}
So, when considering the CPW expansion of correlators with external insertions, of the same type, as in \eqref{eq: four-point 4D at NLO in block expansion}, the contributions associated with the $\bT_{\Delta^\star}$ exchange given by
\begin{equation*}
    (\lambda_{\Delta^\star}^{(0)}\lambda_{\Delta^\star}^{(1)}+\lambda_{\Delta^\star}^{(1)}\lambda_{\Delta^\star}^{(0)})g_{\Delta^\star}(z) + \lambda_{\Delta^\star}^{(1)}\lambda_{\Delta^\star}^{(1)}g_{\Delta^\star}(z)+\lambda_{\Delta^\star}^{(0)}\lambda_{\Delta^\star}^{(0)}\gamma^{(1)}_{\Delta^\star}(g_{\Delta^\star}(z)\log(-z)+\mathrm{g}_{\Delta^\star}(z))=\lambda_{\Delta^\star}^{(1)}\lambda_{\Delta^\star}^{(1)}g_{\Delta^\star}(z)
\end{equation*}
will be either vanishing, if $\lambda^{(1)}_{\Delta^\star}=0$, or of higher $\epsilon_\cT$ order, since $(\lambda^{(1)}_{\Delta^\star})^2\sim\cO(\epsilon_\cT^2)$. In either case, one can conclude that $\bT_{\Delta^\star}$ is not exchanged up to NLO.

Another possible situation arises when
\begin{equation*}
    \lambda^{(0)}_{\Delta^\star}=0\,,\qquad \tilde\lambda^{(0)}_{\Delta^\star}=0\,.
\end{equation*}
In these cases one can impose only the second condition of \eqref{eq: orders of the ope}. This implies that
\begin{equation*}
    \lambda_{\Delta^\star}^{(1)}\tilde\lambda_{\Delta^\star}^{(1)}\sim\cO(\epsilon_\cT)\implies \lambda_{\Delta^\star}^{(1)}\sim\tilde\lambda_{\Delta^\star}^{(1)}\sim\cO(\sqrt{\epsilon_\cT})\,.
\end{equation*}
However, this may simply reflect an unfortunate choice of correlators. By instead considering correlators with different external insertions, one could further constrain the OPE coefficient through the procedure outlined above. On the other hand, if no such possibility arises, the $\epsilon_\cT$ order obtained should be regarded as the correct one. For these cases, one may also note that if the $\Delta^\star$ is smaller than any $\Delta$ of the other exchanged operators, then the $\log(-z)$ contribution is absent. This implies that the exchanged operator must have vanishing anomalous dimension, thus pointing toward protected operators; see discussion in \ref{subsubsec: Relaxing the 3rd assumption}.

These arguments have been presented for $1d$ dCFTs, but they can be extended to higher-dimensional defects and, under appropriate conditions, to general CFTs beyond the context of defect theories.

We conclude this section with a brief discussion on a subtly that can arise when the NLO contributions are obtained from the expansion of the leading order one, without further discussions. In this case we adopt the standard definition of the expansion in terms of the quadratic OPE for which the expansion up to NLO order is given by
\begin{align}
    \Delta&\rightarrow\Delta+\epsilon_\cT\,\gamma^{(1)}_\Delta\nonumber\,,\\
    a^{(0)}_\Delta&\rightarrow a^{(0)}_\Delta+\epsilon_\cT\,a^{(1)}_\Delta\,,\label{eq: NLO expansion of the quadratic OPEs}
\end{align}
with the quadratic OPE coefficients $a_\Delta^{(i)}$ of a given channel. Comparing with the previous discussion these are related by
\begin{equation}
    a_\Delta^{(0)}=\lambda_\Delta^{(0)}\lambda_\Delta^{(0)}\,,\quad a_\Delta^{(1)}=2\epsilon_\cT\,\lambda_\Delta^{(0)}\lambda_\Delta^{(1)}\,\qquad\text{or}\qquad a_\Delta^{(0)}=0\,,\quad a_\Delta^{(1)}=\epsilon_\cT\lambda_\Delta^{(1)}\lambda_\Delta^{(1)}\,.
\end{equation}
With these definitions, \eqref{eq: four-point 4D at NLO in block expansion} would read
\begin{equation}
\label{eq: perturbation of the GFF explicit}
    f^{(0)}_{\bD^4}(z)+\epsilon_\cT f^{(1)}_{\bD^4}(z)=\sum_{\Delta\in \text{GFF}} a_\Delta^{(0)}\,g_{\Delta}(z)+\epsilon_\cT\sum_{\Delta\in \text{GFF}}  a^{(1)}_\Delta g_\Delta(z)+ a_\Delta^{(0)}\gamma^{(1)}_\Delta(g_\Delta(z)\log(-z)+\mathrm{g}_\Delta(z))
\end{equation}
Here, the sum runs over all $\Delta \in \mathrm{GFF}$, \ie over all scaling dimensions of the free theory. We are explicitly perturbing the leading-order data to NLO. The main difference with the previous analysis is that the sum does not include all operators that can formally appear in the OPE.
The limitation of \eqref{eq: perturbation of the GFF explicit} arises when the spectrum of allowed operators changes at NLO. In such cases, to include all possible exchanges, $\Delta$ must be extended to run over all contributions allowed by the OPE, rather than being restricted to $\Delta \in \mathrm{GFF}$. This subtlety does not generate inconsistencies when the dimensions of the higher-order exchanges remain within the set of free-theory scaling dimensions. However, the situation changes when some of the operators appearing at NLO have classical dimensions outside the GFF set. If one does not account for this effect, the analysis that follows is restricted only to a partial sector.

Consider the following example. Suppose there exists an operator $\bT_{\Delta^\star}$, absent at leading order, with dimension $\Delta^\star < \min\{\,\Delta \in \text{GFF}\,\}$. In the expansion \eqref{eq: perturbation of the GFF explicit}, where the sum runs over GFF dimensions, the contribution of $\bT_{\Delta^\star}$ is excluded from the NLO contributions since $\Delta^\star \notin \text{GFF}$. This is equivalent to setting the corresponding CFT data to zero in \eqref{eq: four-point 4D at NLO in block expansion}, thereby restricting the set of operators that can be exchanged.

If the exchanged operators are only multiparticle states built from the external insertions, in the excluded contributions there would be also the exchange of the external field itself.
Such a contribution, however, is only present when it is allowed by the OPE, \eg $\bD \in \bD \times \bD$ and it is related to the explicit breaking of the $\mathbb{Z}_2$ symmetry at NLO. At leading order the sum runs over $\{\Delta \in \mathbb{N} \mid \Delta \geq 2\Delta_{\bD}\}$. Keeping this bound at NLO is equivalent to setting $\lambda_{\bD\bD\bD}=0$ in \eqref{eq: four-point 4D at NLO in block expansion}. Importantly, as we discuss in Sec. \ref{subsubsec: Z2 odd sectors}, this coefficient controls all $\mathbb{Z}_2$-odd exchanges. Thus, setting it to zero automatically decouples these contributions, leaving the NLO analysis to consider only the $\mathbb{Z}_2$-even sector.

\section{Universal sectors}
\label{subsec: Correlators of the displacement multiplet}

The purpose of this section is to identify the operators that can be exchanged in the superdisplacement four-point functions, up to NLO. Building on the considerations of Sec. \ref{subsec: Bootstrapping the epsilon-order of OPE Coefficients}, our goal is to determine whether these correlators exhibit universal behavior. By universal we mean that, independently of the details of the ambient theory, such as the amount of supersymmetry, the four-point function of operators with a given conformal dimension always takes the same functional form, modulo eventual tensor structures.

As we will see, universality ultimately traces back to the constraint that the leading order must correspond to a GFF theory and that the spectrum is restricted to the displacement supermultiplet and its composite operators. As discussed in Sec. \ref{subsec: Bootstrapping the epsilon-order of OPE Coefficients}, one can exploit the leading-order data of the GFF to identify the $\epsilon_\cT$ order of the exchanged operators at NLO. This allows one to reduce the possible operators that can be exchanged in a given four-point function. Indeed, in principle, all operators compatible with symmetry can appear. However, if one can demonstrate that only operators built from the external fields are actually exchanged in the four-point, while all other operators that could in principle appear are excluded, then the correlator becomes independent of the other supermultiplet components, thereby exhibiting universality.

In this regard, we begin by setting some assumptions on the generic theories that we analyze, then we will discuss features that arises when we relax some of them. In particular, we start from theories that are consistent with the requirements
\begin{enumerate}
    \item The $1d$ dCFT is described at leading order by a Generalized Free Field theory.
    \item The only elementary operators are those in $\cD$
    \item $\cD$ does not appear in the OPE of $\cD\times\cD $
\end{enumerate}
The first requirement implies that in all the case we will consider the leading order as completely defined by taking Wick contractions of the operator insertions. Moreover, being a $1d$ dCFT, the $\so(1,2)_{\text{conf}}$ subalgebra will always be present regardless from the case in hand, hence one can always express the four-point functions as a sum of $\sl(2,\mathbb{R})$ blocks.\footnote{This is true even in the case in which one considers superblocks, as they can always be decomposed in a finite sum of conformal blocks.} The second condition implies that we exclude all theories having additional contributions coming from other elementary fields, \eg tilt supermultiplets, that could in principle lead to nontrivial interactions with the displacement multiplet. The last requirement is a simplifying assumption. While it may appear specific, we will see that many theories of interest naturally satisfy this property. The implications of relaxing these assumptions will be discussed at a later stage.

\subsection{Generalized Free Field}
We therefore begin with the analysis by studying the leading-order generalized free field theory, where correlation functions can be computed via Wick contractions of the $n$-point insertions. The lack of interactions makes this framework particularly restrictive.
In fact, the only nonvanishing three-point functions involving two fundamental fields are those in which the third operator is built from the fundamental field itself. In particular we are interested in
\begin{align}
    \vev{\bD(t_1)\bD(t_2)[\bD]^{L=2}_M(t_3)}^{(0)}&\neq 0\,,\\
    \vev{\bD(t_1)\bD(t_2)\bO(t_3)}^{(0)}&=0\label{eq: 3pnt DDO LO}\,,\\
    \vev{\bO(t_1)\bO(t_2)[\bO]^{L=2}_M(t_3)}^{(0)}&\neq0\,.
\end{align}

We use the notation $[\phi]^L_{M}$ for composite operators built from the fundamental operator $\phi$, of length $L$ and with a total of $M$ derivatives. We will not be interested in the particular expression of these operators\footnote{ Even at the level of a GFF computation, generic operators such as $\bD\partial^n_t \bD$ are not primaries in one dimension, and one must construct appropriate linear combinations to obtain them. See \cite{Ferrero:2023znz} for the explicit realizations.}, but rather on the length of these. 
Working in such a basis, which in general is not the same in which the dilatation operator is diagonal, allows in general to have a better control on correlation functions of operator having a well-defined length; see \cite{Ferrero:2023gnu} for a complete discussion.

What matters for our purposes is simply to identify the order of the three-point constants. The task of identifying all these quantities explicitly can be posed when specific cases are considered, or, in any case, pursued at a later stage. From the three-point functions, we can thus conclude that the leading-order OPE coefficients are
\begin{align}
    &\lambda_{\bD\bD[\bD]_M^{L=2}}^{(0)} \sim \cO(1)\nonumber\,,\\
    &\lambda_{\bD\bD\bO}^{(0)} = 0\nonumber\,,\\
    &\lambda_{\bO\bO[\bO]_M^{L=2}}^{(0)} \sim \cO(1)\nonumber\,.
\end{align}
Moreover, given that also the mixed three-point function vanishes
\begin{equation*}
    \vev{\bD(t_1)\bD(t_2)[\bO]^{L=2}_M(t_3)}^{(0)}=0 \implies \lambda_{\bD\bD[\bO]^{L=2}_M}^{(0)}=0\,,
\end{equation*}
we conclude that in the displacement four-point function at leading order, \eqref{eq: block expansion 4D at leading order}, only multiparticle operators of length $L=2$ built out of the displacement can be exchanged. All other exchanges do not appear at leading order. A similar analysis holds also for all the other four-point functions in the multiplet.

Therefore, one can conclude that the sectors are actually decoupled and so the expressions for $f_{\bP^4}^{(0)}(z)$, $f_{\bO^4}^{(0)}(z)$ and $f_{\bD^4}^{(0)}(z)$ are hence universal.\footnote{We stress that these statements are true modulo global symmetries. These, indeed, can introduce tensorial features dressing the functions with indexes. For example, if the displacement is charged under a global symmetry, \eg $\bD\rightarrow\bD^a$ then $f_{\bD^4}(z)\rightarrow f_{\bD^4}^{a_1a_2a_3a_4}(z)$. However, one can always decompose to relate two different four-point functions; see Sec. \ref{subsec: Universality in 1/2 BPS wilson line N=4 SYM}}
Clearly, this is a well-known fact of GFF. In the following, we will look up to which extent this feature is preserved at next-to-leading order. 

\subsection{Next-to-leading order}
We can now proceed with a similar analysis by considering the first nontrivial contributions arising from interactions. Just for reference, we will focus in the following on the four-point of the displacement operator, which, as discussed, is always present in the multiplet as it is related to the breaking of orthogonal translations. However, the same analysis can be performed for each component of a given supermultiplet.

We begin with \eqref{eq: 3pnt DDO LO}, which, in principle, is not guaranteed to vanish also at NLO. At this point, we make use of the starting assumption, namely $\cD \notin \cD \times \cD$, which directly implies that
\begin{equation}
\label{eq: 3point 2D1O}
    \vev{\bD(t_1)\bD(t_2)\bO(t_3)}=0\implies\lambda_{\bD\bD\bO}=0
\end{equation}
at all orders.

However, interactions and symmetry may also allow for additional operators to appear in the OPEs. To verify that no other sector mixes with the displacement four-point function at this order, we must check that all OPE coefficients $\lambda_{\bD\bD\bT}$, where $\bT$ are a generic multiparticle operator built from $\bO$, either vanish or are suppressed as $\cO(\epsilon_\cT)$.

As an example, we focus on the exchange of the multiparticle operators $\bT = [\bO]^{L=2}_M$. 
In the conformal partial wave expansion we can isolate the contribution associated to them by
\begin{equation*}
       \vev{\bD(t_1)\bD(t_2)\bO(t_3)\bO(t_4)}= \frac{(\mathbb{n}_{\bD\bO}C_{\cT})^2}{t_{12}^4t_{34}^{2\Delta_{\bO}}}\Bigl(\sum_{M=0}^{\infty}\lambda_{\bD\bD[\bO]_M^{L=2}}\lambda_{\bO\bO[\bO]_M^{L=2}}\,g_{_{2{\Delta_{\bO}}+M}}(z) +\, ...\Bigr)\,,
\end{equation*}
where the dots represent other exchange of which we are now not interested in. 
When taking into account the expansion in $\epsilon_\cT$ of both sides, as discussed in Sec. \ref{sec: The CFT perspective}, one gets and expression like \eqref{eq: four-point 2D2O at NLO in block expansion}. Given that $\lambda_{\bD\bD[\bO]_M^{L=2}}^{(0)}=0$, the relevant contribution associated to these exchanges at NLO is given by 
\begin{equation*}
    \epsilon_\cT f^{(1)}_{\bD^2\bO^2}(z)\ni\sum_{M=0}^{\infty}\Bigl(\lambda_{\bD\bD[\bO]_M^{L=2}}^{(1)}\lambda_{\bO\bO[\bO]_M^{L=2}}^{(0)}+\lambda_{\bD\bD[\bO]_M^{L=2}}^{(1)}\lambda_{\bO\bO[\bO]_M^{L=2}}^{(1)}\Bigr)g_{_{2{\Delta_{\bO}}+M}}(z)\,.
\end{equation*}
Exploiting the discussion presented in Sec. \ref{subsec: Bootstrapping the epsilon-order of OPE Coefficients} and knowing that $\lambda_{\bO\bO[\bO]_M^{L=2}}^{(0)}\sim \cO(1)$, consistency with the $\epsilon_\cT$ order implies that
\begin{equation*}
    \lambda_{\bD\bD[\bO]_M^{L=2}}^{(1)}\lambda_{\bO\bO[\bO]_M^{L=2}}^{(0)}\sim \cO(\epsilon_\cT)\implies  \lambda_{\bD\bD[\bO]_M^{L=2}}^{(1)}\sim \cO(\epsilon_\cT)\,.
\end{equation*}

We can now look back at the CPW expansion of the displacement four-point function \eqref{eq: four-point 4D at NLO in block expansion}. With the pieces of information that we collected for the OPE coefficients
\begin{equation*}
    \lambda_{\bD\bD[\bO]_M^{L=2}}^{(0)}=0\qquad\text{and}\qquad \lambda_{\bD\bD[\bO]_M^{L=2}}^{(1)}\sim\cO(\epsilon_\cT)\,,
\end{equation*}
the contribution associated to the exchanges $[\bO]_M^{L=2}$ in \eqref{eq: four-point 4D at NLO in block expansion} will be simply given by
\begin{equation}
    \epsilon_\cT f^{(1)}_{\bD^4}(z)\ni\sum_{M=0}^{\infty}\lambda_{\bD\bD[\bO]_M^{L=2}}^{(1)}\lambda_{\bD\bD[\bO]_M^{L=2}}^{(1)}\,g_{_{2{\Delta_{\bO}}+M}}(z)
\end{equation}
where only the quadratic NLO three-point data contributes. Since these contributions are overall of order $\cO(\epsilon_\cT^2)$, we can conclude that the operators $[\bO]_M^{L=2}$ do not get exchanged up to NLO.

One can then use the same logic to show that all other three-point coefficients are $\mathcal{O}(\epsilon_{\mathcal{T}})$, and thus do not contribute to the NLO expansion of the displacement four-point function, as they enter quadratically. A selection of these cases is presented in Table \ref{tab1}. It follows that the only operators which can be exchanged up to NLO in the displacement operator's four-point function are those built from the displacement itself. While this statement is made for the displacement, the same logic applies to all other operators in the displacement multiplet. 

We can therefore conclude that no interaction mixes the various sectors, and consequently, all four-point functions can be discussed independently. As we will later see with explicit examples, this implies a universality in the structure.

Before moving on to cases where some of the assumptions are relaxed, it is worth commenting on the mixing problem. A key advantage of the analysis presented in Sec. \ref{subsec: Bootstrapping the epsilon-order of OPE Coefficients} and in the present discussion is that the identification of universal sectors provides direct control over mixing: it allows us to exclude not only mixing between different components of the same multiplet, but also the exchange of higher-length operators within each sector. As a result, the exchanged operators up to NLO are restricted to multiparticle operators with length $L=2$.

The importance of this observation is twofold. On the one hand, universal sectors capture features that are independent of the specific realization of the theory, on the other, they automatically guarantee the absence of mixing up to NLO.

\begin{table}[h!]
\centering
\begin{tabular}{|c|c|c|c|}
\hline
Operator & Four-point & GFF ($\cO(1)$) & Excluded at NLO\\[0.05cm]
\hline
$[\bO]^{L=2}$ &  $\vev{\bD(t_1)\bD(t_2)\bO(t_3)\bO(t_4)}$ & $\lambda_{\bO\bO[\bO]^{L=2}}^{(0)}$ & $\lambda_{\bD\bD[\bO]^{L=2}}^{(1)}$ \\[0.05cm]
\hline
$[\bD\bO]^{L=2}$ &  $\vev{\bD(t_1)\bD(t_2)\bD(t_3)\bO(t_4)}$ & $\lambda_{\bD\bO[\bD\bO]^{L=2}}^{(0)}$ & $\lambda_{\bD\bD[\bD\bO]^{L=2}}^{(1)}$ \\[0.05cm]
\hline
$[\bD]^{L=k>2}$ &  $\vev{\bD(t_1)\bD(t_2)[\bD]^{L=r}(t_3)[\bD]^{L=s}(t_4)}\vert_{r+s=k}$ & $\lambda_{[\bD]^{L=r}[\bD]^{L=s}[\bD]^{L=k}}^{(0)}\vert_{r+s=k}$ & $\lambda_{\bD\bD[\bD\bO]^{L=k}}^{(1)}$ \\[0.05cm]
\hline
$[\bO]^{L=k}$ &  $\vev{\bD(t_1)\bD(t_2)[\bO]^{L=r}(t_3)[\bO]^{L=s}(t_4)}\vert_{r+s=k}$ & $\lambda_{[\bO]^{L=r}[\bO]^{L=s}[\bO]^{L=k}}^{(0)}\vert_{r+s=k}$ & $\lambda_{\bD\bD[\bO]^{L=k}}^{(1)}$ \\[0.05cm]
\hline
\end{tabular}
\caption{Examples of the identification of OPE coefficients that can be excluded up to NLO, involving two displacement operators and one exchanged operator.}
\label{tab1}
\end{table}

\subsubsection{Relaxing the third assumption}
\label{subsubsec: Relaxing the 3rd assumption}
Much of the discussion so far has relied on the identification of nonvanishing GFF three-point functions. Combined with the vanishing of the three-point function involving two displacement operators and a third one, this allowed us to constrain the minimal order at which the latter interaction can appear. While many of these constraints followed from consistency, the condition in \eqref{eq: 3point 2D1O} was instead imposed as an assumption.

In what follows, we relax this assumption by allowing $\cD \in \cD \times \cD$. 
In this case, \eqref{eq: 3pnt DDO LO} remains valid, but \eqref{eq: 3point 2D1O} no longer holds. We can therefore consider, in general,
\begin{equation*}
    \vev{\bD(t_1)\bD(t_2)\bO(t_3)}^{(1)}\neq0\qquad \vev{\bO(t_1)\bO(t_2)\bO(t_3)}^{(1)}\neq 0\,. 
\end{equation*}
Crucially, given that at the GFF order $\lambda^{(0)}_{\bD\bD\bO}=\lambda^{(0)}_{\bO\bO\bO}=0$ and the expansion up to NLO \eqref{eq: four-point 2D2O at NLO in block expansion}, consistency requires 
\begin{equation}
\label{eq: 3-poinnt vertex}
    \lambda^{(1)}_{\bD\bD\bO}\lambda^{(1)}_{\bO\bO\bO}\sim \cO(\epsilon_\cT) \implies \lambda^{(1)}_{\bD\bD\bO}\sim\lambda^{(1)}_{\bO\bO\bO}\sim \cO(\sqrt{\epsilon_\cT})\,.
\end{equation}
In a holographic setup, this means that diagrams of cubic exchange interactions contribute at the same order as quartic ones. 

In the CFT language, this translates into the fact that a given four-point function also exchanges other fundamental operators belonging to the supermultiplet. Being that these are theory dependent, the resulting four-point functions will in general differ from one theory to another. Notice, however, that the discussion regarding multiparticle operators remains the same as in the previous case. It then follows that the only source of mixing between different sectors is precisely the one described above.

We must also recall that in all this analysis we kept symmetries implicit. In specific situations, when symmetry constraints are also taken into account, one might find additional simplifications. 

Consider, for example, that one finds $\lambda_{\bD\bD\bO}=0$ and $\lambda_{\bD\bD\bD}\neq 0$ as arising from symmetry discussion. In that case, the displacement operator four-point function would again receive no contributions from other sectors. Nevertheless, this situation would still differ from the previous one, precisely because of the additional allowed interaction. In this case, the expansion \eqref{eq: 4point D} will factorize as
\begin{equation}
\label{eq: f1(z) NLO with Z2 odd}
    \tilde f_{\bD^4}^{(1)}(z) =  f^{(1)}_{\bD^4}(z) + (\tilde\lambda_{\bD\bD\bD}^{(1)})^2 \tilde f^{(1)}(z)\,,
\end{equation}
where $\tilde f^{(1)}_{\bD^4}(z) $ is the full NLO function with the factorized four-point function of the previous analysis, $f^{(1)}_{\bD^4}(z) $,  and the additional contribution summarized by  $\tilde f^{(1)}(z)$ with OPE coefficient $\lambda_{\bD\bD\bD}^{(1)}=\sqrt{\epsilon_\cT}\tilde\lambda_{\bD\bD\bD}^{(1)}$. If also $\lambda_{\bD\bD\bD}=0$, the system has again universality, even though it is just for this sector.

Indeed, it is important to stress that these conditions do not necessarily imply the vanishing of $\lambda_{\bO_i\bO_j\bO_k}$. So, even if $\cD\in \cD\times\cD$, when analyzing the three-point functions of the supermultiplet components, together with the constraints imposed by symmetry, we may still find either $\lambda_{\bD\bD\bO}=0$, $\lambda_{\bD\bD\bD}=0$ or both.

We stress that the difference from the earlier case is that, while before all sectors were completely decoupled from one another, here other sectors may still interact through the exchange of fundamental fields.

Although the discussion has been framed around the displacement operator, it applies more generally to any other choice of defect operator in the multiplet. This will be, for example, the case of the $1/2$ BSP Wilson line in $\cN=2$ gauge theories, \cite{Gimenez-Grau:2019hez}, discussed in Sec. \ref{subsec: universal sector in N=2 4d} where it is the superprimary that disentangles from the rest of the multiplet.

\subsubsection{The $\bZ_2$-odd sectors}
\label{subsubsec: Z2 odd sectors}
We aim to better understand the operator content exchanged at NLO when the $\mathbb{Z}_2$ symmetry is broken. As in the $\mathbb{Z}_2$-preserving case, where length $L=2$ operators are exchanged at leading and subleading order, we may expect additional contributions from the $\mathbb{Z}_2$-odd sector. In particular, the fundamental operators associated with nonvanishing three-point functions are naturally expected to appear, but higher-length operators may also be exchanged as a consequence. To keep the discussion clearer, we restrict our discussion to displacement correlators and assume that they receive no contributions from the other sector. In this setup, we therefore allow for a nonvanishing three-point function
\begin{equation}
    \langle \bD(t_1)\bD(t_2)\bD(t_3)\rangle =\frac{(\mathbb{n}_\bD C_\cT)^{3/2}\lambda_{\bD\bD\bD}^{(1)}}{t_{12}^{\Delta_\bD} t_{23}^{\Delta_\bD}t_{13}^{\Delta_\bD}}\,.
\end{equation}
In this case we explicitly restore the normalization, so that the coefficient $\lambda_{\bD\bD\bD}^{(1)}$ denotes the normalized OPE coefficient appearing in the CPW expansion.

Our first goal is to identify the additional $\mathbb{Z}_2$-odd multiparticle operators that may appear up to NLO, and to show that their three-point coefficients at this order are determined by $\lambda_{\bD\bD\bD}$. We first explain why this statement, while natural to expect, is not entirely obvious.

Unlike the $\mathbb{Z}_2$-even case, where contributions can be analyzed via Wick contractions in the free theory, no analogous procedure exists for the $\mathbb{Z}_2$-odd sector. Indeed, the $\mathbb{Z}_2$-odd $n$-point functions vanish in the GFF and become nonvanishing only when interactions are included. Consequently, one cannot directly exploit the GFF analysis to study these three-point functions. The operators appearing in this context are therefore inherently interacting and cannot, in general, be regarded as simple products of fundamental fields, unlike in the GFF case.

Specifically, when studying correlators in the GFF, one can rely on the standard definition for the point-splitting regularization, which provides a straightforward way to define composite operators and compute their three-point functions. This is given by
\begin{equation}
\label{eq: point splitting}
    [\bD]^{L}_{n}(t)= \text{D}_{L,n}(\partial_{t_i})\Bigl(\,\prod_{i=1}^{L}\bD(t_i)-\text{pairwise contractions}\Bigr)\Bigr\vert_{t_i\rightarrow t}\,,
\end{equation}
where $\text{D}_{n,L}(\partial_{t_i})$ collects all derivative contributions, whose precise form is fixed by requiring $[\bD]^{L}_{n}(t)$ to be a primary.
Once again, we are not interested in the explicit form of these operators, also because, as we are about to see, not all of them actually appear in the expansion at NLO. 

In this context, if one naively inserts such operators into $\bZ_2$-odd $n$-point functions and attempts to compute them, one finds that the singularities do not cancel and the result diverges, yielding an ill-defined correlator. As anticipated, this reflects the fact that one is effectively probing the compositeness of the multiparticle operator, which is intrinsically interacting and must therefore be regarded as a single irreducible operator. 

In order therefore to keep a similar structure of \eqref{eq: point splitting}, we have to account for possible mixing with other operators that one would find when dealing with corrections of away from the GFF theory. The extension of \eqref{eq: point splitting} to consider such contribution is given schematically by\footnote{Notice that there is not a unique way to define the regularization which, in general, will be scheme dependent.}
\begin{equation}
\label{eq: point splitting corrected}
    [\bD]^{L}_{n}(t)= \text{D}_{L,n}(\partial_{t_i})\Bigl(\,\prod_{i=1}^{L}\bD(t_i)-\text{nested  OPE contributions}\Bigr)\Bigr\vert_{t_i\rightarrow t}\,,
\end{equation}
where we subtract all the operators that can appear in the nested OPEs so that they compensate all the divergences that can appear from the limits in the point-splitting. 
The advantage of exploiting these expression is that it allows us to easily relate the OPE coefficients of $[\bD]^{L}_{n}(t)$ to those of the OPE of its components and extend \eqref{eq: point splitting} also away from the GFF.
Moreover, one does not actually need the entire OPE, as higher-dimensional operators will just contribute with regular terms. 
In order to make sense of the regularization in \eqref{eq: point splitting corrected}, we focus, for clarity, on the simplest multiparticle operator $[\bD]^{L=3}_{0}(t)$, noting that the generalization to arbitrary $L$ follows similar steps. In this regard, let us first consider the three-point function
\begin{equation}
\label{eq: 3 point DDD3}
    \langle\bD(t_1)\bD(t_2)[\bD]^{L=3}_{0}(t_3)\rangle = \frac{(\mathbb{n}_\bD C_\cT)^{5/2}\lambda^{(1)}_{\bD\bD[\bD]^{L=3}_{n}}}{t_{12}^{-\Delta_{\bD}}\,t_{13}^{ \Delta_{\bD}}\,t_{23}^{ \Delta_{\bD}}}
\end{equation}
We already considered the contribution to appear away from the GFF, so we labeled it by $\lambda^{(1)}_{\bD\bD[\bD]^{L=3}_{n}}$ and we neglected the additional contributions coming from the anomalous dimensions in the denominator, keeping just the lowest NLO contribution. 
Our purpose is now to compute the same three-point function by exploiting \eqref{eq: point splitting corrected}, which specified to this case reads
\begin{equation}
    [\bD]^{L=3}_{0}(t)= \Bigl(\bD(t_1)\bD(t_2)\bD(t_3)-\sum_{i,j,k=1}^{3}\sigma^{ijk}\sum_{\bT}\,\frac{(\mathbb{n}_{\bD}C_\cT)}{t_{ij}^{3\Delta_{\bD}-\Delta_{\bT}}}k_{\bD\bD\bT}\,\bT(t_k)\Bigr)\Bigr\vert_{t_i,t_j,t_k\rightarrow t}\,,
\end{equation}
where $\sigma^{ijk}$ account for all the permutations of the OPE couples and $\bT$ are the $\bZ_2$-odd operators appearing in $\bD\times \bD$. The $k_{\bD\bD\bT}$ coefficients give the contribution that subtract the divergences and $(\mathbb{n}_{\bD}C_\cT)$ is the two-point normalization that comes from the identity exchange. From this expression we can see that only  operators with $\Delta_{\bT}<3\Delta_\bD$ give contributions that are divergent and compensate the one that arises from the point-splitting. 
We consider now the three-point function
\begin{align}
    \langle\bD(t_1)\bD(t_2)[\bD]^{L=3}_{0}(t_3)\rangle =&\Bigl(\langle\bD(t_1)\bD(t_2)\bD(t_3)\bD(t_4)\bD(t_5)\rangle-\\
    &-\sum_{i,j,k=3}^{5}\sigma^{ijk}\sum_{\bT}\,\frac{(\mathbb{n}_{\bD}C_\cT)}{t_{ij}^{3\Delta_{\bD}-\Delta_{\bT}}}k_{\bD\bD\bT}\langle\bD(t_1)\bD(t_2)\bT(t_j)\rangle\Bigr)\Bigr\vert_{t_i\rightarrow t}\nonumber\,.
\end{align}
Under the assumption that this sector is decoupled from the others, we can see that the only contributions in $\bD\times\bD$ that are $\bZ_2$ odd are those coming from the conformal family of $\bD$. By taking the contractions and comparing with \eqref{eq: 3 point DDD3} we identify
\begin{equation}
    \lambda^{(1)}_{\bD\bD[\bD]^{L=3}_{n}} \sim \lambda^{(1)}_{\bD\bD\bD}
\end{equation}
By similar arguments one can study all the higher-length operators. However, these will require a higher number of expansions thus leading to higher powers of $\lambda^{(1)}_{\bD\bD\bD}$ and hence negligible at NLO. Therefore only multiparticle operators that can actually be exchanged in the four-point function up to NLO are the $L=3$ ones. Moreover, since we show that their three-point function is controlled by $\lambda_{\bD\bD\bD}^{(1)}$, by setting this to zero remove all the $\bZ_2$-odd exchange to the four-point function. This is why we can consider the NLO function $f^{(1)}(z)$ as factorized in $\bZ_2$-even and $\bZ_2$-odd parts \eqref{eq: f1(z) NLO with Z2 odd}, with the $\bZ_2$-odd contribution controlled by $\lambda_{\bD\bD\bD}$.

As usual, here we focus on the $\bD$-sector; however, this discussion can be generalized to all the other possibilities and also for the mixed composite multiparticle operators. For example one may expect that if we have mixing with another sector, \eg $\bO\in\bD\times\bD$, then
\begin{equation}
    \langle \bD(t_1)\bD(t_2)\bO(t_3)\sim \lambda^{(1)}_{\bD\bD\bO} \implies \langle \bD(t_1)\bD(t_2)[\bD^2 \bO]_n^{L=3}\rangle\sim\lambda_{\bD\bD [\bD^2 \bO]_n^{L=3}} \propto \lambda^{(1)}_{\bD\bD\bO}\,
\end{equation}
and so on.
We can therefore, in general, expect the four-point function up to NLO to be like
\begin{equation}
\label{eq: 4-point function at NLO with 3-point interactions}
   \langle\bD(t_1)\bD(t_2)\bD(t_3)\bD(t_4)\rangle = f^{(0)}(z)+ \epsilon f^{(1)}(z) + \sum_i(\lambda^{(1)}_{\bD\bD\bO_i})^2 f_{\bO_i}^{(1)}(z) \,,
\end{equation}
where $(\lambda^{(1)}_{\bD\bD\bO_i})^2 f_{\bO_i}^{(1)}(z) $ collect al the contributions coming from the additional interactions.

\subsubsection{Relaxing the second assumption}
In some cases, the theory contains additional fundamental operators, which we denote by $\cT_i$ for some $i\in\mathbb{N}$. One must then consider all possibilities for whether $\cD$ and $\cT_i$ appear in the OPEs $\cD\times\cD$ and $\cT_i\times\cT_j$. By repeating the analysis outlined previously and using arguments similar to those in Sec. \ref{subsec: Bootstrapping the epsilon-order of OPE Coefficients}, one can determine the minimal order at which exchanged operators appear and identify whether a given four-point function exhibits universal features.

Notably, in these enriched scenarios, the analysis applies not only to the superdisplacement but also to the other multiplets that are considered. Given the large number of possible cases, we restrict our discussion to specific examples, such as the $1/2$ BPS Wilson line in AdS$_3\times S^3\times T^4$ in Sec. \ref{subsec: the 1/2 BPS line in AdS3 T3 S4} and the $1/3$ BPS Wilson lines in ABJM in Sec. \ref{Bootstrapping the 1/3 BPS Wilson line in ABJM}.

\subsubsection{Relaxing the first assumption}
For all the cases discussed here and in the following, relaxing the first assumption will not be of particular utility. However, we discuss this possibility, leaving an in-depth study for future works.

First of all, the first thing one could do is to consider still a GFF theory at leading order, but instead of a $1d$ dCFT, one could study a higher-dimensional defect. As Wilson lines provide a realization for line defects, surface defects find many interesting applications. A known example is the $1/2$ BPS surface in $6d$ $(2,0)$ theory at large $N$ \cite{Drukker:2020swu, Drukker:2020atp}. The main philosophy of the discussion presented here may be extended to discuss also these cases. Moreover, the knowledge of this theory up to NLO order provides a stating point to study other surface defects by the identification of the universal sector. 

Another way to relax the first assumptions is to consider the leading order not to be a GFF. In this case, the whole analysis discussed in this section will not apply. However, by studying the new OPE data of the leading order, one may still consider the discussion presented in Sec. \ref{subsec: Bootstrapping the epsilon-order of OPE Coefficients} to study and constrain the NLO order OPE that are allowed by consistency with the strong coupling expansion. 

\subsection{Obtaining the full NLO contributions}
We must point out that, precisely because we are not making any assumption on $\epsilon_\cT$, this approach leaves the overall normalization of $f^{(1)}(z)$ unfixed. Indeed, in all expressions the combination $\epsilon_\cT f^{(1)}(z)$ appears. Hence, if we modify the ansatz by an overall constant factor $a\in\bR$, we can always reabsorb it through a redefinition of the perturbation parameter
\begin{equation*}
    \epsilon_\cT f^{(1)}(z)=\Bigl(\frac{\epsilon_\cT}{a}\Bigr)(a f^{(1)}(z)) = \tilde\epsilon_\cT (a f^{(1)}(z))\,.
\end{equation*}

Therefore, with the above method alone, we cannot determine whether the correct contribution is given by $\epsilon_\cT f^{(1)}(z)$ or by $\tilde\epsilon_\cT (a f^{(1)}(z))$. Without knowing the explicit expression, this redundancy, in principle, would always persist in all applications.

To completely fix the problem, one must compute $\epsilon_\cT$ independently by other means. This can be done, for instance, by exploiting the holographic dual description (when available), or by using purely CFT-based techniques \cite{Drukker:2022pxk,Gabai:2025zcs}.

That said, the results of such an analysis can still be incorporated into the previous discussion. In particular, if we manage to determine the correct normalization in a single case for which universality applies, $\hat\epsilon_\cT \hat f^{(1)}(z)$, we immediately identify the correct combination that can be applied consistently in all other situations. 

Consider the scenario where considering explicitly a theory and we managed to fix $\hat\epsilon_\cT$, by one of the above methods, to be $\hat\epsilon_\cT=\frac{1}{4\pi^2 C_\cT}$. This will be the actual one; see Sec. \ref{subsec: universality ABJM N4 SCSm Th}. The expression for the superprimary four-point function would be
\begin{equation*}
    \vev{\bP(t_1)\bP(t_2)\bP(t_3)\bP(t_4)} = \frac{C^2_\cT}{t_{12}^{2\Delta_{\bP}}t_{34}^{2\Delta_{\bP}}}\Bigl({\hat{f}_{\bP^4}^{(0)}}(z)+\Bigl(\frac{1}{4\pi^2 C_\cT}\Bigr) \hat{f}_{\bP^4}^{(1)}(z)+\cO(\epsilon_\cT^2)\Bigr)\,,
\end{equation*}
where now ${\hat{f}_{\bP^4}^{(0)}}(z)$ and $\hat{f}_{\bP^4}^{(1)}(z)$ are explicitly known functions. Notice that now it is not anymore possible to rescale freely $\hat{f}_{\bP^4}^{(1)}(z)$ as such a factor cannot be absorbed in $\epsilon_\cT$ being everything explicit.

Notice also that the parameter of the expansion must be, for consistency, the same for all the superdisplacement components. This implies that if we consider a four-point function of the $Q$-descendants in the supermultiplet, these  are related by supersymmetry to ${\hat{f}_{\bP^4}^{(0)}}(z)$ and $\hat{f}_{\bP^4}^{(1)}(z)$. 
Thus one can write explicitly the four-point function as given by
\begin{equation}
\label{eq: 4D universal}
    \vev{\bD(t_1)\bD(t_2)\bD(t_3)\bD(t_4)}=\frac{(\mathbb{n}_{\bD}C_{\cT})^2}{t_{12}^{4}t_{34}^{4}}\Bigl(\hat{f}_{\bD^4}^{(0)}(z)+\Bigl(\frac{1}{4\pi^2 C_\cT}\Bigr)\hat{f}_{\bD^4}^{(1)}(z)+\cO(\epsilon_\cT^2)\Bigr)\,.
\end{equation}

What we may notice is that $\epsilon_\cT$ depends only on $C_\cT$ and not $\mathbb{n}_{\bD}$. This is true also for all the other four-point functions in the multiplet. So, $\epsilon_\cT(C_\cT)$ is universal too and can be identified together with $\hat{f}_{\bD^4}^{(1)}(z)$. 

This implies, for example, that if the theory $\cT$ is such that it satisfies all the three conditions, if we have another theory having another line defect, say $\cT'$, also satisfying all the conditions, its displacement four-point function must be specified by
\begin{equation*}
    \vev{\bD(t_1)\bD(t_2)\bD(t_3)\bD(t_4)}=\frac{(\mathbb{n}'_{\bD}C_{\cT'})^2}{t_{12}^{4}t_{34}^{4}}\Bigl(\hat{f}_{\bD^4}^{(0)}(z)+\Bigl(\frac{1}{4\pi^2 C_{\cT'}}\Bigr)\hat{f}_{\bD^4}^{(1)}(z)+\cO(\epsilon_{\cT'}^2)\Bigr)\,,
\end{equation*}
differing from \eqref{eq: 4D universal} just by the different constants $\mathbb{n}'_{\bD}$ and $C_{\cT'}$.

When the defects have the same codimension this identification is typically straightforward, when instead the codimension of the two line defects, $\cT$ and $\cT'$, is different, the equivalence is true upon decomposing the tensor structures; see Sec. \ref{subsec: Universality in 1/2 BPS wilson line N=4 SYM}. 

The same holds also for all the other defect four-point functions when $\cD\notin\cD\times\cD$ and can be generalized for other cases, as discussed before, when relaxing either the second or the third condition. 

It is worth adding a concluding remark regarding the holographic interpretation of dual theory, when available. Recall that we fixed the product $\epsilon_\cT f^{(1)}(z)$ through explicit evaluation. However, due to the freedom of rescaling it may well be that the true parameter matching the holographic analysis is not $\epsilon_\cT(C_\cT)$ itself, but rather $a\,\epsilon_\cT(C_\cT)$ up to a rescaling by some constant $a$. What is guaranteed to match in explicit computations is not the parameter alone but the product of the two contributions $\hat\epsilon_\cT \hat f^{(1)}(z)$.

Having established these universal features, the purpose for the remainder of the article will be twofold: first, to verify these properties in known examples, identifying in such cases universal sectors, and second, to apply them to new cases where, thanks to this argument, we can immediately fix the NLO contributions without relying on any additional technique.

\section{Universal sectors in $1/2$ BPS line defects}
\label{sec: Universal sectors in 1/2 BPS Wilson lines}
In this section, we apply the framework developed so far to identify universal four-point functions in several examples of $1/2$ BPS Wilson lines. To make this universality explicit, we will always work with examples in pairs, comparing two different theories to demonstrate the equivalence, and hence the universal nature, of the identified four-point functions.

\subsection{The $1/2$ BPS Wilson lines in ABJM and in $\cN=4$ CSm theories}
\label{subsec: universality ABJM N4 SCSm Th}
As a first example, we consider line defects of the same codimension. In particular, we take the $1/2$ BPS Wilson line in ABJM theory \cite{Bianchi:2020hsz} and in $\cN=4$ Chern-Simons-matter theories \cite{Pozzi:2024xnu}, under the framework discussed in Sec. \ref{subsec: Correlators of the displacement multiplet}. The defect operators appearing in the two theories are characterized by the preserved superconformal algebra on the line. More precisely, they are classified by the quantum numbers of the bosonic subalgebra $\so(2,1)_{\text{conf}}\oplus\su(3)_{\cR}\oplus\u(1)_{J_0}$ for the ABJM line, and $\so(2,1)_{\text{conf}}\oplus\su(2)_{\cR}\oplus\u(1)_{J_0}$ for the $\cN=4$ CSm one \cite{Agmon:2020pde}.

We can now begin by verifying that all the required conditions are satisfied.
First, let us recall that in both cases the only elementary operators that appear are those belonging to the displacement supermultiplet. Schematically, these are specified by
\begin{equation}
\label{eq: D in ABJM and N=4 SCSM}
\begin{array}{lllll}
\cD\text{ in ABJM}&\qquad  [\ft{3}{2}]^{(0,0)}_{\frac{1}{2}} \longrightarrow &  [2]^{(1,0)}_{1} \longrightarrow &  \left[\ft{5}{2}\right]_{\frac{3}{2}}^{(0,1)} \longrightarrow &
  [3]_{2}^{(0,0)}\; , \\
  &&&& \\
\cD\text{ in }\cN =4 \text{ CSm}&\qquad  & [1]_{1}^{(0)} \longrightarrow & \left[1\right]_{\frac{3}{2}}^{(1)} \longrightarrow & [1]_{2}^{(0)}\; ,
\end{array}
\end{equation}
where the charge assignment is schematically written as $[j_0]_{\Delta}^{(\mathcal{R})}$, with $(\mathcal{R})$ standing for the Dynkin labels of the considered representation and $j_0$ the Abelian charge. For the two cases the labels for the R-symmetries refer to the $\su(3)$ for the line in ABJM and $\su(2)$ for the $\cN=4$ CSm case. The arrows represent schematically the action of the supercharges in the respective theories.

It follows, from the discussion in Sec. \ref{subsec: Correlators of the displacement multiplet}, that both theories satisfy the three assumptions we required. Hence, in both cases, when considering the four-point function of a single defect operator, no operators from the other component of the multiplets is exchanged. Consequently, the four-point functions are universal up to NLO and hence the correlation function of one theory can be directly mapped to that of the other.

This observation is particularly useful when bootstrapping the four-point functions in $\cN=4$ CSm. Indeed, imposing standard constraints such as crossing symmetry and consistency conditions does not suffice to completely fix the correlator, essentially due to the reduced supersymmetry that limits the problem. By contrast, once one notices the universal feature, one can bootstrap the full function up to NLO without resorting to the usual ansatz-driven bootstrap procedure.

For concreteness, let us consider the displacement four-point function \eqref{eq: 4point D}. We focus on this case because the displacement operator is neutral under the R-symmetry in both theories, which makes the identification particularly transparent. The discussion, however, applies more generally to correlation functions of compatible operators, namely operators with the same conformal dimension but transforming in different representations, \eg the tilt operators. In such cases, one must first choose an appropriate embedding in order to place the correlators on the same footing as discussed in the next sections.

In the presence of chiral and antichiral superdisplacement multiplets, the displacement operators are naturally organized into complex combinations that make the $U(1)$ symmetry manifest. From this perspective, while the two displacement operators differ by their Abelian charge, this distinction plays no role in the $\bD \times \bar\bD$ OPE, which involves only neutral combinations. With this in mind, we can now look at the four-point function that reads
\begin{align}
\label{eq: 4D in ABJM}
\langle\mathbb{D}(t_1)\bar{\mathbb{D}}(t_2)\mathbb{D}(t_3)\bar{\mathbb{D}}(t_4)\rangle_{\text{ABJM}}=&
\frac{(12 C_{\text{ABJM}})^2}{t_{12}^4 t_{34}^4}\, 
\Bigl[1+z^4+\epsilon_{\text{ABJM}}\Bigl(-16 -2z -\frac{7z^2}{3} -2z^3-16z^4
\cr
+\Bigl(6 - \frac{16}{z}+&6 z^4 -16z^5\Bigr)\log(1-z) + 2z^4 (8z-3)\log(-z)\Bigr)+\cO(\epsilon_{\text{ABJM}}^2)\Bigr]\,,
\end{align}
for the line in ABJM. For the line in $\cN=4$ CSm (SCSm) theories the four-point function will be, besides overall theory-specific constants, the very same expression
\begin{align}
\langle\mathbb{D}(t_1)\bar{\mathbb{D}}(t_2)\mathbb{D}(t_3)\bar{\mathbb{D}}(t_4)\rangle_{\text{SCSm}}=&
\frac{(6 C_{\text{SCSm}})^2}{t_{12}^4 t_{34}^4}\, 
\Bigl[1+z^4+\epsilon_{\text{SCSm}}\Bigl(-16 -2z -\frac{7z^2}{3} -2z^3-16z^4
\cr
+\Bigl(6 - \frac{16}{z}+ &6 z^4 -16z^5\Bigr)\log(1-z) + 2z^4 (8z-3)\log(-z)\Bigr)+\cO(\epsilon_{\text{SCSm}}^2)\Bigr].
\end{align}
In \cite{Pozzi:2024xnu} the same conclusion was reached by observing that certain $\cN=4$ CSm theories can be obtained from ABJM through specific orbifold constructions. Here we strengthen this result by showing that it indeed holds for all such $1/2$ BPS line in $\osp(4\vert4)$ theories, that also satisfy the three requirements outlined in Sec. \ref{subsec: Correlators of the displacement multiplet}.

Moreover, as discussed, we can also identify the perturbation parameter $\epsilon_\cT(C_\cT)$. This has been explicitly evaluated for ABJM in different ways \cite{Bianchi:2020hsz,Drukker:2022pxk}, giving
\begin{equation}
\label{eq: epsilon ABJM}
    \epsilon_{\text{ABJM}} = \frac{1}{4\pi^2 C_\text{ABJM}}\,,
\end{equation}
where $C_\text{ABJM}$ is superdisplacement and superprimary normalization. Also in $\cN=4$ CSm theories the expansion parameter has been computed in \cite{Pozzi:2024xnu} and, after matching the conventions, one has
\begin{equation}
    \epsilon_{\text{SCSm}} = \frac{1}{4\pi^2 C_\text{SCSm}}\,,
\end{equation}
where now $C_\text{SCSm}$ is the normalization of the superfield and superprimary of the $1/2$ BPS defect of $\cN=4$ CSm theories. For both of them we have, as anticipated
\begin{equation}
\label{eq: epsilon universal}
    \epsilon_\cT(C_\cT) = \frac{1}{4\pi^2 C_\cT}\,,
\end{equation}
where $\cT$ can be seen either as $\cT=\text{ABJM}$ or $\cT=\text{SCSm}$.

In this case, we have preformed the identification for the displacement operator, but a similar analysis applies also for all the other four-point functions of defect operators in \eqref{eq: D in ABJM and N=4 SCSM} that have the same conformal dimension.

\subsection{The $1/2$ BPS Wilson line in $\cN=4$ SYM}
\label{subsec: Universality in 1/2 BPS wilson line N=4 SYM}
We now turn to another interesting case, namely the exploration of universality across defect lines of different codimensions. In particular, we consider the $1/2$ BPS Wilson line in ABJM and in $\cN=4$ SYM, building on the results obtained in \cite{Bianchi:2020hsz,Giombi:2017cqn, Liendo:2018ukf, Ferrero:2023gnu, Ferrero:2023znz}. The Wilson line in $\cN=4$ SYM preserves the bosonic subalgebra $\so(2,1)_{\text{conf}}\oplus\so(3)_{\text{rot}}\oplus \mathfrak{so}(5)_\cR$.\footnote{In \cite{Agmon:2020pde} $\so(2,1)_{\text{conf}}\oplus\su(2)_{\text{rot}}\oplus \mathfrak{usp}(4)_\cR$. At the level of algebra this is equivalent given that, we recall, $\mathfrak{usp}(4)\simeq\so(5)_{R}$ and $\su(2)_{\text{rot}}\simeq\so(3)_{\text{rot}}$}

In this case, the dual holographic description is well understood, and much of the discussion can indeed be supported by direct arguments from the bulk perspective. In this regard, one can perform very similar considerations to those presented in Sec. \ref{sec:The holographic perspective}. Here, however, we restrict the discussion to a purely CFT-based analysis. 

We begin by collecting the superdisplacement multiplets associated with the two $1/2$ BPS lines.
\begin{equation}
\begin{array}{llllll}
\cD\text{ in ABJM} &\qquad & [\ft{3}{2}]^{(0,0)}_{\frac{1}{2}} \longrightarrow &  [2]^{(1,0)}_{1} \longrightarrow &  \left[\ft{5}{2}\right]_{\frac{3}{2}}^{(0,1)} \longrightarrow &
  [3]_{2}^{(0,0)}\; , \\
  &&&&& \\
\cD\text{ in } \text{ SYM} &\qquad & & [0]_{1}^{(0,1)} \longrightarrow & \left[1\right]_{\frac{3}{2}}^{(1,0)}  \longrightarrow & [2]^{(0,0)}_{2}  \; , \\
\end{array}
\end{equation}
As previously introduced, the Dynkin labels in round and square brackets refer to the representations of $\su(3)_{\cR}$ and $\u(1)_{j_0}$ for the Wilson line in ABJM, while they correspond to $\so(5)_{R}$ and $\so(3)_{\text{rot}}$ for the Wilson line in SYM.\footnote{See \cite{Agmon:2020pde} for a detailed discussion of the representations.}
Crucially, by studying the OPE relations for the superdisplacement multiplet \cite{Ferrero:2023znz, Ferrero:2023gnu}, one finds again that $\cD \notin \cD \times \cD$, so all the required conditions are satisfied. 

However, in contrast to the previous case, there is no straightforward identification between the two theories, even for the displacement operator. Being different codimensional defects, the displacements transform under different rotation groups. To enable a meaningful comparison, one must first rearrange the expressions.

We therefore make the $SO(2)$ structure explicit by rewriting the displacements of the line in ABJM as 
\begin{equation*}
\mathbb{D}(t)=\frac{1}{\sqrt{2}}(\mathbb{D}^1(t)+i\,\mathbb{D}^2(t)) \qquad \text{and}\qquad\bar{\mathbb{D}}(t)=\frac{1}{\sqrt{2}}(\mathbb{D}^1(t)-i\,\mathbb{D}^2(t))\,,
\end{equation*}
such that $\mathbb{D}^\ii(t)$ transforms in the fundamental of $SO(2)$. The four-point function can thus be expressed as
\begin{equation}
\label{eq: from U(1) to SO(2)}
    \langle\mathbb{D}(t_1)\bar{\mathbb{D}}(t_2)\mathbb{D}(t_3)\bar{\mathbb{D}}(t_4)\rangle_{\text{ABJM}} = \frac{1}{4}\sum \sigma_{\ii\ji\ki\li}\langle\mathbb{D}^{\text{i}}(t_1){\mathbb{D}}^\ji(t_2)\mathbb{D}^\ki(t_3){\mathbb{D}}^\li(t_4)\rangle_{\text{ABJM}}\,,
    \end{equation}
where $\ii,\ji,\ki,\li = 1,2$ and $\sigma_{\ii\ji\ki\li}=\delta_{\ii\ji}\delta_{\ki\li}+\delta_{\ii\li}\delta_{\ji\ki}-\delta_{\ii\ki}\delta_{\ji\li}$, and the correlators on the right-hand side can be written as
\begin{equation}
\label{eq: 4D in  SO(2)}
    \langle\mathbb{D}^{\text{i}}(t_1){\mathbb{D}}^\ji(t_2)\mathbb{D}^\ki(t_3){\mathbb{D}}^\li(t_4)\rangle_{\text{ABJM}} = \frac{\cC_{\text{ABJM}}^2}{t_{12}^4t_{34}^4}F^{\ii\ji\ki\li}(z)\,,
\end{equation}
for some function $F^{\ii\ji\ki\li}(z)$. For convenience, here we have absorbed all the overall factors in $\cC_{\text{ABJM}}$. 

Following the analysis in \cite{Giombi:2017cqn}, it is convenient to reexpress \eqref{eq: 4D in SO(2)} in terms of the singlet $F_S(z)$, the symmetric traceless $F_T(z)$, and the antisymmetric $F_A(z)$ contributions associated with the two pairs of indices. This is done by taking the decomposition
\begin{equation}
\label{eq: FSO2}
    F^{\ii\ji\ki\li}(z) =\delta^{\ii\ji}\delta^{\ki\li} \,F_S(z)+(\delta^{\ii\ji}\delta^{\ki\li}-\delta^{\ii\li}\delta^{\ji\ki}) F_A(z) + (\delta^{\ii\ki}\delta^{\ji\li}+\delta^{\ii\li}\delta^{\ji\ki}-\,\delta^{\ii\ji}\delta^{\ki\li})F_T(z)\,.
\end{equation}

Now we move to the four-point insertions in the $1/2$ BPS $\cN=4$ SYM line. The correlator of the displacement, for this case, reads
\begin{equation}
\label{eq: four-point displacement SO(3)}
    \langle\mathbb{D}^{i}(t_1){\mathbb{D}}^{j}(t_2)\mathbb{D}^{k}(t_3){\mathbb{D}}^{l}(t_4)\rangle_{\text{SYM}} = \frac{\cC_{\text{SYM}}^2}{t_{12}^4t_{34}^4}G^{ijkl}_{SO(3)}(z)\,,
\end{equation}
where $i,j,k,l=1,2,3$ and all the overall constant and the superprimary normalization $C_{SYM}$ have been collected in $\cC_{SYM}$.
Also in this case $G^{ijkl}_{SO(3)}(z)$ can be decomposed in singlet $G_S(z)$, symmetric traceless $G_T(z)$ and antisymmetric $G_A(z)$ contributions, thus leading to
\begin{equation}
\label{eq: GSO3}
    G^{ijkl}_{SO(3)}(z) =\delta^{ij}\delta^{kl} \,G_S(z)+(\delta^{ij}\delta^{kl}-\delta^{il}\delta^{jk}) G_A(z) + (\delta^{ik}\delta^{jl}+\delta^{il}\delta^{jk}-2/3\,\delta^{ij}\delta^{kl})G_T(z)\,.
\end{equation}
Crucially, in order to compare the two expressions we need to decompose \eqref{eq: GSO3} into $SO(2)$ representations. This can be done by using the standard embedding of $SO(2)$ into $SO(3)$, realized as rotations about the third axis. This leads to the expression
\begin{equation}
\label{eq: GSO2}
    G^{\ii\ji\ki\li}_{SO(2)}(z) =\delta^{\ii\ji}\delta^{\ki\li} \,(G_S(z)+1/3 G_T(z))+(\delta^{\ii\ji}\delta^{\ki\li}-\delta^{\ii\li}\delta^{\ji\ki}) G_A(z) + (\delta^{\ii\ki}\delta^{\ji\li}+\delta^{\ii\li}\delta^{\ji\ki}-\,\delta^{\ii\ji}\delta^{\ki\li})G_T(z)\,,
\end{equation}
which is precisely on the form of \eqref{eq: FSO2}, upon the identifications 
\begin{equation*}
    F_S(z)=G_S(z)+1/3\,G_T(z)\,,\qquad F_T(z)=G_T(z)\,,\quad \text{and}\quad F_A(z)=G_A(z)\,.
\end{equation*}

Based on the preliminary considerations, we expect universal features up to NLO, which implies a matching between the two four-point functions $F^{\ii\ji\ki\li}(z)=G^{\ii\ji\ki\li}_{SO(2)}(z)$. 

We can verify this by substituting \eqref{eq: 4D in  SO(2)} in \eqref{eq: from U(1) to SO(2)} with the identification \eqref{eq: GSO2}, thus obtaining
\begin{align}
\label{eq: 4D in ABJM and SYM }
    \langle\mathbb{D}(t_1)\bar{\mathbb{D}}(t_2)\mathbb{D}(t_3)\bar{\mathbb{D}}(t_4)\rangle_{\text{ABJM}} =& \frac{\cC_{\text{ABJM}}^2}{t_{12}^4t_{34}^4}\Bigl[G_S^{(0)}(z)-G_A^{(0)}(z)+\frac{G_T^{(0)}(z)}{3}+\nonumber\\
    +&\epsilon_{\text{ABJM}}\Bigl(G_S^{(1)}(z)-G_A^{(1)}(z)+\frac{G_T^{(1)}(z)}{3}\Bigr)+\cO(\epsilon_{\text{ABJM}}^2)\Bigr]\,.
\end{align}
This expression for the four-point of the displacement in the ABJM line is fully specified by the functions obtained by the analysis for the line in SYM. 

The last step is thus to explicitly evaluate the four-point functions with the results obtained in \cite{Giombi:2017cqn, Liendo:2018ukf} and compare the outcome with the bootstrap result of \cite{Bianchi:2020hsz}. 

At leading order they read
\begin{equation*} 
G^{(0)}_S(\chi) = 1 + \frac{2}{3}G^{(0)}_T(\chi)\, , \quad 
G^{(0)}_T(\chi) = \frac{1}{2}\left(\chi^2 + \frac{\chi^2}{(1-\chi)^2}\right)\, , \quad 
G^{(0)}_A(\chi) = \frac{1}{2}\left(\chi^2 - \frac{\chi^2}{(1-\chi)^2}\right)\, ,
\end{equation*}
and at the next-to-leading order they are found to be
\begin{align}
{G_S^{(1)}(\chi)}=&
-\frac{2\left(24 \chi ^8-90 \chi ^7+125 \chi ^6-76 \chi ^5+125 \chi ^4-306 \chi ^3+438 \chi ^2-288 \chi +72\right)}{9
   (\chi -1)^4}
\cr
&
-\frac{4  \left(4 \chi ^6-\chi ^5-6 \chi +12\right)}{3 \chi }\log (1-\chi)
\nonumber\\
&
+\frac{4  \chi ^4 \left(4 \chi ^6-21 \chi ^5+45 \chi ^4-50 \chi ^3+30 \chi ^2-6 \chi +2\right)}{3 (\chi -1)^5}\log ( \chi )\label{eq: GS of SO(3)}\ , 
\\[0.3cm]
{G_T^{(1)}(\chi)}=&
-\frac{\left(48 \chi ^4-198 \chi ^3+313 \chi ^2-230 \chi+115\right) \chi ^4}{6 (\chi -1)^4}
- (8 \chi -5) \chi ^4 \log ( 1-\chi )
\nonumber\\
&
+\frac{\left(8 \chi ^6-45 \chi ^5+105 \chi ^4-130 \chi ^3+90 \chi ^2-30 \chi+10\right) \chi ^4 }{(\chi -1)^5}\log (\chi )
\ ,\label{eq: GT of SO(3)} \\[0.3cm]
{G_A^{(1)}(\chi)} =&
-\frac{(\chi -2) \left(48 \chi^6-90 \chi ^5+91 \chi ^4+4 \chi ^3-17 \chi ^2+18 \chi -6\right) \chi }{6 (\chi -1)^4}
\cr
&
-\left(8 \chi ^5-3 \chi ^4+2\right) \log ( 1-\chi )
\nonumber\\
&
+\frac{(\chi -2) \left(8 \chi ^4-27 \chi^3+41 \chi ^2-28 \chi +14\right) \chi ^5 }{(\chi -1)^5}\log ( \chi )\ ,
\end{align}
where we kept $\chi=\frac{z}{z-1}$ for keeping the expression concise.\footnote{\label{foot1}Compared to \cite{Giombi:2017cqn,Liendo:2018ukf} we have rescaled these function by a factor of 2 to have the discussion more straightforward. See \eqref{eq: matching SYM four-points}
}

By substituting explicitly these expressions in \eqref{eq: 4D in ABJM and SYM } one obtains the displacement four-point function \eqref{eq: 4D in ABJM}. We can thus conclude, also in this case, that the $1/2$ BPS line in SYM has universal features.

We end this section with the explicit check that, through this analysis, we obtain not only the NLO function but indeed also the full NLO contribution, including the correct expansion parameter. From this perspective, let us consider the full expressions in the form of \eqref{eq: 4D universal}, where we explicitly take $\epsilon_\cT$ as given in \eqref{eq: epsilon universal}. These are
\begin{align}
    \langle\mathbb{D}^i(t_1){\mathbb{D}}^j(t_2)\mathbb{D}^k(t_3){\mathbb{D}}^l(t_4)\rangle_{\text{ABJM}} &= \frac{\cC_{\text{ABJM}}^2}{t_{12}^4t_{34}^4}\Bigl(F_{(0)}^{ijkl}(z)+\Bigl(\frac{1}{4\pi^2 C_{\text{ABJM}}}\Bigr)F_{(1)}^{ijkl}(z)+...\Bigr)\,,\\
    \langle\mathbb{D}^{i}(t_1){\mathbb{D}}^{j}(t_2)\mathbb{D}^{k}(t_3){\mathbb{D}}^{l}(t_4)\rangle_{\text{SYM}} &= \frac{\cC_{\text{SYM}}^2}{t_{12}^4t_{34}^4}\Bigl(G^{ijkl}_{(0)}(z)+\Bigl(\frac{1}{4\pi^2 C_{\text{SYM}}}\Bigr)G_{(1)}^{ijkl}(z)+...\Bigr)\,.\label{eq: four-point SYM displacement}
\end{align}
For notational convenience, we use the subscript to denote the order of the expansion and set $G^{ijkl}(z) = G^{ijkl}_{SO(2)}(z)$. From the discussion above, we have the identifications 
\begin{equation*}
    F^{ijkl}_{(0)}(z) = G^{ijkl}_{(0)}(z)\qquad\text{and}\qquad F^{ijkl}_{(1)}(z) = G^{ijkl}_{(1)}(z)\,.
\end{equation*}

In the case of $\cN=4$ SYM, the bremsstrahlung function at strong coupling is known, \cite{Correa:2012at,Drukker:2011za}. We can thus explicitly write the two-point constant $C_{\text{SYM}}$ in terms of the 't~Hooft coupling which is given by
\begin{equation*}
C_{\text{SYM}} =2B(\lambda_{\text{SYM}})= \frac{\sqrt{\lambda_{\text{SYM}}}}{2\pi^2} \, .
\end{equation*}
Accordingly, Eq. \eqref{eq: four-point SYM displacement} can be written explicitly as
\begin{equation*}
    \langle\mathbb{D}^{i}(t_1){\mathbb{D}}^{j}(t_2)\mathbb{D}^{k}(t_3){\mathbb{D}}^{l}(t_4)\rangle_{\text{SYM}} = \frac{\cC_{\text{SYM}}^2}{t_{12}^4t_{34}^4}\Bigl(G^{ijkl}_{(0)}(z)+\frac{1}{2\sqrt{\lambda_{\text{SYM}}}}\,G_{(1)}^{ijkl}(z)+...\Bigr)\,,
\end{equation*}
which, when evaluated explicitly, matches the expressions in \cite{Giombi:2017cqn, Liendo:2018ukf} which are written equivalently in terms of $\tilde{G}_{(1)}^{ijkl}(z)$, under the following redefinition:
\begin{equation}  
\label{eq: matching SYM four-points}
\frac{1}{2\sqrt{\lambda_{\text{SYM}}}}\,G_{(1)}^{ijkl}(z)=\frac{1}{\sqrt{\lambda_{\text{SYM}}}}\frac{G_{(1)}^{ijkl}(z)}{2}=\frac{1}{\sqrt{\lambda_{\text{SYM}}}}\,\tilde{G}_{(1)}^{ijkl}(z)\,.
\end{equation}

As argued at the end of Sec. \ref{subsec: Correlators of the displacement multiplet}, this reflects the fact that, through the identification of the universal sectors, we are matching the entire NLO contribution rather than isolating the expansion parameter $\epsilon_\cT$ itself. The latter can always be rescaled by a constant, which can then be absorbed into the definition of the NLO function, as illustrated in the example above.

\subsection{The $1/2$ BPS line defects in $\cN=2$ gauge theories}
\label{subsec: universal sector in N=2 4d}
 
We now turn to the case of $1/2$ BPS Wilson lines in $\cN=2$ gauge theories, investigated in \cite{Gimenez-Grau:2019hez}, where we examine the possible presence of universal sectors in the displacement supermultiplet four-point functions. These theories are characterized by the preserved $\osp(4^*\vert 2)$ subalgebra of the ambient $\su(2,2\vert 2)$ superconformal algebra in $4d$. Accordingly, the defect operators are classified by the quantum numbers of the preserved bosonic subalgebra, namely $\so(2,1)_{\text{conf}},\oplus \su(2)_{\text{rot}} \oplus \su(2)_{\cR}$.

Among the examples studied so far, this is the first case in which the third condition is not satisfied, since the explicit analysis shows that $\cD \in \cD \times \cD$. 

Nevertheless, as discussed in Sec. \ref{subsec: Correlators of the displacement multiplet}, one may still expect that some universal subsectors can be identified if, by symmetry constraints, some of the three-point functions of the individual components vanish. Indeed, it turns out that this is the case, leaving room for universal structures to emerge.

As before, we begin by analyzing the displacement supermultiplets. To make the comparison clearer, we shall again draw a parallel with the $1/2$ BPS Wilson line in $\cN=4$ SYM. From \cite{Agmon:2020pde}, the displacement multiplets are specified by
\begin{equation}
\begin{array}{lllll}
\cD\text{ in } \text{ SYM}\qquad & [0]_{1}^{(0,1)} \longrightarrow & \left[1\right]_{\frac{3}{2}}^{(1,0)}  \longrightarrow & [2]^{(0,0)}_{2}  \; , \\
  &&&& \\
\cD\text{ in } \cN=2 \text{ theories}\qquad & [0]_{1}^{(0)} \longrightarrow & \left[1\right]_{\frac{3}{2}}^{(1)}  \longrightarrow & [2]^{(0)}_{2}  \; ,
\end{array}
\end{equation}
 where, also in this case, the arrows represent the $Q$ actions and the Dynkin labels refer to the corresponding preserved algebra. For the displacement in SYM this is the same as in Sec. \ref{subsec: Universality in 1/2 BPS wilson line N=4 SYM}, while for the $\osp(4^*\vert2)$ one, the only difference is on the R-charge Dynkin labels associated with $\su(2)_{\cR}$. 
 
 Hence, we define the defect operators as
\begin{equation*}
    [0]_1^{(0)}\rightarrow \bP(t)\,,\qquad [1]_{\frac{3}{2}}^{(1)}\rightarrow\mathbb{\Lambda}_{\a}^a(t)\,,\qquad [2]_{2}^{(0)}\rightarrow\bD^i (t)\,.
\end{equation*}
Given that the displacement enters on the three-point functions we cannot be generic on the sector to study, but we have to identify the ones that do not mix. In particular, one may notice that the three-point functions
\begin{align}
    \langle \bP(t_1)\bP(t_2)\mathbb{\Lambda}_\alpha^a(t_3)\rangle &= 0\,,\\
    \langle \bP(t_1)\bP(t_2)\bD^i(t_3)\rangle &= 0\,,
\end{align}
vanish at all orders. This constraints are given purely by symmetry preserving argument. Thus, even if the superdisplacement appears in its OPE and its three-point is nonvanishing, one can still find some vanishing contributions by studying the three-point functions of the components.  In this case, a dual theory is not known, but the constraint would imply that, in any Lagrangian description of the holographic dual, the fluctuations associated with $\bP(t)$ admit no cubic interactions with the duals to $\mathbb{\Lambda}^a_\a(t)$ and $\mathbb{D}^i(t)$.

Although $\mathbb{Z}_2$-odd contributions may arise, as the three-point function of $\bP$ is not generically zero, we shall not pursue their analysis here and will only comment on them at the end. We thus set as a working hypothesis $\lambda_{\bP\bP\bP}=0$.

We now use the methodology developed up so far to obtain the contributions up to NLO of the four-point function
\begin{equation}
\label{eq: 4 P in N=2}
        \langle\mathbb{P}(t_1){\mathbb{P}}(t_2)\mathbb{P}(t_3){\mathbb{P}}(t_4)\rangle = \frac{C_{\Phi}^2}{t_{12}^2 t_{34}^2}\Bigl(F^{(0)}(z)+\Bigl(\frac{1}{4\pi^2 C_{\Phi}}\Bigr)F^{(1)}(z)+...\Bigr)\,,
\end{equation}
where $C_{\Phi}$ is the two-point function normalization of the superdisplacent multiplet.

As in the previous case, we begin from the known expressions for the line in $\cN=4$ SYM and we deduce the result from them.

We start then from the superprimary four-point function of SYM, given by
\begin{equation*}
        \langle\mathbb{P}^{a}(t_1){\mathbb{P}}^{b}(t_2)\mathbb{P}^{c}(t_3){\mathbb{P}}^{d}(t_4)\rangle_{\text{SYM}} = \frac{C_{\text{SYM}}^2}{t_{12}^2 t_{34}^2}\Bigl(G^{abcd}_{(0)}(z)+\Bigl(\frac{1}{4\pi^2 C_{\text{SYM}}}\Bigr)G_{(1)}^{abcd}(z)+...\Bigr)\,,
\end{equation*}
with $a,b,c,d=1,...,5$ where now $G^{abcd}_{(0)}(z)$ and $G^{abcd}_{(1)}(z)$ are the leading and next-to-leading order of $G^{ijkl}_{SO(5)}(z)$ defined as
\begin{equation}
\label{eq: so(5) function}
    G^{abcd}_{SO(5)}(\chi) =\delta_{ab}\delta_{cd} \,G_S(\chi)+(\delta_{ab}\delta_{cd}-\delta_{ad}\delta_{bc}) G_A(\chi) + (\delta_{ac}\delta_{bd}+\delta_{ad}\delta_{bc}-2/5\,\delta_{ab}\delta_{cd})G_T(\chi)\,.
\end{equation}
In the present case, $G_S(z)$, $G_T(z)$ and $G_A(z)$ are the singlet, symmetric traceless and  antisymmetric contributions of $G^{abcd}_{SO(5)}(z)$. 

In order to extract the full trace, we define the projector\footnote{The normalization can be either fixed so that it matches the GFF results or by group theory arguments, \eg $\cP_{ijkl}\delta^{ij}\delta^{kl}=1$ and so on.}
\begin{equation*}
    \cP_{abcd}=\frac{1}{35}(\delta_{ab}\delta_{cd}+\delta_{ac}\delta_{bd}+\delta_{ad}\delta_{bc})\,.
\end{equation*}
By contracting with \eqref{eq: so(5) function}, we just get
\begin{equation*}
    F(z) = \cP_{abcd}\,G^{abcd}_{SO(5)}(z) = G_S(z)+\frac{8}{5}G_T(z)
\end{equation*}
and so \eqref{eq: 4 P in N=2} can be simply rewritten explicitly as
\begin{equation}
\label{eq: 4 P explicit in GS GT}
        \langle\mathbb{P}(t_1){\mathbb{P}}(t_2)\mathbb{P}(t_3){\mathbb{P}}(t_4)\rangle = \frac{C_{\Phi}^2}{t_{12}^2 t_{34}^2}\Bigl[G_S^{(0)}(z)+\frac{8}{5}G_T^{(0)}(z)+\Bigl(\frac{1}{4\pi^2 C_{\Phi}}\Bigr)\Bigl(G_S^{(1)}(z)+\frac{8}{5}G_T^{(1)}(z)\Bigr)+...\Bigr]\,.
\end{equation}
The expression \eqref{eq: 4 P explicit in GS GT} makes manifest the relation between the Wilson line in $\cN=2$ gauge theories and the one in SYM. This allows us to rely on the explicit evaluation of the corresponding functions of the line in SYM \cite{Giombi:2017cqn,Liendo:2018ukf} to compute the four-point function in the present case. These are given by
\begin{equation*}
G_S^{(0)}(\chi)=1+\tfrac{2}{5} G_T^{(0)}(\chi) \ ,  
\qquad G_T^{(0)}(\chi)=\tfrac{1}{2}\Big[\chi^2 +\frac{\chi^2}{(1-\chi)^2}\Big]\,,\qquad 
G_A^{(0)}(\chi)=\tfrac{1}{2}\Big[\chi^2 -\frac{\chi^2}{(1-\chi)^2}\Big]\,.
\end{equation*}
for the leading order and\footnote{Also in this case we have rescaled by a factor of $2$; see footnote \ref{foot1} and the corresponding discussion.}
\begin{align}
&G_S^{(1)}(\chi) = -\frac{2 \left(\chi ^4-4 \chi ^3+9 \chi ^2-10 \chi +5\right)}{5 (\chi -1)^2}
+\frac{\chi ^2 \left(2 \chi ^4-11 \chi ^3+21 \chi ^2-20 \chi +10\right)}{5 (\chi -1)^3}\log(\chi)\cr
&~~~~~~~~~\qquad \qquad \qquad \qquad ~~~~~~~~~~~~~~~~~~\ \ \ \ -\frac{2 \chi ^4-5 \chi ^3-5 \chi +10}{5\chi }\log(1-\chi)\ ,\nonumber\\
&G_T^{(1)}(\chi) = -\frac{\chi ^2 \left(2 \chi ^2-3 \chi +3\right)}{2(\chi -1)^2}
+\frac{\chi ^4 \left(\chi ^2-3 \chi +3\right)}{(\chi -1)^3}\log(\chi) -\chi ^3 \log(1-\chi) \ , \label{GSTA}   \\
&G_A^{(1)}(\chi) =\frac{\chi  \left(-2 \chi ^3+5 \chi ^2-3 \chi +2\right)}{2 (\chi -1)^2}
+\frac{\chi ^3 \left(\chi ^3-4 \chi ^2+6 \chi -4\right)}{(\chi -1)^3}\log(\chi)
-(\chi ^3-\chi ^2-1)\log(1-\chi)\,,\nonumber 
\end{align}
for NLO. Again we have kept the expression in terms of $\chi=\frac{z}{z-1}$ just for conciseness. 

By the explicit evaluation, we conclude that the final expression for the superprimary four-point correlation function up to leading order is 
\begin{align}
        \langle\mathbb{P}(t_1){\mathbb{P}}(t_2)\mathbb{P}(t_3)&{\mathbb{P}}(t_4)\rangle = \frac{C_{\Phi}^2}{t_{12}^2 t_{34}^2}\Bigl[1+\chi^2+\frac{\chi^2}{(\chi-1)^2}+\Bigl(\frac{1}{4\pi^2 C_{\Phi}}\Bigr)\Bigl(-\frac{4(1-\chi+\chi^2)^2}{(\chi-1)^2}+\nonumber\\
        &+(2-\frac{4}{\chi}+2\chi^2-4\chi^3)\log(1-\chi)+\frac{2\chi^2(2-4\chi+9\chi^2-7\chi^3+2\chi^4)}{(\chi-1)^3}\log(\chi)\Bigr)\Bigr]\,\nonumber
\end{align}
which in terms of $z$ simply reds\footnote{Notice indeed that \eqref{eq: 4 P in N=2} is braiding invariant $g(z/(z-1))=g(z)$ }
\begin{align}
\label{eq: 4 P explicit in GS GT}
        \langle\mathbb{P}(t_1){\mathbb{P}}(t_2)\mathbb{P}(t_3)&{\mathbb{P}}(t_4)\rangle = \frac{C_{\Phi}^2}{t_{12}^2 t_{34}^2}\Bigl[1+z^2+\frac{z^2}{(z-1)^2}+\Bigl(\frac{1}{\pi^2 C_{\Phi}}\Bigr)\Bigl(-\frac{(1-z+z^2)^2}{(z-1)^2}+\nonumber\\
        &+(\frac{1}{2}-\frac{1}{z}+\frac{z^2}{4}-z^3)\log(1-z)+\frac{z^2(2-4z+9z^2-7z^3+2z^4)}{2(z-1)^3}\log(-z)\Bigr)\Bigr]\,.
\end{align}
Having the expression for the superprimary four-point function explicit, it is interesting to perform then a direct comparison with the results obtained in \cite{Gimenez-Grau:2019hez}. Mapping to our convention the NLO ansatz
\begin{equation*}
    H_{0}^{(1)}(\chi) = R_0^{(1)}(\chi)\log(\chi)+\frac{\chi^2}{(\chi-1)^2}R_0^{(1)}(\chi)\log(1-\chi)+Q_0^{(1)}(\chi)\,,
\end{equation*}
the explicit bootstrap procedure allowed us to reach the expression 
\begin{align}
   H_0^{(1)}(\chi) =& \frac{c_2(1-\chi+\chi^2)^2}{(\chi-1)^2}+\Bigl(c_1 \chi+c_2(\chi-1)^2(\chi+\frac{3\chi+2}{2\chi})\Bigr)\log(1-\chi)-\nonumber\\
   &-\Bigl(c_1\frac{\chi^2}{\chi-1}+c_2 \frac{z^4(7+\chi(-7+2\chi}{2(\chi-1)^3}\Bigr)\log(\chi)\nonumber
\end{align}
for two unfixed parameters $c_1 $ and $ c_2$. Exploiting the analysis above, the universal sector allows one to fix, by comparison,, the previously undetermined bootstrap parameters, leading to $c_1 = c_2 = -4$. With this identification, the theory can be completely solved up to NLO, as all conformal data become determined. For completeness, we report the anomalous dimensions evaluated for these fixed values of the constants
\begin{align}
    \gamma^{[0,0]}&=4-\Delta-\Delta^2\nonumber\,,\\
    \gamma^{[1,0]} &= 2-\Delta-\Delta^2\nonumber\,.
\end{align}
Notice that, according to the argument above, this result holds universally for any line. Moreover, the analysis shows that the coefficients $c_1$ and $c_2$ are theory independent, with the only theory-specific input in the correlator being $C_\Phi$. As in the case of the $\cN=4$ CSm theories, the presence of unfixed parameters is simply a consequence of an insufficient number of constraints being available or employed to fully constrain the bootstrap ansatz.

We recall, however, that we have set as an initial condition the superprimary three-point function to vanish. The reason why consistency with the bootstrap result is nonetheless obtained follows from the discussion at the end of Sec. \ref{subsec: Bootstrapping the epsilon-order of OPE Coefficients}. In the bootstrap procedure, the four-point function was computed by perturbing the GFF data to NLO as in \eqref{eq: NLO expansion of the quadratic OPEs} and \eqref{eq: perturbation of the GFF explicit}. As discussed, under certain circumstances, this procedure automatically decouples all contributions arising from the $\mathbb{Z}_2$-odd sector. Indeed, in this case $\min\{\Delta\in\text{GFF}\}=2$ in \eqref{eq: perturbation of the GFF explicit}, and excluding $\Delta_{\bP}=1$ implicitly sets $\lambda_{\bP\bP\bP}=0$. Given that the three-point function controls also all the $\bZ_2$-odd OPEs, it follows that, effectively, the four-point function considered in the bootstrap is restricted to the $\mathbb{Z}_2$-even contributions, in accordance with what is shown. 
It is possible, however, that the three-point coefficient vanishes for other reasons. We leave this, along with the study of these additional features of the correlator, for future work.

\subsection{The $1/2$ BPS line in $AdS_3\times S^3 \times T^4$}
\label{subsec: the 1/2 BPS line in AdS3 T3 S4}
We conclude this section with a brief analysis of the cases studied in \cite{Bliard:2024bcz, Correa:2021sky}, focusing on the comparison between the bootstrap and holographic results of \cite{Bliard:2024bcz} and the universal sector analysis. In these setups, both the second and third assumptions of Sec. \ref{sec: The CFT perspective} fail as one finds $\cD \in \cD \times \cD$ as well as the presence of a tilt supermultiplet $\cT$, with $\cD \in \cT \times \cT$ but $\cT \notin \cD \times \cD$. The bootstrap analysis further shows that the displacement four-point function matches precisely the structure in \eqref{eq: f1(z) NLO with Z2 odd}. For the remainder of this discussion, however, we restrict our discussion to the simplified case where all three-point functions are set to zero, which corresponds to a dual configuration with purely Ramond-Ramond (R-R) background flux \cite{Bliard:2024bcz}. We note, however, that the full expression remains useful for further studies, as it provides a representative example of its universality class. In particular, it can be applied to other setups where universality arises, provided the operator content is consistent with that of the present case.

By setting $\lambda_{\bD\bD\bD}=0$, the four-point function can be derived directly by exploiting universality as discussed in the previous examples. Concretely, we focus again on the displacement four-point function and, analogously to the analysis of Sec. \ref{subsec: universal sector in N=2 4d}, we take the full trace of \eqref{eq: four-point displacement SO(3)} in order to match the representation of the correlators. 

We introduce then the projector
\begin{equation*}
    \cP_{ijkl}=\frac{1}{15}(\delta_{ij}\delta_{kl}+\delta_{ik}\delta_{jd}+\delta_{id}\delta_{jk})\,.
\end{equation*}
By contracting with \eqref{eq: four-point displacement SO(3)}, one obtains
\begin{equation*}
    f^{(1)}(\chi) = \cP_{ijkl}\,G^{(1)\,ijkl}_{SO(3)}(\chi) = G_S(\chi)+\frac{4}{3}G_T(\chi)\,,
\end{equation*}
where $G_S(z)$ and $G_T(z)$ are those defined in \eqref{eq: GS of SO(3)} and \eqref{eq: GT of SO(3)}. The explicit evaluation gives
\begin{align}
f^{(1)}(\chi) &=- \frac{4(1 - \chi + \chi^2)}{(1 - \chi)^4} (12-36\chi+25\chi^2+10\chi^3+25\chi^4-36\chi^5+12 \chi^6)\nonumber
\\
&  -\frac{8\chi^4}{(1-\chi)^5}
(2 - 6 \chi + 20 \chi^2 - 30 \chi^3 + 25 \chi^4 - 
 11 \chi^5 + 2 \chi^6)\log(\chi)\nonumber
 \\
 &  - \frac{8}{\chi}
(2 - \chi  - 
 \chi^5 + 2 \chi^6)\log(1-\chi) \label{eq: f three-point to zero}\,,
\end{align}
which is consistent with the bootstrap results, up to an overall normalization whose precise expression cannot be predicted by symmetry considerations. 
In this perspective, the key advantage of exploiting the universal sector is that it grants access to the complete NLO structure, including the expansion parameter. This, in turn, enables a direct match with the holographic analysis and allows us to extract the coupling dependence of the two-point constant of the supermultiplet, related to the bremsstrahlung function. 
In this regard, we consider the identity between the NLO defect CFT result obtained above and its holographic counterpart,
\begin{equation*}
\Bigl(\frac{1}{4\pi^2 C_\Phi}\Bigr) f^{(1)}(\chi) = \frac{1}{g\,\pi} f^{(1)}_{\text{bulk}}(\chi)\,,
\end{equation*}
where $f^{(1)}_{\text{bulk}}(\chi)$ denotes the result obtained in \cite{Bliard:2024bcz} from the holographic computation in the case where only the R-R flux is turned on, $g$ is the bulk coupling constant and $f^{(1)}(\chi)$ is that in \eqref{eq: f three-point to zero}. By the explicit comparison one finds that
\begin{equation*}
    f^{(1)}(\chi)= 4 f^{(1)}_{\text{bulk}}(\chi)\,,
\end{equation*}
thus leading to 
\begin{equation*}
    C_\Phi=\frac{g}{\pi}\,.
\end{equation*}

\section{Universality from the holographic description}
\label{sec:The holographic perspective}

As discussed, Wilson lines can be consistently studied within the framework of supersymmetric defect CFT$_1$, treating the line as a defect in a given SCFT$_d$. 
In this section, we briefly comment on the holographic evidence supporting the analysis of the previous sections. The discussion applies only to cases that admit a string realization of the Wilson line \cite{Maldacena:1998im, Rey:1998ik}. However, within these setups, the relation between different types of Wilson loops becomes more transparent and intuitive. 

The discussion will largely follow the standard procedure outlined in \cite{Giombi:2017cqn,Bianchi:2020hsz}, while keeping the reference to the specific theory as general as possible. For clarity, however, we focus on line defects in $3d$ theories \cite{Ouyang:2015bmy, Drukker:2008zx, Berenstein:2008dc, Chen:2008bp, Mauri:2017whf, Mauri:2018fsf}, although the analysis extends straightforwardly to other codimensions, for instance in $4d$ theories \cite{Giombi:2017cqn}, leading to analogous conclusions.

In this perspective, we begin with the M-theory background $\text{AdS}_4 \times S^7/\mathbb{Q}$, where $\mathbb{Q}$ denotes a generic discrete quotient of $S^7$, such as $\mathbb{Q}=\mathbb{Z}_k$ \cite{Aharony:2008ug, Hosomichi:2008jb, Aharony:2008gk} or $\mathbb{Q}=\mathbb{Z}_{rk}\times \mathbb{Z}_{r}$ \cite{Gaiotto:2008sd, Hosomichi:2008jd}. As known, the ten-dimensional type IIA description emerges by shrinking the M-theory circle, which can be realized in practice by taking the limit $k\to\infty$. For instance, in the standard case one finds
\begin{equation*}
S^7/\mathbb{Z}_{k\to\infty} \;\longrightarrow\; \mathbb{C}\mathrm{P}^3 \, .
\end{equation*}
More generally, we denote by $\cM$ the resulting internal manifold obtained in 
the limit. The corresponding type IIA background then takes the form $\text{AdS}_4 \times \cM$ specified by
\begin{equation*}
    ds^2 = R^2 (ds^2_{\text{AdS}_4}+4 ds^2_{\cM})\,,
\end{equation*}
leaving implicit all quantities that are strictly dependent on the theories or fluxes that can be present, assuming they do not change the following considerations.
Employing the standard description, we consider a Poincaré patch for the AdS$_4$ metric 
\be 
ds^2_{\text{AdS}_4} = \frac{dz^2 + dx^i dx^i }{z^2} \,,
\ee
where, specialized to the Euclidean, we have $x^i = (x^0,x^1,x^2)$ and $z$ for the radial coordinate. The bosonic contribution of the Polyakov action for the superstring reads

\begin{equation*}
S_B =\frac{1}{2} T   \int d^2\sigma \sqrt{h}\,   h^{\mu\nu} \Big[  \frac{1}{z^2}  \left(\partial_\mu x^i\partial_\nu x^i+\partial_\mu z\partial_\nu z\right)
+   4\,G_{MN}^{\cM}\,{\partial_\mu Y^M\partial_\nu Y^N } \Big]  \, ,
\end{equation*}
where $T$ is the effective tension, $\sigma^\mu = (t,s)$ are the Euclidean worldsheet coordinates and we have split the embedding AdS coordinates from the compact ones. The $1/2$ BPS solution which has the
Wilson line as boundary condition is achieved by considering the solution corresponding to the minimal AdS$_2$ surface specified by
\begin{equation*}
    z = s \,,\qquad x^0 = t\,,\qquad x^i = 0\,,
\end{equation*}
with all other coordinates vanishing.

The insertion of operators in the line can now be mapped to the fluctuations of the transverse coordinates. From the identification within the AdS/CFT dictionary \cite{Aharony:2008ug}, we have for bosons $m^2=\Delta(\Delta-1)$ and for fermions $m^2 = (\Delta-\frac{1}{2})^2$.
We can therefore expand around the $1/2$ BPS solution of AdS$_{2}$ by considering a Nambu-Goto action for the string worldsheet. We define the AdS$_{4}$ transverse coordinates as $X = \frac{1}{\sqrt{2}}(x^{1} + ix^{2})$. These are free to fluctuate, while the $z$ and $x^0$ are static. The AdS$_{4}$ metric can be casted as \cite{Giombi:2017cqn,Bianchi:2020hsz} 
\begin{equation}
\label{eq:AdS2 embedding}
ds^{2}_{AdS_{4}} = \frac{\left(1 + \frac{1}{2}\vert X \vert^{2}\right)^{2}}{\left(1 - \frac{1}{2}\vert X \vert^{2}\right)^{2}}ds^{2}_{AdS_{2}} + \frac{2\,dXd\bar{X}}{\left(1 - \frac{1}{2}\vert X \vert^{2}\right)^{2}}\,,
\end{equation} 
where the AdS$_{2}$ metric is
\begin{equation*}
ds^{2}_{AdS_{2}}  = \frac{1}{s^{2}}( ds^{2}+dt^{2} )\,.
\end{equation*}
Compactly we can just write 
\begin{equation*}
ds^{2}_{AdS_{4}}  = g_{\mu\nu}(X,\bar X)d\sigma^\mu d\sigma^\nu\,.
\end{equation*}
This parametrization makes manifest the $SU(1,1)$ line conformal symmetry, and endows the transverse directions $X,\bar{X}$ with a $U(1)_{j_{0}}$ charge.
The contribution for the compact manifold can be stated implicitly as
\begin{equation*}
    ds^2_{\cM}=g_{ab}(w,\bar{w})d\bar{w}^a dw^b\,,
\end{equation*}
where $\{w\}$ and $\{\bar{w}\}$ are generic parametrization of the metric of the compact manifold $g_{ab}(w,\bar{w})$ which can be, for example, $\mathbb{CP}^3$ for the ABJM case. These quantities are chosen such that they preserves a subgroup of the isometries of $\cM$ that match with the preserved R-symmetry expected from the dual description. \footnote{For example, see chapter 5 of \cite{Bianchi:2020hsz}}
The Nambu-Goto action can be written compactly as
\begin{equation*}
S_B = T   \int d^2\sigma \sqrt{ \det\Big[  \frac{\left(1 + \frac{1}{2}\vert X \vert^{2}\right)^{2}}{\left(1 - \frac{1}{2}\vert X \vert^{2}\right)^{2}}g^{AdS_2}_{\mu\nu}+ \frac{2\,\partial_\mu X\partial_\nu\bar{X}}{\left(1 - \frac{1}{2}\vert X \vert^{2}\right)^{2}} + 4\,g_{\mu\nu}(w,\bar{w}) \Big] } \, .
\end{equation*}
where $g^{AdS_2}_{\mu\nu}=\frac{1}{s^2}\delta_{\mu\nu}$ and $g_{\mu\nu}(w,\bar w)$ is the pullback of $g_{ab}(w,\bar w)$. This is interpreted as a fundamental string stretching from the center of AdS$_4$ to the boundary, located at $z=0$, where the CFT$_1$ lives. By expanding in terms of the fluctuations one obtains various terms which schematically are
\begin{equation*}
    S_B = T   \int d^2\sigma \sqrt{g^{\text{AdS$_2$}}}\Bigl(L_{2}+L_{4X}+L_{4w}+L_{2X,2w}+...\Bigr)\,,
\end{equation*}
where $L_{2}$ is the free term, and the other are the quartic interactions, \cite{Bianchi:2020hsz}. If we focus on the four-point function of the transverse fluctuations up to NLO, only the contribution from $L_{4X}$ is needed and explicitly it reads
\begin{equation*}
   L_{4X} =  2|X|^4+|X|^2  \,(g^{\mu\nu}\partial_\mu X \partial_\nu \bar{X}) -\frac{1}{2}\,(g^{\mu\nu}\partial_\mu X \partial_\nu X) \,(g^{\rho\kappa}\partial_\rho \bar X \partial_\kappa \bar X)\,.
\end{equation*}

Crucially, $L_{4X}$ is structurally independent of the specifics of the compact manifold. Therefore, the evaluation of the Witten diagrams for the four-point functions of the transverse fluctuations $X$ and $\bar X$ always yields the same result, as the only relevant contribution to the connected part comes from $L_{4X}$.

From the defect CFT perspective, this implies that the displacement insertions share the same structural form irrespective of the bulk theory. Moreover, since supersymmetry relates the four-point function of the displacement to those of the other operators in the supermultiplet, the functional form of all such correlators is thereby constrained, up to NLO, and different correlation functions can be related as in the previous sections; see Sec. \ref{subsec: universality ABJM N4 SCSm Th}.

While the discussion above has been developed for defects in $3d$ theories, the same reasoning applies, modulo symmetry considerations, to different line defects in higher-dimensional setups, such as the $1/2$ BPS Wilson line in $\cN=4$ SYM \cite{Giombi:2017cqn}. The main difference lies in the tensorial structure of the transverse fluctuations $x^i$ related to the different embedding of AdS$_2$ in AdS$_5$, instead of AdS$_4$ as for the cases discussed above.
 However, the nature of the $L_{4x}$ interaction, discussed in \cite{Giombi:2017cqn}, will be qualitatively equivalent to that discussed here. Therefore, the four-point interactions can be identified leading to an analysis analogous to that presented in Sec. \ref{subsec: Universality in 1/2 BPS wilson line N=4 SYM}.


\section{The $1/2$ BPS line in $\cN=2$ Chern-Simons-matter theories}
\label{sec: The 1/2 BPS line in N=2 Chern-Simons matter theories}
We now apply the approach for the identification of universal sectors, developed in Secs. \ref{sec: The CFT perspective} and \ref{subsec: Correlators of the displacement multiplet}, to the study of $1/2$ BPS superconformal lines in $3d$ $\cN=2$ theories. These cases are relevant in the context of $1/2$ BPS Wilson lines in $\cN=2$ Chern-Simons-matter theories \cite{Bianchi:2018scb, Mauri:2018fsf}. Moreover, since the preserved symmetry algebra in this case is a subalgebra of that preserved by the $1/6$ BPS Wilson line in ABJM \cite{Bianchi:2018scb}, the two setups are closely related. In fact, the identification of universal sectors can be exploited also to gain insights into the $1/6$ BPS Wilson line at strong coupling. However, for the $1/6$ BPS line one must also take into account the additional defect supermultiplet, following an analysis similar to that outlined in Sec. \ref{subsec: Correlators of the displacement multiplet}. 

As already noted, once the universal sectors are identified, the entire discussion of constructing an ansatz and performing the bootstrap can be bypassed. We therefore begin by defining the displacement supermultiplet, its four-point functions, and the corresponding OPE rules. After that, once the appropriate (super)conformal block expansion has been identified, the conformal data up to next-to-leading order can be extracted directly by projection, exploiting the orthogonality of the (super)blocks. Since the analysis proceeds in close analogy with the more supersymmetric cases of $1/2$ BPS lines, namely, those in ABJM and $\cN=4$ CSm theories \cite{Bianchi:2020hsz,Pozzi:2024xnu},we shall, for brevity, outline only the essential points.


\subsection{Supermultiplet}
The identification of the displacement supermultiplet builds from the observation that broken generators of the bulk theory are related to local defects section by the Ward identity \eqref{Ward-id}. In this case the $1/2$ BPS line breaks the superalgebra $\osp(2\vert4)$ thus preserving half of the supercharges \eqref{eq: preserved supercharges}. For the broken ones, $\mathbf{Q}=-{}^-Q_+$ and $\bar{\mathbf{Q}}=-i{}^+Q_-$, one can associate the fermionic insertions 
\begin{equation*}
    [\gQ,\cW]= \int dt\,\cW[\mathbb{\Lambda}(t)]\,,\qquad [\bar\gQ,\cW]=\int dt\,\cW[\bar\blambda(t)]\,.
\end{equation*}
Associated to the broken orthogonal translations, $\mathbf{P}=P_2+iP_1$ and $\bar{\mathbf{P}}=P_2-iP_1$, one instead defines the displacement operators
\begin{equation*}
    [\gP,\cW]= \int dt\,\cW[\bD(t)]\,,\qquad [\bar\gP,\cW]=\int dt\,\cW[\bar\bD(t)]\,.
\end{equation*}
By exploiting the preserved superalgebra, one can read out the charges thus obtaining the displacement multiplet
\begin{equation}
\label{eq: N=2 SCSm superdisp}
    L\bar A\left[-\ft{3}{2}\right]_{\frac{3}{2}}\,:\qquad\left[-\ft{3}{2}\right]_{\frac{3}{2}} \longrightarrow   [-1]_{2}\,.
\end{equation}
Here, the operators are labeled by $[j_0]_{\Delta}$, where $\Delta$ is the scaling dimension and $j_0$ is the charge under the preserved Abelian R-symmetry, \eqref{eq: J0 charge in N=2 SCSM line}. To refer to the entire multiplet, we will also use the notation of \cite{Agmon:2020pde}, indicating whether the multiplet is long $L$ ($\bar L$) or short at threshold $A$ ($\bar A$) with respect to $Q$ ($\bar Q$).
The arrow represents schematically the action of the supercharge $Q$, \eqref{eq: preserved supercharges}, and explicitly this is given by
\begin{equation}
\label{eq: susy relation for multiplet components}
    \{Q,\blambda\}=2\bD\,,\qquad[Q,\bD]=0\,.
\end{equation}
Exploiting super-Jacobi identities, one obtains the one involving the action of $\bar Q$, which would correspond to the opposite direction of the arrow in \eqref{eq: N=2 SCSm superdisp}. Explicitly these relations are
\begin{equation*}
    [ \bar Q,\bD]=\partial_t\blambda\qquad\{\bar Q,\blambda\}=0\,.
\end{equation*}
\subsection{Superfield realization and correlation function in superspace}
\label{subsec: Superfield realization and correlation function in superspace}
In order to study correlation functions of the displacement supermultiplet \eqref{eq: N=2 SCSm superdisp}, we are interested in providing a superfield realization of the superdisplacement. So, we first proceed by introducing the superspace coordinates
\begin{align}
    y= t + \theta\Bar{\theta}\,, \qquad 
    \bar{y} &= t - \theta \Bar{\theta},
\end{align}
and, correspondingly, the covariant derivatives 
\begin{equation*}
    D = \partial_\theta + \Bar{\theta}\partial_t\,,\qquad  \Bar{D} = \Bar{\partial}_{\bar\theta} + {\theta}\partial_t\,,
\end{equation*}
such that $D \bar y=0$ and $\bar{D}  y=0$. 
This allows us to introduce the component expansion of generic (anti)chiral superfields obeying the conditions  $\bar{D}\cD(y,\theta) = 0$ and $D\bar \cD(\bar y,\bar\theta)=0$. In our case, this simply reads
\begin{align}
    \cD(y,\theta) &= \blambda(y) +2\theta\bD(y)\,,\nonumber\\
    \bar\cD(y,\theta) &= \bar\blambda(y) +2\bar\theta\bar\bD(y)\,\nonumber.
\end{align}
The coefficients have been consistently chosen by consistency with the supersymmetry action \eqref{eq: susy relation for multiplet components}. The two-point function is given by
\begin{equation*}
\langle \cD(y_{1},\theta_{1}) \bar{\cD}(\bar y_{2},\bar \theta_{2})\rangle = \frac{C_{\cD}}{\langle 1\bar{2} \rangle^{2\Delta}},
\label{2pt-superfield-N=2}
\end{equation*}
where we defined the chiral invariant distance
\begin{equation*}
    \vev{ij}= y_i-\bar y_j-2\theta_i\bar \theta_j\,.
\end{equation*}
Expanding in terms of the superfield component we obtain 
\begin{equation*}
    \vev{\blambda(t_1)\bar\blambda(t_2)}=\frac{C_\blambda}{t_{12}^3}\,,\qquad \vev{\bD(t_1)\bar\bD(t_2)}=\frac{C_\bD}{t_{12}^4}\,,
\end{equation*}
where $C_\blambda = C_\cD$ and $C_\bD=3C_\blambda/2=3C_\cD/2$. The four-point function is defined as in \eqref{eq: generic four-point of the superdisplacement}. In this case, chirality prevents the existence of nilpotent invariants \cite{Bianchi:2020hsz, Fitzpatrick:2014oza}, while still allowing for two inequivalent orderings of the insertions. This arises because, in one-dimensional theories, operator insertions are intrinsically ordered along the line. Therefore, we consider
\begin{equation}
\label{4pnt-generic-1}
\langle \cD(y_{1},\theta_{1}) \bar{\cD}(\bar y_{2},\bar \theta_{2}) \cD(y_{3},\theta_{3}) \bar{\cD}(\bar y_{4},\bar \theta_{4})\rangle = \frac{C_{\cD}^{2}}{\langle 1\bar{2} \rangle^{3} \langle 3\bar{4} \rangle^{3}}f(\mathcal{Z})
\end{equation}
and
\begin{equation}
\label{4pnt-generic-2}
\langle \cD(y_{1},\theta_{1}) \bar{\cD}(\bar y_{2},\bar \theta_{2}) \bar{\cD}(\bar y_{3},\bar \theta_{3}) \cD(y_{4},\theta_{4})\rangle = -\frac{C_{\cD}^{2}}{\langle 1\bar{2} \rangle^{3} \langle 4\bar{3} \rangle^{3}}h(\mathcal{X})\,,
\end{equation}
where,  $f(\cZ)$ and $h(\mathcal{X})$ are functions of the cross-ratios of the chiral distances
\begin{equation}
\label{eq: super cross ratio}
\cZ = \frac{\langle 1\bar{2} \rangle \langle 3\bar{4} \rangle}{\langle 1\bar{4} \rangle \langle 3\bar{2} \rangle}, \qquad \mathcal{X} = \frac{\langle 1\bar{2} \rangle \langle 4\bar{3} \rangle}{\langle 1\bar{3} \rangle \langle 4\bar{2} \rangle}\,.
\end{equation}
By expanding in the fermionic coordinates, one can read the bosonic part of these cross-ratios, which are those introduced in \eqref{eq: cross ratios}, which we report here for clarity
\begin{equation*}
    z=\frac{t_{12}t_{34}}{t_{14}t_{32}},\qquad \chi=\frac{t_{12}t_{34}}{t_{13}t_{24}}\,.
\end{equation*}

By expanding \eqref{4pnt-generic-1} and \eqref{4pnt-generic-2} in the Gra\ss mann variables one can match the correlators of the superdisplacement components to the term of the expansion. For the superprimary one simply has

\begin{align}
\label{eq: 4point function superprimaries N=2 SCSm}
    \langle \blambda(t_{1})\bar{\blambda}(t_{2})\blambda(t_{3})\bar{\blambda}(t_{4}) \rangle = \frac{C_{\cD}^{2}}{t_{12}^{3}t_{34}^{3}}f(z)\,,
    \qquad 
   \langle \blambda(t_{1})\bar{\blambda}(t_{2})\bar{\blambda}(t_{3})\blambda(t_{4}) \rangle
= \frac{C_{\cD}^{2}}{t_{12}^{3}t_{34}^{3}}h(\chi)\,,
\end{align}
where  $t_{1}<t_{2}<t_{3}<t_{4}$.
The displacement four-point function is instead related to the superprimary by
\begin{equation}
\label{eq: four-point of the displacement in 1/2 BPS N=2 as derivatives}
    \langle \bD(t_{1})\bar{\bD}(t_{2})\bD(t_{3})\bar{\bD}(t_{4}) \rangle = \frac{(3C_{\cD}/2)^{2}}{t_{12}^{4}t_{34}^{4}}\frac{1}{9}\Bigl(9f(z)-5zf'(z)+z^2f''(z)-(z^2f'(z)+z^3f''(z))\Bigr)
\end{equation}
and, as expected, it has the shape of \eqref{eq: function of displacement from primary}.

Crucially, in $1d$, although the operator insertions are ordered, it is still possible to exchange the second and fourth insertions (or, equivalently, the first and third), a process referred to as braiding. This is essentially related to a sequence of exchanging insertions, applying symmetrization, and finally using cyclicity after having identified the point at infinity, see \cite{Liendo:2018ukf}. This allows one to relate different configurations, such as
\begin{equation}
\label{eq: braiding corr}
    \langle \blambda(t_{1})\bar{\blambda}(t_{2})\bar{\blambda}(t_{3})\blambda(t_{4}) \rangle= \langle \blambda(t_{1}){\blambda}(t_{4})\bar{\blambda}(t_{3})\bar\blambda(t_{2}) \rangle\,,
\end{equation}
thus leading to
\begin{equation}
\label{eq: braiding 2}
    (1-\chi)^{2\Delta} h(\chi)=\chi^{2\Delta}h(1-\chi)\,,
\end{equation}
where, in this case, $\Delta=3/2$. 

\subsection{OPEs and block expansions}
\label{subsec: OPEs and block expansions}
In order to properly study the correlation function of the displacement multiplet we study in the following the OPE. Due to the similarities with the more supersymmetric cases, the $1/2$ BPS line in ABJM and $\cN=4$ CSm theories \cite{Bianchi:2020hsz,Pozzi:2024xnu}, here we summarize concisely the OPE of this case.

First of all, given the correlators \eqref{4pnt-generic-1} and \eqref{4pnt-generic-2} we can see that we have different OPEs to take into account and those are the chiral-antichiral $\cD\times\bar\cD$ and the chiral-chiral $\cD\times\cD$. In \cite{Bianchi:2018scb}, the selection rules governing the $1/6$ BPS defect in ABJM theory were systematically derived. The analysis revealed that, in the chiral-antichiral OPE of $\mathfrak{su}(1,1|1)$, the only allowed contributions are those of the identity operator and long multiplets. Thus we have
\begin{equation}
\label{eq: chiral-antichiral ope}
    L\bar A[-\ft{3}{2}]_{\frac{3}{2}}\times A\bar L[\ft{3}{2}]_{\frac{3}{2}}\sim \cI+LL[0]_\Delta\,.
\end{equation}
For the chiral-chiral OPE, one instead can constrain it by chirality. This simply gives
\begin{equation*}
    L\bar A[-\ft{3}{2}]_{\frac{3}{2}}\times L\bar A[-\ft{3}{2}]_{\frac{3}{2}}\sim L\bar A[-3]_{3}+L\bar L[-\ft{5}{2}]_{\Delta}\,.
\end{equation*}
So in the superprimary OPE the exchanged operators are
\begin{equation*}
     [-\ft{3}{2}]_{\frac{3}{2}}\times [-\ft{3}{2}]_{\frac{3}{2}}\sim [-3]_{3}+\bar Q[-\ft{5}{2}]_{\Delta}\,.
\end{equation*}
We thus see that, contrary to the previous case, these are not the full multiplets being exchanged, but rather only a single component from each supermultiplet.
Having identified all the OPE of interest, we can now discuss the partial wave expansion. For the chiral-antichiral case, long multiplets contribute through all allowed superdescendants. Accordingly, the four-point correlator admits a decomposition in superconformal blocks. These are obtained by acting with the Casimir of the preserved superconformal algebra on the four-point function \eqref{casimir equation}. The expansions read
\begin{equation*}
    f(z) = 1+\sum_{\Delta>0}a_\Delta G_{\Delta}(z)\,,\qquad   h(\chi) = 1+\sum_{\Delta>0}b_\Delta G_{\Delta}(\chi)\,,
\end{equation*}
where the superblocks solving the differential equations are
\begin{equation}
\label{eq: N=2 SCSm blocks}
    G_\Delta(z) = (-z)^\Delta\,_2F_1(\Delta,\Delta,2\Delta+1;z)\qquad G_\Delta(\chi) = \chi^\Delta\,_2F_1(\Delta,\Delta,2\Delta+1;\chi)\,.
\end{equation}
These relations can be rewritten in the CPW expansion just by projecting on the $1d$ conformal blocks. This leads to expressions like 
\begin{equation*}
    G_\Delta(z)  = g_\Delta(z)+\frac{\Delta}{2(1+2\Delta)}g_{\Delta+1}(z)\,.
\end{equation*}
The quadratic OPE coefficients are 
\begin{equation*}
    a_\Delta= \lambda_{\blambda\bar\blambda\bT_\Delta}\lambda_{\blambda\bar\blambda\bT_\Delta}\,,\qquad b_\Delta= \lambda_{\blambda\bar\blambda\bT_\Delta}\lambda_{\bar\blambda\blambda\bT_\Delta}\,,
\end{equation*}
where $\bT_\Delta$ is the exchanged operator of dimension $\Delta$. One can further relate the different OPE coefficients by applying parity to the three-point function \cite{Bianchi:2020hsz}. In the case of fermionic external operators, this yields to
\begin{equation}
\label{eq: parity of the only OPEs}
    \lambda_{\blambda\bar\blambda\bT_\Delta}=(-1)^{P_{\bT}+1}\lambda_{\bar\blambda\blambda\bT_\Delta}\,,
\end{equation}
thus leading to the relation
\begin{equation}
\label{eq: parity of the OPEs}
    a_\Delta=(-1)^{P_{\bT}+1}b_\Delta\,,
\end{equation}
where $P_{\bT}$ is the parity of the exchanged operator.
For example, when considering the exchange of multiparticle operators of the form $\bT_{\Delta=3+n} = [\blambda\bar{\blambda}]_n$, one finds that $P_{\bT} = n$.

This leads to a  functional equivalence between the expressions for the correlators. Indeed, the combination $a_{3+n}G_{3+n}(z)$ can be mapped to $b_{3+n}G_{3+n}(\chi)$ under $z \to \chi$. This happens precisely because of the factor $(-1)^{3+n}$ in the definition of the superblock $G_{3+n}(z)$ \eqref{eq: N=2 SCSm blocks}, that compensates the corresponding sign in the parity relation \eqref{eq: parity of the OPEs}. Since this argument relies only on parity, the property remains valid perturbatively \cite{Bianchi:2020hsz}.

While for the first case in \eqref{eq: 4point function superprimaries N=2 SCSm} the superblock expansion discussed above is sufficient for all the relevant channels, for the second one we must also take into account the chiral-chiral OPE. Indeed, given the expression \eqref{eq: braiding corr} and \eqref{eq: braiding 2} we get the conformal block expansion
\begin{equation*}
    h(1-\chi) =\text{b}_{3}\,g_{3}(1-\chi)+\sum_{\Delta>3}\text{b}_{\Delta}\,g_{\Delta}(1-\chi)\,,
\end{equation*}
where the conformal blocks are given by the standard one-dimensional expression \cite{Dolan:2003hv,Dolan:2011dv}
\begin{equation*}
    g_\Delta(\chi)=\chi^\Delta\,_2F_1(\Delta,\Delta,2\Delta;\chi)\,.
\end{equation*}

\subsection{Universal sectors}

We now turn to the conditions that characterize the presence of universal sectors. The analysis of the explicit OPEs shows that the displacement multiplet is never exchanged, so all the requirements discussed in Sec. \ref{sec: The CFT perspective} are satisfied. We therefore conclude that the superprimary and the displacement sectors remain decoupled up to NLO. The identification of the NLO four-point function then becomes straightforward, allowing us to bypass the standard bootstrap procedure of stating and fixing an ansatz.
This case is very similar to that discussed in Sec. \ref{subsec: universality ABJM N4 SCSm Th}. The four-point function of the superprimary can be obtained directly from that of the first superdescendant of the $\cN=4$ CSm  Wilson line superprimary \cite{Pozzi:2024xnu}.
 This leads to
\begin{align}
\langle\mathbb{\Lambda}(t_1)\bar{\mathbb{\Lambda}}(t_2)\mathbb{\Lambda}(t_3)\bar{\mathbb{\Lambda}}(t_4)\rangle=&
\frac{C_{\cD}^{2}}{t_{12}^3 t_{34}^3}\, 
\Bigl[1-z^3+\epsilon_{\cD}\Bigl(-9+\frac{z}{2}-\frac{z^2}{2}+9z^3 +\nonumber\\
+&\Bigl( 5-\frac{9}{z}-5z^3+9z^4\Bigr)\log(1-z) +(5-9z)z^3\log(-z)\Bigr)+\cO(\epsilon_\cD^2)\Bigr]\label{eq: 4-point superprimary N=2 CSm}.
\end{align}
The displacement four-point function again takes the same structural form as in \eqref{eq: 4D universal}, and explicitly it reads
\begin{align}
\langle\mathbb{D}(t_1)\bar{\mathbb{D}}(t_2)\mathbb{D}(t_3)\bar{\mathbb{D}}(t_4)\rangle=&
\frac{(3C_{\cD}/2)^{2}}{t_{12}^4 t_{34}^4}\, 
\Bigl[1+z^4+\epsilon_{\cD}\Bigl(-16 -2z -\frac{7z^2}{3} -2z^3-16z^4
\cr
+\Bigl(6 - \frac{16}{z}+ &6 z^4 -16z^5\Bigr)\log(1-z) + 2z^4 (8z-3)\log(-z)\Bigr)+\cO(\epsilon_\cD^2)\Bigr].
\end{align}
In this case, the overall coefficient is fixed explicitly by the analysis in Sec. \ref{subsec: Superfield realization and correlation function in superspace} and the expansion parameter is again linked to the two-point normalization as in \eqref{eq: epsilon universal} by
\begin{equation*}
    \epsilon_\cD=\frac{1}{4\pi^2 C_\cD}\,.
\end{equation*}

\subsection{Extracting the conformal data}
\label{subsec: Extracting the conformal data}
Exploiting the (super)block expansion and the knowledge of the leading and subleading strong coupling four-point functions, we can now extract the conformal data.
\subsubsection{Leading-order data}
As discussed in Sec. \ref{subsec: Correlators of the displacement multiplet} the exchanged operators at leading order are those of the multiparticle sates: $[\blambda\bar\blambda]^{L=2}_n$ in the chiral-antichiral channels and  $[\blambda\blambda]^{L=2}_n$ in the chiral-chiral one, all having dimensions $\Delta=3+n$. We can thus extract the leading-order data by exploiting the orthogonal relation for the blocks and superblocks discussed in the Appendix \ref{app: blocks and superblocks properties}. 
We thus get 
\begin{equation*}
    a_n^{(0)}=\frac{2^{-7-2n}\sqrt{\pi}\,(n+1)(n+2)\Gamma(n+6)}{\Gamma(n+\ft{7}{2})}
\end{equation*}
and the $b^{(0)}_n$ are related to these just by the relation \eqref{eq: parity of the OPEs}. In the case of the chiral-chiral channel one finds 
\begin{align}
    \text{b}^{(0)}_n &= \frac{2^{-5-2n}\sqrt{\pi}\,(n+1)(n+2)\Gamma(n+5)}{\Gamma(n+\ft{5}{2})} &n\quad\text{odd}&\,,\\
    \text{b}^{(0)}_n &= 0 &n\quad\text{even}&\,.
\end{align}
\subsubsection{Next-to-leading-order data}
We now turn to the next-to-leading order of the expansion, beginning with the chiral-chiral channel. In this case, the anomalous dimensions are extracted from the logarithmic contributions of the NLO term in the correlation function \eqref{eq: 4-point superprimary N=2 CSm}. Explicitly, they are given by
\begin{equation*}
    \gamma^{(1)}_n = \frac{1}{a^{(0)}_n}\oint \frac{dz}{2\pi i}\omega(z)\Bigl((5-9z)z^3 \Bigr)G_{-3-n}(z)\,,
\end{equation*}
where we used the orthogonality relations defined in \eqref{eq: orthogonality}. By taking the residue one obtains
\begin{equation*}
    \gamma^{(1)}_n =-5-6n-n^2\,.
\end{equation*}
From this explicit expression we highlight a crucial feature revealed by the universal sector procedure: the mild growth of the anomalous dimension with $n$. This behavior is in fact a standard requirement imposed in the bootstrap approach when fixing the ansatz up to NLO \cite{Liendo:2018ukf, Gimenez-Grau:2019hez, Bianchi:2020hsz, Pozzi:2024xnu, Bliard:2024und}. In the present discussion, it represents a universal feature that is automatically inherited in all the cases discussed.
The growth of the anomalous dimension with $\Delta$ is holographically related to the nature of the interaction: the more irrelevant the interaction, the faster the growth \cite{Heemskerk:2009pn, Fitzpatrick:2011dm}. The fact that, within the universal sector, the same scaling behavior is observed indicates that, up to NLO, the corresponding bulk interactions are effectively equivalent.

By the explicit evaluation we find that, as expected, the next-to-leading OPE coefficients are consistent with the derivative rule \cite{Heemskerk:2009pn, Fitzpatrick:2011dm,Alday:2014tsa}
\begin{equation}
\label{eq: general OPE NLO}
    a^{(1)}_n=\partial_n(a^{(0)}_n\gamma_n^{(1)})\,,
\end{equation}
which explicitly reads
\begin{equation*}
    a^{(1)}_n= a_n^{(0)}\Bigl[-11-3n +\gamma_n^{(1)}\Bigl(\psi(n+6)-\psi(n+\ft{7}{2})-2\log2+\frac{1}{n+2}\Bigr)\Bigr]\,,
\end{equation*}
where $\psi(z)=\Gamma'(z)/\Gamma(z)$.

Exploiting the relation under the map $z\rightarrow\chi$ discussed in Sec. \ref{subsec: OPEs and block expansions} and the braiding relation \eqref{eq: braiding 2}, one can obtain the anomalous dimensions for the operators exchanged in the chiral-chiral channel. These are given by
\begin{equation*}
   \upgamma_n^{(1)} = \frac{1}{\text{b}_n^{(0)}}\oint \frac{d\chi }{2\pi i}\rho(\chi)\Bigr[\frac{(1-\chi)^3}{\chi^3}\Bigl( 5-\frac{9}{\chi}-5\chi^3+9\chi^4\Bigr)\Bigr]\,g_{-2-n}(1-\chi)\,,
\end{equation*}
and the explicit evaluation leads to
\begin{align}
    \upgamma^{(1)}_n&=-4-5n-n^2&n\quad\text{odd}&\nonumber\,,\\
    \upgamma^{(1)}_n&=0&n\quad\text{even}&\nonumber\,.
\end{align}
Also for this case we find agreement with the expression
\begin{equation*}
    \text{b}_n^{(1)}=\partial_n(\text{b}_n^{(0)} \upgamma^{(1)}_n)
\end{equation*}
and, explicitly, the OPE coefficients read
\begin{align}
    \text{b}_n^{(1)}&=\text{b}_n^{(0)} \Bigl[-9-3n+ \upgamma^{(1)}_n\Bigl(\psi(n+5)-\psi(n+\ft{5}{2})-2\log 2+\frac{1}{n+2}\Bigr)\Bigr]&n\quad\text{odd}&\,,\nonumber\\
    \text{b}_n^{(1)}&=0&n\quad\text{even}&\nonumber\,.
\end{align}

\section{The $1/3$ BPS Wilson line in ABJM}
\label{Bootstrapping the 1/3 BPS Wilson line in ABJM}
We conclude with a brief discussion of the $1/3$ BPS Wilson line in ABJM, leaving a complete analysis for future work. Our main aim is to emphasize that the considerations above can also be applied to the study of other four-point functions beyond the superdisplacement correlator. An in-depth weak-coupling analysis of the explicit realization of these Wilson lines, including the relevant multiplets and their properties, was carried out in \cite{Drukker:2022txy}.

Here, we focus on the strong-coupling regime, analyzed through the framework of universal sectors. In this case, the $1/3$ BPS line preserves the bosonic subalgebra $\so(2,1)_{\text{conf}}\oplus\su(2)_\cR\oplus\u(1)_{J_0}\oplus\u(1)^2$, see \cite{Agmon:2020pde, Drukker:2022txy}. In terms of representations, they can be obtained by decomposing the $1/2$ BPS displacement multiplet of ABJM, yielding
\begin{equation*}
L\bar{A}_{1}[\ft{3}{2}]_{\frac{1}{2}}^{(0,0)}\rightarrow  L\bar{A}_{1}[1]_{1}^{(0)} \oplus  L\bar{A}_{1}[\ft{1}{2}]^{(0)}_{\frac{1}{2}}\,.
\end{equation*}
The first term corresponds to the $1/3$ BPS displacement multiplet, while the second is a $1/3$ BPS supermultiplet associated with the broken supercurrents. As discussed in \cite{Drukker:2022txy}, there are actually two sets of supermultiplets associated with the broken supercurrents, referred to as the tilt and tlit multiplets, which should be regarded as separate contributions with different normalizations. For brevity, in the following we focus on just one of these.

We notice from the outset that the subalgebra preserved is consistent with the one preserved by the $1/2$ BPS line in $\osp(4\vert4)$ theories. The crucial difference is that, in this case, we have further fundamental defect contributions coming from the additional breakings. In terms of the components, the decomposition explicitly reads
\begin{equation*}
\label{eq: 1/2 BPS in 1/3 BPS}
\begin{array}{llllll}
&\mathbb{F}:\left[\frac{3}{2}\right]_{\frac{1}{2}}^{(0,0)}\rightarrow & \quad  &\quad &\quad \mathbbm{F}  : [\frac{1}{2}]^{(0)}_{\frac{1}{2}} \\
&\quad\;\downarrow \quad &\quad & \quad &\quad \downarrow \quad \\
&\mathbb{O}^a :\left[2\right]_{1}^{(1,0)}\rightarrow & \mathbbm{R}: [1]_{1}^{(0)} \quad &\oplus &\quad \mathbbm{O}^r:[\frac{1}{2}]^{(1)}_{1} \\
&\quad\;\downarrow \quad &\downarrow & \quad &\quad \downarrow \quad \\
&\mathbb{\Lambda}_a:\left[\frac{5}{2}\right]_{\frac{3}{2}}^{(0,1)}\rightarrow & \mathbbm{V}_r: [1]_{\frac{3}{2}}^{(1)} \quad &\oplus &\quad \mathbbm{U}:[\frac{1}{2}]^{(0)}_{\frac{3}{2}} \\
&\quad\;\downarrow \quad &\downarrow & \quad &\quad  \quad \\
&\mathbb{D}:\left[3\right]_{2}^{(0,0)}\rightarrow & \mathbbm{D}:[1]_{2}^{(0)} \quad &\quad &\quad
\end{array}
\end{equation*}
where the vertical lines stand for the action of the preserved supercharges corresponding to the considered theory and, for notation convenience, we also included the corresponding operators ($a=1,2,3$ and $r=1,2$).
As in the order case, the four-point of the tilt and of the displacement (anti)chiral multiplet will be specified by the top components' four-point functions of $\bR$ ($\bar\bR$) and ${\mathbbm{F}}$ ($\bar{\mathbbm{F}}$) respectively.

We now investigate the presence of the universal sector in both the displacement and tilt four-point functions. Focusing first on the displacement supermultiplet, one can verify exploiting the same arguments of Sec. \ref{subsec: Correlators of the displacement multiplet} that the displacement sector does not mix with the tilt sector and, moreover, that the individual defect operators within the multiplet do not mix among themselves. It follows that the corresponding four-point function exhibits universal features. Furthermore, since the algebraic structure coincides with that of the $1/2$ BPS line in $\osp(4\vert 4)$, the analysis leads to the same expressions of \cite{Pozzi:2024xnu}. 

We now turn to the analysis of the four-point function of the tilt supermultiplet, which is determined by that of its superprimary
\begin{equation}
\label{eq: 1/3 BPS 4point function}
    \langle \mathbbm{F}(t_1)\bar{\mathbbm{F}}(t_2)\mathbbm{F}(t_3)\bar{\mathbbm{F}}(t_4)\rangle = \frac{C_{\cT}^2}{t_{12}t_{34}}f(z)\,,
\end{equation}
where $C_\cT$ denotes the two-point normalization of the tilt supermultiplet, which is distinct from that of the displacement multiplet. Nevertheless, if the theory is controlled by a single perturbation parameter describing deviations from the GFF, one may still expect $C_\cT$ to be related to the two-point normalization of the displacement supermultiplet, as indeed is the case \cite{Drukker:2022txy}.
The relevant selection rules follow from the same argument presented in Sec. \ref{subsec: OPEs and block expansions} for \eqref{eq: chiral-antichiral ope}, thus leading to
\begin{equation*}
    L\bar A[\ft{1}{2}]_{\frac{1}{2}}^{(0)}\times A\bar L[-\ft{1}{2}]_{\frac{1}{2}}^{(0)}\sim \cI+LL[0]_\Delta^{(0)}\,.
\end{equation*}
In this case we must consider the mixing also with the displacement sector. This can be again studied by looking at mixed four-point functions of tilt and the superdisplacement insertions. In order to show that multiparticle operators do not enter one can look, for example, at
\begin{equation}
    \langle \bD(t_1)\bar\bD(t_2)\bF(t_3)\bar\bF(t_4)\rangle^{(1)} \implies \lambda_{\bD\bar\bD[\bD\bar\bD]^{L=2}_M}^{(0)} \lambda_{\bF\bar\bF[\bD\bar\bD]^{L=2}_M}^{(1)}\sim \cO(\epsilon)\,,
\end{equation}
and similarly at the other mixed contributions. At this level of discussion, one could in principle introduce distinct expansion parameters, not merely related by overall factors but characterizing different deviations from the GFF in the various sectors. For simplicity, we assume that all these parameters are related, \eg through additional multiplicative factors, so that deviations in all sectors are treated on the same footing, thereby avoiding configurations in which part of the correlators are expanded while others remain at the GFF level.

We can thus identify the universal four-point function and extract the leading and next-to-leading OPE data. In this case, the expansion up to NLO reads
\begin{align}
\label{eq: four-point tilt}
    \hspace{-.4cm}\langle \mathbbm{F}(t_1)\bar{\mathbbm{F}}(t_2)\mathbbm{F}(t_3)\bar{\mathbbm{F}}(t_4)\rangle=\frac{C_{\cT}^2}{t_{12}t_{34}}\Bigl[1-z+ \frac{1}{4\pi^2 C_{\cT}}\Big(z-1+z (3-z) \log (-z)-\frac{(1-z)^3 }{z}\log (1-z)\Big)\Bigr]\,.
\end{align}
This is equivalent to the expansion for the superprimary of the $1/2$ BPS Wilson line in ABJM \cite{Bianchi:2020hsz}.
As in the previous case, the expansion is in terms of superblocks. However, as mentioned above, given the similarity with the lines in $\cN=4$ CSm theories, the supercasimir equation will be essentially the one discussed in \cite{Pozzi:2024xnu}, so one can exploit the relations defined therein. Therefore we consider the expansion
\begin{equation*}
   f(z)=1+\sum_{\Delta>0}c_\Delta G_\Delta(z)\qquad\text{with}\qquad G_\Delta(z)=(-z)^\Delta \,_2F_1(\Delta,\Delta,2\Delta+2)
\end{equation*}
and, from \eqref{eq: four-point tilt}, we can extract the leading-order data, thus obtaining
\begin{equation*}
    c^{(0)}_n = \frac{2^{-2(1+n)}\sqrt{\pi}\,(2+n)\Gamma(n+1)}{\Gamma(n+\ft{3}{2})}\,.
\end{equation*}
As before, we characterize the supermultiplets contributions by $n$, \ie the total number of derivatives of the multiparticle operators $[\mathbbm{F}\bar{\mathbbm{F}}]_n$, with the corresponding dimension given by $\Delta = 1 + n$.
The NLO order data can be obtained again from the $\log(-z)$ contribution of \eqref{eq: four-point tilt}. The anomalous dimensions read
\begin{equation*}
    \gamma^{(1)}_n=-3-3n-n^2
\end{equation*}
and the OPE data at NLO are found to be
\begin{equation*}
    c^{(1)}_n= c^{(0)}_n\Bigl[ -3-2n+\gamma_n^{(1)}\Bigl( \psi(n+1)-\psi(n+\ft{3}{2})-2\log2+ \frac{1}{n+2} \Bigr)\Bigr]\,,
\end{equation*}
consistently with the general expression \eqref{eq: general OPE NLO}.

We end this discussion with a last analysis on the other four-point function that can be studied which is
\begin{equation*}
    \langle \mathbbm{F}(t_1)\bar{\mathbbm{F}}(t_2)\bar{\mathbbm{F}}(t_3){\mathbbm{F}}(t_4)\rangle=\frac{C_{\cT}^2}{t_{12}t_{34}}h(\chi)\,.
\end{equation*}
Up to NLO this is given by
\begin{equation}
\label{eq: four-point tilt chi}
        \langle \mathbbm{F}(t_1)\bar{\mathbbm{F}}(t_2)\bar{\mathbbm{F}}(t_3){\mathbbm{F}}(t_4)\rangle=\frac{C_{\cT}^2}{t_{12}t_{34}}\Bigl[1-\chi+ \frac{1}{4\pi^2 C_{\cT}}\Big(\chi-1+\chi (3-\chi) \log (\chi)-\frac{(1-\chi)^3 }{\chi}\log (1-\chi)\Big)\Bigr]\,
\end{equation}
and it is related by the mapping $z\rightarrow\chi$ to the other four-point function \eqref{eq: four-point tilt}.
As discussed in Sec. \ref{sec: The 1/2 BPS line in N=2 Chern-Simons matter theories}, these correlation functions admit two distinct channels. For the chiral-antichiral expansion, the analysis is equivalent to the discussion presented above, whereas in the chiral-chiral channel one must consider the OPE\footnote{See \cite{Pozzi:2024xnu} for further details}
\begin{equation*}
    L\Bar{A}[\ft{1}{2}]_{\frac{1}{2}}^{(0)}\times L\Bar{A}[\ft{1}{2}]_{\frac{1}{2}}^{(0)}\sim L\Bar{A}[1]_{1}^{(0)}+L\Bar{A}[1]^{(1)}_{\frac{3}{2}}+L\Bar{L}[1]^{(0)}_{\Delta}.
\end{equation*}
One can see that the quantum numbers of $L\Bar{A}[1]_{1}^{(0)}$ are consistent with those of the displacement supermultiplet hence, in principle, it could appear in the exchange. However, this expansion must be related by an analogous relation of \eqref{eq: braiding corr} to the one where the second and fourth insertions are swapped, \ie braiding. Thus, consistency with \eqref{eq: four-point tilt} and \eqref{eq: four-point tilt chi} allows one to exclude the superdisplacement in the possible exchanges. One can see this explicitly by considering the expansion up to NLO 
\begin{equation}
\label{eq: four-point titl chiral-chiral}
    \langle \mathbbm{F}(t_1)\bar{\mathbbm{F}}(t_2)\bar{\mathbbm{F}}(t_3){\mathbbm{F}}(t_4)\rangle=\frac{C_{\cT}^2}{t_{14}t_{23}}\Bigl(h^{(0)}(1-\chi)+ \frac{1}{4\pi^2 C_{\cT}}h^{(1)}(1-\chi)+\text{h}^{(1)}(1-\chi)\Bigr)\,,
\end{equation}
where, as discussed for \eqref{eq: f1(z) NLO with Z2 odd}, we collected all the contributions controlled by the three-point exchanges in $\text{h}^{(1)}(1-\chi)$. 
If now one considers the braiding relation for \eqref{eq: four-point titl chiral-chiral}, one would obtain an extra term, related to $\text{h}^{(1)}(1-\chi)$, which is not present in \eqref{eq: four-point tilt chi}, and thus leading to the conclusion that this extra term must vanish. 

More specifically, another way to verify that the superdisplacement components do not contribute in this four-point function is by studying the three-point function of two tilt superprimaries and one component of $\cD$ ($\bar\cD$). First of all, we can exclude the appearance of $\mathbb{V}_r$ given that
\begin{equation*}
    \langle \mathbbm{F}(t_1)\mathbbm{F}(t_2) \bar{\mathbbm{V}}^r(t_3)\rangle=0\,,
\end{equation*}
by symmetry preserving argument. The other nontrivial three-point functions are given by
\begin{equation*}
    \langle \mathbbm{F}(t_1)\mathbbm{F}(t_2) \bar{\mathbbm{R}}(t_3)\rangle\,,\qquad  \langle \mathbbm{F}(t_1)\mathbbm{F}(t_2) \bar{\mathbbm{D}}(t_3)\rangle\,.
\end{equation*}
Requiring invariance under the $\mathbb{Z}_2$ parity transformation $t \rightarrow -t$ imposes a selection rule on the parity of the third operator in the three-point function containing two fermions. This is essentially what has been discussed in \eqref{eq: parity of the only OPEs}, which applied to the present case leads to the expression 
\begin{equation*}
    \lambda_{\mathbbm{F} \mathbbm{F} \bT_\Delta}=(-1)^{P_{\bT}+1}\lambda_{\mathbbm{F} \mathbbm{F} \bT_\Delta}\,.
\end{equation*}
This constrains the three-point function to be nonvanishing only for parity-odd contributions, as also observed in Sec. \ref{sec: The 1/2 BPS line in N=2 Chern-Simons matter theories}. Specializing to the cases $\bT_\Delta = \bar{\mathbbm{R}}$ and $\bT_\Delta = \bar{\mathbbm{D}}$, which both have even parity, this immediately implies the vanishing of the corresponding three-point function coefficients, thus obtaining
\begin{equation*}
    \lambda_{\mathbbm{F} \mathbbm{F} \bar{\mathbbm{R}}}=0\,,\qquad  \lambda_{\mathbbm{F} \mathbbm{F} \bar{\mathbbm{D}}}=0\,,\qquad  \lambda_{\mathbbm{F} \mathbbm{F} \bar{\mathbbm{V}}}=0\,.
\end{equation*}
By recalling that these three-point coefficients also control the associated mixed multiparticle states, setting them to zero removes all such additional contributions, as discussed in Sec. \ref{subsubsec: Z2 odd sectors}. 
We thus find consistency with the results obtained from the braiding and we can conclude that the superdisplacement contributions are not exchanged.

It should be noted, however, that this discussion does not, in principle, exclude all supermultiplets with the quantum numbers of $L\Bar{A}[1]_{1}^{(0)}$. Potential contributions associated with the composite operators of the tilt, encoded in $h^{(0)}(1-\chi)$ and $h^{(1)}(1-\chi)$, could in principle appear. Nevertheless, by arguments analogous to those discussed above, these contributions are absent, as they correspond to parity-even sectors; see Sec. \ref{sec: The 1/2 BPS line in N=2 Chern-Simons matter theories} and \cite{Bianchi:2020hsz}.

One can thus proceed to extract the conformal data. This analysis leads to results equivalent to those obtained in the chiral-chiral channel for the superprimary of the $1/2$ BPS Wilson line in ABJM \cite{Bianchi:2020hsz}.

Under similar considerations, this analysis could be extended to the study of all other four-point functions involving displacement, tilt and tlit multiplets. We do not pursue this here and simply note it as a possible direction for further study.

\subsection*{Acknowledgments}
I am grateful to Lorenzo Bianchi, Eduardo Casali and Andrea Mattiello for discussions and to  Diego Trancanelli for discussions and comments on the draft.
I would like to thank the organizers of the COST CA22113 meeting: "Recent Developments in Quantum Field Theory", where part of this work was presented and in particular Kiril Hristov also for hosting my visit at Sofia University. I thank COST Action CA22113 and BNSF, grant “Competition for Financial Support for Basic Research Projects - 2024”, for partial support.
\appendix

\section{The $\osp(2\vert 4)$ algebra and its $1/2$ BPS subalgebra}
\label{app:algebra}

We collect here details about our conventions, the symmetries of the the bulk $\cN=2$ theory and the one preserved and broken by the $1/2$ BPS Wilson lines, working in Euclidean space $\mathbb{R}^3$. Some results about the defect algebra were already obtained \cite{Bianchi:2018scb} for the $1/6$ BPS Wilson line in ABJM. In this case we fix our conventions and explicitly show how the breaking pattern from the $\cN=2$ CSm theory occurs.

The three-dimensional $\cN=2$ superconformal algebra is $\mathfrak{osp}(2\vert 4)$. Its bosonic subalgebra consists of the three-dimensional conformal algebra $\mathfrak{so}(1,4)$ and of the R-symmetry algebra $\mathfrak{so}(2)_{R}$. The conformal generators are the rotations $M^{\mu\nu}$, the translations $P^\mu$, the special conformal transformations $K^\mu$ and the dilations $D$, with $\mu,\nu=0,1,2$ and algebra given by
\bea
&&[M^{\mu\nu},M^{\rho\sigma}]=\delta^{\sigma\mu}M^{\nu\rho}-\delta^{\sigma\nu}M^{\mu\rho}+\delta^{\rho\nu}M^{\mu\sigma}-\delta^{\rho\mu}M^{\nu\sigma},\cr
&&[P^\mu,M^{\nu\rho}]=\delta^{\mu\nu}P^{\rho}-\delta^{\mu\rho}P^{\nu},\cr
&&[K^\mu,M^{\nu\rho}]=\delta^{\mu\nu}K^{\rho}-\delta^{\mu\rho}K^{\nu},\cr
&&[P^\mu,K^\nu] = -2\delta^{\mu\nu}D-2 M^{\mu\nu},\cr
&&[D,P^\mu]=P^\mu,\qquad [D,K^\mu]=-K^\mu.
\eea
The R-symmetry generator is $R_{IJ}=-R_{JI}$, with $I,J=1,2$, and obey
\bea
[R_{IJ},R_{KL}] = 0\,,
\eea
as expected from $\so(2)$.
The fermionic generators $Q_{I\alpha}$ and $S^\alpha_I$, with spinorial indices $\alpha=\pm$, satisfy
\bea
\label{fermionic-anticom}
&& \{Q_{I\alpha}, Q_{J\beta}\} = 2i\delta_{IJ} (\gamma^\mu)_{\alpha\beta} P_\mu, \cr
&& \{S_I^\alpha, S_J^\beta\} = 2i\delta_{IJ} (\gamma^\mu)^{\alpha\beta} K_\mu, \cr
&& \{Q_{I\alpha}, S_J^\beta\} = \delta_{IJ} \left((\gamma^{\mu\nu})_\alpha{}^\beta M_{\mu\nu} + 2\delta_\alpha^\beta  D\right) + 2\delta_\alpha^\beta R_{IJ}, 
\eea
with $(\gamma^{\mu})_\alpha{}^\beta$ being the Pauli matrices satisfying $\{\gamma^\mu,\gamma^\nu\}=2\delta^{\mu\nu}$ and $\gamma^{\mu\nu}= \frac{1}{2}(\gamma^\mu\gamma^\nu-\gamma^\nu\gamma^\mu)=i\epsilon^{\mu\nu\rho}\gamma^\rho$.

The remaining commutation relations are
\bea
[D,Q_{I\alpha}]=\frac{1}{2} Q_{I\alpha},\qquad &&\qquad [D,{S_I}^\alpha]=-\frac{1}{2}{S_I}^\alpha, \cr
[M^{\mu\nu},Q_{I\alpha}]=-\frac{1}{2}(\gamma^{\mu\nu})_\alpha^\beta Q_{I\beta}, \quad \quad && \quad \quad [M^{\mu\nu},{S_I}^\alpha]=\frac{1}{2}(\gamma^{\mu\nu})^\alpha_{\,\beta} S_I^\beta,\cr
[K^{\mu},Q_{I\alpha}]=i(\gamma^{\mu})_{\alpha\beta} {S_I}^\beta, \quad \quad && \quad \quad [P^{\mu},{S_I}^\alpha]=-i(\gamma^{\mu})^{\alpha\beta} Q_{I\beta}\cr
[R_{IJ},Q_{K\alpha}] = \delta_{IK}Q_{J\alpha}-\delta_{JK}Q_{I\alpha},\quad\quad && \quad\quad [R_{IJ},S_{K}^\alpha] = \delta_{IK}S_{J}^\alpha-\delta_{JK}S_{I}^\alpha,
\eea
where spinorial indexes are raised/lowered with $\epsilon^{\alpha\beta}$ and $\epsilon_{\alpha\beta}$, such that $ \epsilon^{+-}=\epsilon_{-+} = 1$. Contrary to other more supersymmetric cases, \textit{e.g.} \cite{Bianchi:2020hsz,Pozzi:2024xnu}, the $1/2$ BPS Wilson line, thus breaking half of the supercharges, does not break the $\so(2)$ R-symmetry. First of all, in order to meet the definitions of \cite{Agmon:2020pde}, we define the combinations
\begin{equation*}
    {}^{\pm}Q_{\alpha}=\frac{1}{\sqrt{2}}\Bigl({{Q_{1\alpha}}} \pm i\, {Q_{2\alpha}} \Bigr)\,,\qquad {}^{\pm}{{S^{\alpha}}}=\frac{1}{\sqrt{2}}\Bigl(S^{\alpha}_{1} \pm i\, S^{\alpha}_{2} \Bigr)\,,
\end{equation*}
where, to avoid any confusion with spinorial indexes, we use an up-left index convention for the particular combinations. 
In terms of ${}^{\pm}Q_{\alpha}$ and ${}^{\pm}S^{\alpha}$ the anti-commutation relations above now read 
\bea
&& \{{}^{\pm}Q_{\alpha}, {}^{\mp}Q_{\beta}\} = 2i (\gamma^\mu)_{\alpha\beta} P_\mu, \cr
&& \{{}^{\pm}S^{\alpha}, {}^{\mp}S^{\beta}\} = 2i (\gamma^\mu)^{\alpha\beta} K_\mu, \cr
&& \{{}^{\pm}Q_{\alpha}, {}^{\mp}S^{\beta}\} =  \left((\gamma^{\mu\nu})_\alpha{}^\beta M_{\mu\nu} + 2\delta_\alpha^\beta  D\right) \mp 2\delta_\alpha^\beta R, 
\eea
where now we defined $R=i\,R_{12}$ such that 
\begin{equation*}
    [R,{}^{\pm}Q_{\alpha}]=\pm {}^{\pm}Q_{\alpha}\,,\qquad [R,{}^{\pm}S^{\alpha}]= \pm{}^{\pm}S^{\alpha}
\end{equation*}

The insertion of a 1/2 BPS Wilson line breaks the $\mathfrak{osp}(2\vert 4)$ of the bulk theory down to $\mathfrak{su}(1,1|1)$. The $\mathfrak{su}(1,1)$ generators are those of the one-dimensional conformal group, {\it i.e.} $\{ D,P\equiv P_0,K\equiv K_0\} $, satisfying
\begin{equation}
 [P,K]=-2 D, \qquad [D,P]=P, \qquad [D,K]=-K.
\end{equation}
As discussed, the R-symmetry $\so(2)\simeq\mathfrak{u}(1)$ is fully preserved. As in \cite{Pozzi:2024xnu} we choose $(\gamma^{\mu})_{\alpha}^{\;\;\beta} = (\sigma^{z},\sigma^{x}, \sigma^{y})_{\alpha}^{\;\;\beta}$ as a basis and therefore $(\gamma^{\mu})_{\alpha\beta}=(\sigma^1, -\sigma^3, i\mathbb{1})$ and $(\gamma^{\mu})^{\alpha\beta}=(-\sigma^1, \sigma^3, i\mathbb{1})$. From the anticommutation relations one can identify the preserved fermionic charges, which are
\begin{equation}
\label{eq: preserved supercharges}
 Q\equiv {}^{+}Q_{+}, \qquad S\equiv i\,{}^{+}S^{-},  \qquad  \bar{Q} \equiv -i\,{}^{-}Q_{-}, \qquad  \bar{S} \equiv  {}^{-}S^{+},
\end{equation}
which obey the anticommuation relations
\begin{align}
 \{Q , \bar{Q}\} &= 2 P ,\qquad  \{S, \bar{S}\} = 2 K,\cr
 \{Q,\bar{S}\}&= 2\left( D -J_{0} \right),\cr
 \{\bar{Q},S\}&= 2\left(D + J_{0} \right) ,
\end{align}
with 
\bea
\label{eq: J0 charge in N=2 SCSM line}
 J_{0}= {R}-i M_{12}\,,
\eea
generating the $\u(1)_{j_0}$ obtained by combining the contributions of the preserved rotations $\so(2)_{\text{rot}}$ and the $\so(2)$ R-symmetry.
The mixed bosonic/fermionic commutation relations are
\begin{align}
\label{mixedFB1}
[D,{Q}]&=\frac{1}{2}{Q},\quad  \quad [D,{\bar{Q}}]=\frac{1}{2}{\bar{Q}}, \quad \quad [D,{S}]=-\frac{1}{2}{S},\quad  \quad [D,\bar{S}]=-\frac{1}{2}\bar{S},\cr
[K,{Q}]&={S},\quad  \quad [K,{\bar{Q}}] ={\bar{S}}, \quad \quad [P,{S}]=-{Q},\quad  \quad [P,\bar{S}]=-\bar{Q},
\end{align}
and
\begin{align}
\label{mixedFB2}
[{R}, Q] &=Q ,\qquad  [M^{12}, Q] = -i\frac{1}{2} Q, \qquad [J_{0},Q]=\frac{1}{2}Q \ec
\cr
[{R}, S] &=S,\qquad  [M^{12}, S] = -i\frac{1}{2} S ,\qquad [J_{0},S]=\frac{1}{2}S \ec
\cr
[R, \bar{Q}] &=- \bar{Q},\qquad  [M^{12}, \bar{Q}] = i\frac{1}{2} \bar{Q} ,\qquad [J_{0},\bar{Q}]=-\frac{1}{2}\bar{Q} \ec
\cr
[R, \bar{S}] &=-\bar{S},\qquad  [M^{12}, \bar{S}] = i\frac{1}{2} \bar{S}^{a} ,\qquad [J_{0},\bar{S}]=-\frac{1}{2}\bar{S},
\end{align}
finding complete agreement with the analysis in \cite{Bianchi:2018scb}, where also the classification of representations is reported. We then move to the identification of quadratic Casimir of the algebra, which is given by
\begin{equation}
\label{Casimir diff}
    \mathfrak{C}=D^2-\frac{1}{2}\{K,P\}+\frac{1}{4}[\Bar{S},Q]+\frac{1}{4}[S,\Bar{Q}]-J_0^2\;.
\end{equation}
When acting on highest weight states of the $\su(1,1\vert 1)$ it has eigenvalue
\begin{equation}
    \mathfrak{c}_{\Delta,j_0}=\Delta^2-j_0^2\,.
\end{equation}
With the (anti)commutation relations in hand, we can define the differential action of the preserved generators which is given by
\begin{align}
    & P = -\partial_t, \qquad
    D = -t \partial_t-\frac{1}{2}\theta\deltetha-\frac{1}{2}\Bar{\theta}\deltethabar-\Delta\;,\cr
    & K = -t^2 \partial_t -t\theta \deltetha-t\Bar{\theta}\deltethabar-2 t\Delta-2 j_0\theta\bar\theta\;,\cr
    & Q = \deltetha-\Bar{\theta}_a\partial_t, \qquad 
    \Bar{Q} = \deltethabar-\theta\partial_t\;,\cr
    & S = (t+\theta\Bar{\theta})\deltetha- t\theta\Bar{\theta}\partial_t-2(\Delta-j_0)\Bar{\theta}\;,\cr
    & \Bar{S} = (t-\theta\Bar{\theta})\deltethabar-t\theta\Bar{\theta}\partial_t-2(\Delta+j_0){\theta}\;,\cr
    & J_0=\frac{1}{2}(-\theta\deltetha+\bar\theta\deltethabar+j_0)\,,\cr
    & R= -\theta\deltetha+\Bar{\theta}\deltethabar,
\end{align}
where $\partial_t=\frac{\partial}{\partial t}$, $\partial_\theta=\frac{\partial}{\partial\theta}$ and similarly $\partial_{\bar\theta}=\frac{\partial}{\partial\bar\theta}$.
\subsection{Blocks and superblocks}
\label{app: blocks and superblocks properties}
Superblocks are obtained by solving the supercasimir equations
\begin{equation}
\label{casimir equation}
\left( \mathfrak{D}_{1,2} - \mathfrak{c}_{\Delta,0} \right)\langle \cD(y_{1},\theta_{1}) \bar{\cD}(\bar y_{2},\bar \theta_{2}) \cD(y_{3},\theta_{3}) \bar{\cD}(\bar y_{4},\bar \theta_{4})\rangle=0\,,
\end{equation}
where the Casimir is defined as acting on the first and second insertion by
\begin{equation}
\label{Casimir diff}
    \mathfrak{D}_{1,2}=D_{\text{s}}^2-\frac{1}{2}\{K_{\text{s}},P_{\text{s}}\}+\frac{1}{4}[\Bar{S}_{\text{s}},Q_{\text{s}}]+\frac{1}{4}[S_{\text{s}},\Bar{Q}_{\text{s}}]\;,
\end{equation}
where the operators are taken as the sum, \eg $D_s=D_1+D_2$ with $D_i$ acting on the $i$-th insertion; see \cite{Pozzi:2024xnu, Fitzpatrick:2014oza}. Given that only chargeless long multiplets are exchanged in this channel, the eigenvalue is $\mathfrak{c}_{\Delta,0} =\Delta^2$.
By expanding in the Grassmann variables, one  can identify the second-order differential equation
\begin{equation}
\label{differential equation}
z (z-1)\Bigl(\partial_zF_{\Delta}(z)+z\, \partial_z^2F_{\Delta}(z)\Bigr) = \Delta^2 F_{\Delta}(z)\;,
\end{equation}
which, neglecting the shadow contributions, is solved by the superblocks
\begin{equation}
   G_\Delta(z)= (-z)^{\Delta } \, _2F_1(\Delta ,\Delta ,2 \Delta +1;z)\,.
\end{equation}
The orthogonality are given by the expressions
\begin{equation}
\label{eq: orthogonality}
    \oint \frac{dz}{2\pi i}\omega(z) G_{n}(z)G_{-m}(z) = \delta_{n,m},
    \qquad \oint \frac{d\chi }{2\pi i}\rho(\chi) g_{n}(1-\chi)\,g_{1-m}(1-\chi) = \delta_{n,m}\,,
\end{equation}
where we also included the relation for the conformal block used in \ref{subsec: Extracting the conformal data}. The densities are $\omega(z) = \tfrac{1}{(1-z)z}$ and $\rho(\chi) = -\tfrac{1}{(1-\chi)^2}$ and the integration contours wind counterclockwise around $z=0$ and $\chi=1$.

\bibliographystyle{utphys2}
\bibliography{refs.bib}

\end{document}